\documentclass{aa} 
\usepackage{graphicx}
\usepackage{txfonts}
\begin{document}
   \title{A VLT/NACO Survey for Triple and Quadruple Systems among Visual Pre-Main Sequence Binaries\thanks{Based on observations collected at the  European Southern Observatory, Chile, program numbers 070.C-0701 and 073.C-0379.}}

   \subtitle{}

   \author{S. Correia\inst{1}
          \and
          H. Zinnecker\inst{1}
          \and 
          Th. Ratzka\inst{2}
          \and 
          M.F Sterzik\inst{3}
          }

   \offprints{S. Correia}

   \institute{Astrophysikalisches Institut Potsdam, An der Sternwarte 16, D-14482 Potsdam, Germany\\
              \email{scorreia@aip.de, hzinnecker@aip.de}
         \and
             Max-Planck-Institut f\"{u}r Astronomie, K\"{o}nigstuhl 17, D-69117 Heidelberg, Germany\\
             \email{ratzka@mpia.de}
         \and
             European Southern Observatory, Casilla 19001, Santiago 19, Chile\\
             \email{msterzik@eso.org}
             }

   \date{Received 5 May 2006 / Accepted 28 August 2006}

\abstract
{}
{This paper describes a systematic search for high-order multiplicity among wide visual Pre-Main Sequence (PMS) binaries.
}
{We conducted an Adaptive Optics survey of a sample of 58 PMS wide binaries from various star-forming regions, 
which include 52 T Tauri systems with mostly K- and M-type primaries, with the NIR instrument NACO at the VLT.
}
{Of these 52 systems, 7 are found to be triple (2 new) and 7 quadruple (1 new). The new close companions are most likely 
physically bound based on their probability of chance projection and, for some of them, on their position on a color-color diagram.
The corresponding degree of multiplicity among wide binaries (number of triples and quadruples divided by 
the number of systems) is 26.9\,$\pm$7.2\% in the projected separation range $\sim$\,0\farcs07-12$^{\prime\prime}$, 
with the largest contribution from the Taurus-Auriga cloud. We also found that this degree of multiplicity is twice in Taurus 
compared to Ophiuchus and Chamaeleon for which the same number of sources are present in our sample.  
Considering a restricted sample composed of systems at distance 140-190\,pc, the degree of multiplicity is  
26.8\,$\pm$8.1\%, in the separation range 10/14 AU - 1700/2300 AU (30 binaries, 5 triples, 6 quadruples). 
The observed frequency agrees with results from previous multiplicity surveys within the uncertainties, 
although a significant overabundance of quadruple systems compared to triple systems is apparent.
Tentatively including the spectroscopic pairs in our restricted sample and comparing the multiplicity fractions to those 
measured for solar-type main-sequence stars in the solar neighborhood leads to the conclusion that both the ratio of triples 
to binaries and the ratio of quadruples to triples seems to be in excess among young stars.
Most of the current numerical simulations of multiple star formation, and especially smoothed particles hydrodynamics simulations, 
over-predict the fraction of high-order multiplicity when compared to our results. 
The circumstellar properties around the individual components of our high-order multiple systems tend to favor mixed systems 
(i.e. systems including components of wTTS and cTTS type), which is in general agreement with previous studies of disks in binaries, 
with the exception of Taurus, where we find a preponderance of similar type of components among the multiples studied. 
}
{}

\titlerunning{A VLT/NACO Survey for Triple and Quadruple PMS Systems}
\authorrunning{S. Correia et al.}

   \maketitle


\section{Introduction}
It has long been recognized that the formation of binary or multiple stars is an efficient way to solve one of the most prominent problems in 
the theory of star formation, namely the "angular momentum problem" (Mestel \& Spitzer\,\cite{Mestel_Spitzer1956}, Larson\,\cite{Larson_1972}, 
Mouschovias\,\cite{Mouschovias_1977}, Simon et al.\,\cite{Simon_etal1995}, Bodenheimer\,\cite{Bodenheimer_1995}, Larson\,\cite{Larson_2002}).
Yet, although we have a general understanding of the formation of single low-mass stars (e.g. Shu et al.\,\cite{Shu_etal_1987}), the theory of the origin 
of binary and multiple low-mass stars is not quite settled (for a review see Tohline\,\cite{Tohline_2002}). 
The favoured mechanism for the formation of binary and multiple stars is the collapse and fragmentation of molecular cloud cores, 
either fragmentation during dynamical collapse (e.g., Boss\,\cite{Boss_1986}) or from a quasistatic situation (i.e. disk fragmentation e.g., Bonnell\,\cite{Bonnell_1994}) 
or filament fragmentation (e.g. Zinnecker\,\cite{Zinnecker_1991}, Bonnell \& Bastien\,\cite{Bonnell_Bastien1991}). Prompt fragmentation following 
the collision of two Jeans masses has also been suggested (Pringle\,\cite{Pringle_1991}). 
In addition, further processes likely to occur during the formation of binary and multiple stars like circumbinary disk accretion,
interaction with circumstellar disks and dynamical interactions are nowadays being investigated in detail through numerical
simulations (Bate et al.\,\cite{Bate_etal2003}, Sterzik \& Durisen\,\cite{Sterzik_Durisen2003}, \cite{Delgado_Donate_etal2004}, Goodwin et al.\,\cite{Goodwin_etal2004}). 
The scope of these studies is mainly to obtain statistical properties of the stellar systems resulting from these evolutionary calculations 
for direct comparison with  observations; such as, relative frequency of single, binary and multiple stars, the properties of multiple stars 
and brown dwarfs, and the initial mass function.

Several multiplicity surveys among T Tauri stars using high-angular resolution techniques have been carried out over the last 
decade, with the main outcome that binary stars are much more common in several of the young star-forming T-associations 
compared to the main-sequence stars in the solar-neighborhood (Leinert et al.\,\cite{Leinert_etal1993}, Ghez et al.\,\cite{Ghez_etal1993}, Ghez et al.\,\cite{Ghez_etal1997a}, 
Duch\^{e}ne\,\cite{Duchene_1999}). 
In addition, it seems that a comparable or even higher frequency of binary and multiple stars is found at earlier evolutionary stages 
(e.g. Reipurth 2000, Duch\^{e}ne et al.\,\cite{Duchene_etal_ppv} and references therein). 
Paradoxically, although studies of high-order multiple star (N$>$2) systems might provide important constraints about the process of 
multiple star formation (Batten\,\cite{Batten_1973}, Fekel\,\cite{Fekel_1981}, Sterzik \& Tokovinin\,\cite{Sterzik_Tokovinin_2002}, 
Tokovinin \& Smekhov\,\cite{Tokovinin_Smekhov_2002}, see also Tokovinin\,\cite{Tokovinin_2001}), 
relatively little has been done observationally in a systematic way regarding higher-order multiplicity among Pre-Main Sequence (PMS) stars. An exception is 
Koresko (\cite{Koresko_2002}) who observed the close environment of 14 PMS binaries in southern star-forming regions with high-angular 
resolution techniques and found that half of them are in fact hierarchical triples or present resolved circumstellar structures consistent with 
marginally resolved companions. 
In the last years, an increasing number of high-angular resolution observations reports the discovery of new companions in known multiple 
T Tauri systems. Examples are T Tau itself (Koresko\,\cite{Koresko_2000}, K\"{o}hler et al.\,\cite{Koehler_etal2000a}), VW Cha (Brandeker et al. \cite{Brandeker_etal2001}), 
V773 Tau (Duch\^{e}ne et al.\,\cite{Duchene_etal2002}), AS 353 (Tokunaga et al.\,\cite{Tokunaga_etal2004}). 
Some of these systems present close to edge-on circumstellar disks, more easily detectable in multiple systems and with current imaging techniques like AO. 
The only known cases of such disks are HK Tau B (Stapelfeldt et al.\,\cite{Stapelfeldt_etal1998}, 
Koresko\,\cite{Koresko_1998}), HV Tau C (Monin \& Bouvier\,\cite{Monin_Bouvier_2000}, Stapelfeldt et al.\,\cite{Stapelfeld_etal2003}), 
and LkH$\alpha$ 263\,C (Chauvin et al.\,\cite{Chauvin_etal2002}, Jayawardhana et al.\,\cite{Jayawardhana_etal2002}). 
Also, it has been noted that approximately half of the known T Tauri spectroscopic binaries are in 
fact hierarchical triples with an outlier companion (Melo\,\cite{Melo_2003}, Sterzik et al.\,\cite{Sterzik_etal2005}).
All the above suggests that with the current observational techniques we are able to reveal a large enough number of high-order multiple 
PMS stars to allow for meaningful statistical analysis of their occurrence and properties.
 
We report here about a high-angular resolution survey which focuses on the N$>$2 multiplicity among T Tauri stars. 
We searched in a sample of 58 wide PMS binaries extracted from the list of Reipurth \& Zinnecker (\cite{Reipurth_Zinnecker1993},
hereafter RZ93) to discover additional close companions using the Adaptive Optics (AO) near-infrared instrument NAOS/CONICA 
(NACO, Rousset et al.\,\cite{Rousset_etal2002}) attached to the telescope UT4/Yepun of the Very Large Telescope (VLT).
The rationale behind this effort was the following\,: if one considers a wide 2\,arcsec PMS binary (corresponding to $\sim$\,300\,AU at 150\,pc, 
the distance of the most prominent star forming regions), then a close ($\sim$\,0.1-0.2\,arcsec) companion could still exist without violating the hierarchical 
triple system stability criterion (Harrington\,\cite{Harrington_1975}, Eggleton  \& Kiseleva\,\cite{Eggleton_Kiseleva_1995}, Mardling \& Aarseth\,\cite{Mardling_Aarseth_2001}). 

This paper is organized as follows\,: in Sect.\,\ref{sect:sample} we present the sample of wide PMS binaries and the AO near-infrared 
observations obtained at VLT with NACO in Sect.\,\ref{sect:obs_data_red}. Candidate visual companions to these binaries detected in the images 
are presented in Sect.\,\ref{sect:results} and physical companions are identified on both colors and statistical grounds. This allows us, in Sect.\,\ref{sect:multiplicity_stat}, 
to derive the frequency of triples/quadruples systems among wide young binaries and to compare this quantity to previous surveys 
(Sect.\,\ref{sect:multiplicity_stat_compa_surveys}), to recent results of numerical simulations of multiple star formation (Sect.\,\ref{sect:multiplicity_stat_compa_theory}), 
as well as to what is found among main-sequence multiples (Sect.\,\ref{sect:MS_mult} and \ref{sect:total_mult}). 
We also discuss the properties of these triples/quadruples systems in terms of dynamical stability (Sect.\,\ref{sect:syst_stability}) 
and relative disk evolutions (Sect.\,\ref{sect:disk_evolution}). 
Finally, we summarize our main results in Sect.\,\ref{sect:summary}.


\begin{table*}
\scriptsize
\caption{Observed sample of wide PMS binaries.}
\begin{center}
\renewcommand{\arraystretch}{0.5}
\setlength\tabcolsep{5pt}
\begin{tabular}{l@{\hspace{1mm}}c@{\hspace{1mm}}l@{\hspace{2mm}}l@{\hspace{1mm}}r@{}l@{\hspace{3mm}}l@{\hspace{1mm}}r@{\hspace{2mm}}r@{\hspace{2mm}}r
@{\hspace{4mm}}r@{\hspace{2mm}}r@{\hspace{1mm}}c
@{\hspace{0mm}}c@{\hspace{2mm}}l@{\hspace{2mm}}l@{\hspace{0mm}}c}
\hline\noalign{\smallskip}

Name & HBC & Other & \multicolumn{1}{c}{R.A.} & \multicolumn{2}{c}{Decl.} &  Cloud & D & V & K & 
\multicolumn{2}{r}{RZ93$^c$} & add. &
SB & SpT1 & SpT2 & Ref  \\

 &  & name & \multicolumn{3}{c}{[J2000.0]} &   & [pc] & &  & 
 $\rho$ & PA & comp.$^b$ &
 & & & (SpT) \\
\noalign{\smallskip}
\hline
\noalign{\smallskip}
\object{LkH$\alpha$ 262/263} ..      &  8+9 & 		& 02 56 08.4 & +   & 20 03 40  &  MBM 12  	&    275   & 14.6  &  9.5  & 4.1 & 57 & X &  & M0 & M0+M4 & 1 \\
\noalign{\smallskip}
\object{J 4872} ...............      		  & 	& 		& 04 25 17.6 & +   & 26 17 51  &  Tau-Aur  	&    142   & 13.0  &  8.6  & 3.5 & 231 & X &  & K9 & M1 & 2 \\
\noalign{\smallskip}
\object{FV Tau} .............        		  & 386+387 & 		& 04 26 53.6 & +   & 26 06 55  &  Tau-Aur  &    142   & 15.4  &  7.4   & 0.7 & 289 & X &  & K5+K6 & M3+M5 &  1 \\
\noalign{\smallskip}
\object{UX Tau} .............        		  & 43+42 & 		& 04 30 04.0 & +   & 18 13 49  &  Tau-Aur  	&    142   & 10.7  &  7.6    & 2.7 & 270 & X &  & K4 & M2+M3 &  1 \\
\noalign{\smallskip}
\object{DK Tau} .............        		  & 45 & 		& 04 30 44.3 & +   & 26 01 25  &  Tau-Aur &    142   & 12.6  &  7.1    & 2.4 & 119 & &  & K7/K9 & K7/M1 & 3,2 \\
\noalign{\smallskip}
\object{HK Tau} .............        		  & 48 & 		& 04 31 50.6 & +   & 24 24 18  &  Tau-Aur  &    142   & 15.0  &  8.6    & 2.4 & 172 & &  & M1 & M2 & 4, 2  \\
\noalign{\smallskip}
\object{LkH$\alpha$ 266} .........   	  & 51+395 & V710 Tau & 04 31 57.8 & +   & 18 21 37  &  Tau-Aur  	&    142   & 14.6  &  8.5    & 3.2 & 177 & &  & M0.5 & M2 & 3  \\
\noalign{\smallskip}
\object{GG Tau} .............        		  & 54 & 		& 04 32 30.3 & +   & 17 31 41  &  Tau-Aur  &    142   & 12.3  &  7.4    & 1.4 & 185 & X &  & K7+M0.5 & M5.5+M7.5 &  1 \\
\noalign{\smallskip}
\object{UZ Tau} .............        		  & 52+53	& 		& 04 32 43.0 & +   & 25 52 32  &  Tau-Aur  	&    142   & 12.9  &  7.4    & 3.6 & 275 & X & 1 & M1 & M2+M2 & 1 \\
\noalign{\smallskip}
\object{HN Tau} .............        		  & 60+406 & 		& 04 33 39.3 & +   & 17 51 53  &  Tau-Aur  &    142   & 13.7  &  8.4    & 3.2 & 217 & &  & K5 & M4 & 3 \\
\noalign{\smallskip}
\object{IT Tau} ...............      		  & 	&  Haro 6-26 & 04 34 13.9 & +   & 26 11 42  &  Tau-Aur  	&    142   & 14.9  & 7.9    & 2.4 & 224 & &  &  K3 & M4 & 2 \\
\noalign{\smallskip}
\object{L1642-1} ............        		  & 413 & 	EW Eri & 04 35 02.3 &$-$& 14 13 41  &  L1642  &    140   & 13.7  &  7.7    & 2.7 & 349 & &  &  K7 &  & 5 \\
\noalign{\smallskip}
\object{RW Aur} ............         		  & 80+81 & 		& 05 07 49.6 & +   & 30 24 05  &  Tau-Aur  	&    142   & 10.3  &  7.0    & 1.4 & 254 & & 2 & K1: & K5: & 4, 1  \\
\noalign{\smallskip}
\object{CO Ori} ..............       		  & 84 & 		& 05 27 38.3 & +   & 11 25 39  &  Orion &    460   & 10.6  &  6.5    & 2.0 & 280 & &  & F8:e & & 4  \\
\noalign{\smallskip}
\object{AR Ori} .............        		  & 155 & 		& 05 35 54.1 &$-$& 05 04 14  &  Orion  &    460   & 13.9  &  9.8    & 2.0 & 249 & &  & K7/M0 & & 4  \\
\noalign{\smallskip}
\object{LkH$\alpha$ 336} ........    	  & 190+516 & 		& 05 54 20.1 & +   & 01 42 56  &  L1622 &    460   & 14.4  &  9.2    & 5.8 & 95 & &  & K7 & M0.5 &  1 \\
\noalign{\smallskip}
\object{CGH$\alpha$ 5/6} ........    	  & 	& 		& 07 31 37.4 &$-$& 47 00 22  &  Gum Neb. 	&    450   & 14.2  &  9.1    & 11.1 & 317 & &  & K7 & K2/K5 & 6  \\
\noalign{\smallskip}
\object{PH$\alpha$ 14} ............  	  & 554 & 		& 08 08 33.8 &$-$& 36 08 10  &  Gum Neb. 	&    450   & 15.8  &10.3    & 0.6 & 164 & &  & M2: & & 4,7  \\
\noalign{\smallskip}
\object{PH$\alpha$ 30} ............  	  & 	& SPH 42	& 08 12 05.6 &$-$& 35 31 45  &  Gum Neb. 	&    450   & 15.1  &12.2    & 0.7 & 345 & &  & B2e & & 8 \\
\noalign{\smallskip}
\object{vBH 16} .............        		  & 	& vdBH 16 & 08 27 39.0 &$-$& 51 09 50  &  Gum Neb. 	&    450   & 15.6  &  9.3    & 4.5 & 330 & &  &  &  \\
\noalign{\smallskip}
\object{PH$\alpha$ 51} ............. 	  & 561& 		& 08 15 55.3 &$-$& 35 57 58  &  Gum Neb. 	&    450   & 15.9  &11.1    & 0.6 & 185 & &  & K7/M0 &  & 8 \\
\noalign{\smallskip}
\object{HD 76534} .........          	  & 	& He 3-225 &  08 55 08.7 &$-$& 43 28 00  & Vela R2$^d$ &   830$^d$    &   8.0  &  7.8    & 2.1 & 303 & &  & B2 & B3 & 14 \\
\noalign{\smallskip}
\object{SX Cha} .............        		  & 564 & T3		& 10 55 59.9 &$-$& 77 24 41  &   Cha I  		&    160   & 14.7  &  8.7    & 2.2 & 288 & &  &M0.5 & & 4  \\
\noalign{\smallskip}
\object{Sz 15} .................     		  & 	&  T19 & 11 05 41.5 &$-$& 77 54 44  &   Cha I  		&    160   &   13.2     &10.6    & 10.6 & 28 & &  & K5/K6 & & 7 \\
\noalign{\smallskip}
\object{ESO H$\alpha$ 281} .....      & 	& 		& 11 07 04.0 &$-$& 76 31 45  &   Cha I  		&    160   &   13.7$^a$     &  9.7   & 1.7 & 167 & &  &  M0.5 & M4.5 & 10 \\
\noalign{\smallskip}
\object{Sz 19} ................      		  & 245 & DI Cha & 11 07 20.7 &$-$& 77 38 07  &   Cha I &    160   & 10.9  &  6.2    & 4.6 & 203 & &  & G2 & & 4  \\
\noalign{\smallskip}
\object{VV Cha} ............         		  & 573 & 	T27	& 11 07 28.4 &$-$& 76 52 12  &   Cha I &    160   & 14.8  &  9.5    & 0.8 & 14 & &  & M1.5 & M3 & 10 \\
\noalign{\smallskip}
\object{VW Cha} ............         		  &  575 & 	T30/T31	& 11 08 01.8 &$-$& 77 42 29  &   Cha I  &    160   & 12.6  &  7.0    & 16.7 & 221 & X & 3 &  K5/K7 & K7 &  1 \\
\noalign{\smallskip}
\object{Glass I} ...............     		  & 	& Ced 111 IRS 4	 & 11 08 15.4 &$-$& 77 33 54  &   Cha I  		&    160   & 13.3  &  6.9    & 2.4 & 105 & &  &  K3 & G5e & 11 \\
\noalign{\smallskip}
\object{CoD $-$29$\degr$8887} ... & 	& TWA 2 & 11 09 14.0 &$-$& 30 01 39  &   TW Hya 	&     50    & 11.1  &  6.7    & $<$0.8 & 40 & &  & M2e & M2 & 12 \\
\noalign{\smallskip}
\object{Sz 30} ..................    		  & 	& T39 & 11 09 12.3 &$-$& 77 29 12  &   Cha I  		&    160   & 13.2  &  9.0    & 1.1 & 22 & &  & M0.5 & M2 & 10 \\
\noalign{\smallskip}
\object{Hen 3-600} ..........        	  & 	&  TWA 3	& 11 10 28.9 &$-$& 37 32 05  &   TW Hya 	&     50    & 12.1  &  6.8    & 1.4 & 230 & & 4 & M3e & M3.5 & 12 \\
\noalign{\smallskip}
\object{Sz 41} .................     		  & 588 & T51 & 11 12 24.5 &$-$& 76 37 06  &   Cha I  &    160   & 11.6  & 8.0    & 1.9 & 164 & & 5 & K3.5  &  & 9  \\
\noalign{\smallskip}
\object{CV Cha} .............        		  & 247+589 & 	T52	& 11 12 27.8 &$-$& 76 44 22  &   Cha I  &    160   & 11.0  & 6.9    & 11.4 & 105 & & 6 & G8 & G: & 4  \\
\noalign{\smallskip}
\object{Sz 48} ..................	  	  & 	& 		& 13 00 53.2 &$-$& 77 09 10	&    Cha II		&    178   & 17.4 & 9.5     & 1.5 & 232 & &  & K7/M0 & M0 & 10   \\
\noalign{\smallskip}
\object{BK Cha} ..............		  & 	& 		& 13 07 09.3 &$-$& 77 30 24	&    Cha II		&    178   & 15.2   & 8.4     & 0.8 & 329 & &  & K5/K7 & M0.5 & 10 \\
\noalign{\smallskip}
\object{Sz 60} ..................		  & 	& 		& 13 07 23.4 &$-$& 77 37 23	&    Cha II		&    178   &  17.1 &  9.5    & 3.4 & 280 & &  & M1 & M4 & 13 \\
\noalign{\smallskip}
\object{Sz 62} ..................     		  & 	& 		& 13 09 50.7 &$-$& 77 57 24  &   Cha II 		&    178   & 15.6  & 9.1    & 1.1 & 261 & &  & M2 & M3.5 & 10  \\
\noalign{\smallskip}
\object{Herschel 4636} ....          	  & 	& Hen 3-949 & 13 57 44.1 &$-$& 39 58 45  & NGC 5367 	&    630   &   9.7  & 7.2     & 3.7 & 214 & &  & B3 & & 14 \\
\noalign{\smallskip}
\object{ESO H$\alpha$ 283} ......     & 	& 		& 15 00 29.6 &$-$& 63 09 46  & Circinus 	&    700   &  15.5  &10.3   & 2.1 & 242 & &  &  &  \\
\noalign{\smallskip}
\object{Sz 65} .................. 		  & 597 & 		& 15 39 27.7 &$-$& 34 46 17	&    Lup I &    190   & 12.7  &  8.0    & 6.4 & 98 & &  & M0 & & 15  \\
\noalign{\smallskip}
\object{Sz 68} ..................		  & 248 & 	HT Lup & 15 45 12.9 &$-$& 34 17 31	&    Lup I &    190   & 10.4  &  6.5    & 2.6 & 295 & X &  & K2 & M6 &  1  \\
\noalign{\smallskip}
\object{HO Lup} ..............		  & 612 & 		& 16 07 00.6 &$-$& 39 02 19	&    Lup III &    190   & 13.0  &  8.6    & 1.5 & 36 & &  & K7/M0 & M2 & 10  \\
\noalign{\smallskip}
\object{Sz 101} ................		  & 	& 		& 16 08 28.4 &$-$& 39 05 32	&    Lup III	&    190   & 15.5  &  9.4    & 0.8 & 305 & &  & M2.5 & M3.5 & 10  \\
\noalign{\smallskip}
\object{Sz 108} ................		  & 620 & 		& 16 08 42.7 &$-$& 39 06 18	&    Lup III &    190   & 13.1  &  8.8    & 4.2 & 25 & &  & M1 & & 15  \\
\noalign{\smallskip}
\object{Sz 120} ................		  & 	    & HD 144965 & 16 10 10.6 &$-$& 40 07 44  &    Lup III	&    190   &   7.1	  &  6.2    & 2.6 & 142 & &  & B5  & & 15  \\
\noalign{\smallskip}
\object{WSB 3} ...............		  & 	& 		& 16 18 49.5 &$-$& 26 32 53	&    Oph	&    160   & 15.0 &  9.3    & 0.6 & 162 & &  & M3  & & 16  \\
\noalign{\smallskip}
\object{WSB 11} .............		  & 	& 		& 16 21 57.3 &$-$& 22 38 16	&    Oph	&    160   & 18.5$^a$  &10.1   & 0.5 & ... & &  &  &  \\
\noalign{\smallskip}
\object{WSB 20} .............		  & 257 & 	Haro 1-4	& 16 25 10.5 &$-$& 23 19 14	&    Oph	&    160   & 13.4  &  7.5    &$ <$1.0 & 23 & &  & K6/K7 & & 4  \\
\noalign{\smallskip}
\object{WSB 28} .............		  & 	& ISO-Oph 27 & 16 26 20.7 &$-$& 24 08 48	&    Oph	&    160   &  14.7$^a$ &  9.5    & 5.1 & 358 & &  & M3 & M7 & 17 \\
\noalign{\smallskip}
\object{SR 24} .................	 	  & 262 & 	WSB 42    & 16 26 58.8 &$-$& 24 45 37	&    Oph 	&    160   &  15.9  &  7.1    & 5.2 & 348 & X &  & K2: & M0.5 &   1  \\
\noalign{\smallskip}
\object{Elias 2-30} ..........		  & 	&  SR 21		& 16 27 10.2 &$-$& 24 19 16	&    Oph	&    160   & 14.1  &  6.7    & 6.4 & 175 & &  & G2.5 & M4 & 17  \\
\noalign{\smallskip}
\object{WSB 46} .............		  & 641+640 & 	ROXs 20	& 16 27 15.1 &$-$& 24 51 39	&    Oph	&    160   &   14.9 &  9.4    & 10.3 & 301& &  & M0/M3  & M1/M4 &18 \\
\noalign{\smallskip}
\object{Haro 1-14c} .........	  	  & 644+267 & 		& 16 31 04.4 &$-$& 24 04 32	&    Oph	&    160   & 12.3  &  7.8    & 12.9 & 122 & & 7 & K3 & M0 & 4 \\
\noalign{\smallskip}
\object{ROXs 43} .............		  & 	& NTTS 162819-2423 & 16 31 20.1 &$-$& 24 30 05	&    Oph	&    160   & 10.6  &  6.7    & 4.3 & 13 &  X & 8 & G0 & K3 &  1 \\
\noalign{\smallskip}
\object{Elias 2-49} ...........		  & 	& HD 150193 & 16 40 17.9 &$-$& 23 53 45  &    Oph	&    160   &   8.9	  &  5.5    & 1.1 & 227 & & & A1/A3Ve &  &  19 \\
\noalign{\smallskip}
\object{HBC 652} ............		  & 652 & V2507 Oph & 16 48 18.0 &$-$& 14 11 15	 &    L162		&    160   & 13.5  &  7.5    & 8.7 & 322 & &  & K4/K5 & & 4  \\
\noalign{\smallskip}
\object{LkH$\alpha$ 346} ........... & 275 & 	TH$\alpha$ 27-4 & 17 11 03.9 &$-$& 27 22 57	 &    B59		&    160   & 14.3 &   8.1   & 5.0 & 329 & &  & K7 & M5 &  1  \\
\noalign{\smallskip}
\hline
\end{tabular}
\end{center}
\label{Tab:sample}
\begin{minipage}[position]{18cm}
Note\,: HBC = catalog entry number in Herbig \& Bell\,\cite{Herbig_Bell_1988}.\\
$^a$\,: R magnitude. \\   
$^b$\,: Additional visual companion(s) discovered since RZ93 (see references in Table\,\ref{Tab:systems_parameters} or in the text). \\   
$^c$\,: Astrometry of the wide binary as quoted in RZ93 with $\rho$ in arcsec and PA in degree (only one entry for the possible triples quoted in RZ93). \\
$^d$\,: see Herbst\,\cite{Herbst_1975} and van den Ancker et al.\,\cite{Van_den_Ancker_etal1998}. \\
  Spectral type references\,: 
  (1) see Table\,\ref{Tab:systems_accretion}.
  (2) Duch\^{e}ne\,\cite{Duchene_etal1999}.  
  (3) Hartigan et al.\,\cite{Hartigan_etal1994}. 
  (4) Herbig \& Bell\,\cite{Herbig_Bell_1988}.
  (5) Sandell et al.\,\cite{Sandell_etal_1987}. 
  (6) Reipurth \& Pettersson\,\cite{Reipurth_Pettersson1993}. 
  (7) Kim et al.\,\cite{Kim_etal2005}.
  (8) Pettersson\,\cite{Pettersson_1987}.
  (9) Luhman\,\cite{Luhman_2004}. 
  (10) Brandner \& Zinnecker\,\cite{Brandner_Zinnecker_1997}. 
  (11) Feigelson \& Kriss\,\cite{Feigelson_Kriss_1989}.
  (12) Torres et al.\,\cite{Torres_etal2003}.
  (13) Hughes \& Hartigan\,\cite{Hughes_Hartigan_1992}.
  (14) Viera et al.\,\cite{Viera_etal2003}.
  (15) Hughes et al.\,\cite{Hughes_etal1994}.
  (16) Meyer et al.\,\cite{Meyer_etal1993}.
  (17) Prato et al.\,\cite{Prato_etal2003}.
  (18) Luhman \& Rieke\,\cite{Luhman_Rieke_1999}.
  (19) Gray \& Corbally\,\cite{Gray_Corbally_1998}.
 \\
Distance references\,: 
  Chamaeleon (I\,: 160\,pc $\pm$ 15\,pc, II\,: 178\,pc $\pm$ 18\,pc, Whittet et al.\,\cite{Whittet_etal1997}), 
  Gum Nebula (450\,pc, Kim et al.\,\cite{Kim_etal2005}), 
  Lupus (190\,pc $\pm$ 27\,pc, Wichmann et al.\,\cite{Wichmann_etal1998}), 
  Ophiuchus (160\,pc, Whittet\,\cite{Whittet_1974}), 
  Taurus-Auriga (142\,pc $\pm$ 14\,pc, Wichmann et al.\,\cite{Wichmann_etal1998}), 
  TW Hya (50\,pc $\pm$ 12\,pc, Mamajek\,\cite{Mamajek_2005}), 
  MBM12 (275\,pc, Luhman\,\cite{Luhman_2001}), 
  L1642, aka MBM20 (112-160\,pc, Hearty et al.\cite{Hearty_etal2000}, 140\,pc adopted), 
  NGC\,5367, Circinus, L162 and B59 (values reported in RZ93 adopted).
\\
SB notes\,: 
  (1) UZ Tau A\,: 19.1-days single-lined SB (SB1) (Mathieu et al.\,\cite{Mathieu_etal1996}), converted to double-lined SB (SB2) by Prato et al.\,\cite{Prato_etal2002}. 
  (2) RW Aur A\,: suspected SB1 with P=2.77-days (Gahm et al.\,\cite{Gahm_etal1999}, Petrov et al.\,\cite{Petrov_etal2001}, 
  	see also Alencar et al.\,\cite{Alencar_etal2005}).
  (3) VW Cha A\,: possible SB2 (Melo\,\cite{Melo_2003}).
  (4) TWA 3A\,: suspected SB2 (Torres et al.\,\cite{Torres_etal2003}).
  (5) Sz 41 A\,: probable SB (Reipurth et al.\,\cite{Reipurth_etal2002}) with period of about 125 days (best candidate out of five radial-velocity variables). 
  (6) CV Cha\,: possibly SB, variable radial-velocity (Reipurth et al.\,\cite{Reipurth_etal2002}).  
  (7) Haro 1-14c\,: 591-days SB1 (Reipurth et al.\,\cite{Reipurth_etal2002}), converted to SB2 by Simon \& Prato\,\cite{Simon_Prato_2004}.
  (8) ROXs 43 A\,: 89-days SB1 (Mathieu et al.\,\cite{Mathieu_etal1989}).
\end{minipage}
\end{table*}

\begin{table*}
\scriptsize
\caption{Observation Log.}
\begin{center}
\renewcommand{\arraystretch}{0.5}
\setlength\tabcolsep{5pt}
\begin{tabular}{l@{\hspace{2mm}}l@{\hspace{2mm}}r@{\hspace{4mm}}r@{$\times$}l@{\hspace{1mm}}r@{$\times$}l@{\hspace{1mm}}r@{$\times$}lr@{\hspace{1mm}}r@{\hspace{2mm}}c@{\hspace{2mm}}r@{\hspace{1mm}}r@{\hspace{1mm}}r@{\hspace{4mm}}c@{\hspace{2mm}}cc@{\hspace{1mm}}r@{\hspace{1mm}}r}
\hline\noalign{\smallskip}
Name & Obs. date & Cam &  \multicolumn{6}{c}{Int. time [s]} && WFS-Dic. & seeing$^a$ & \multicolumn{3}{c}{FWHM [mas]}  && Airmass  & \multicolumn{3}{c}{Strehl ratio [\%]} \\
\cline{4-9}  \cline{13-15}  \cline{18-20}\\ 
  &  &  & \multicolumn{2}{c}{[FeII]} & \multicolumn{2}{c}{H$_2$} & \multicolumn{2}{c}{Br$\gamma$} && & [arcsec]  & [FeII]  & H$_2$ & Br$\gamma$ && & [FeII]  & H$_2$ & Br$\gamma$ \\
\noalign{\smallskip}
\hline
\noalign{\smallskip}
LkH$\alpha$ 262/263 ..      & 2002 Nov 14 & S27 & 15 & 2.7 & 15 & 1.5 & 15 & 1.4 && VIS-VIS  & 1.0 	
&  129 &  89    &   88   && 1.43 & 4 & 14 & 12 \\
\noalign{\smallskip}
J 4872 ...............      		  & 2002 Nov 13 & S13  	& 30 & 1.1 & 30 & 1.6 & 30 & 1.6 && VIS-VIS  & 0.8 	
&   72   &  73   &   72   && 1.62 & 18 & 42 & 40  \\
\noalign{\smallskip}
FV Tau$^b$ .............             & 2002 Nov 13 & S27  	& 9 & 47 & 9 & 19 & 9 & 19 && IR-N20C80 & 0.8 	
&  121  &  89   &  81   && 1.59 &  7  & 19 & 13  \\
\noalign{\smallskip}
UX Tau .............        		  & 2002 Oct 22 & S13  	& 60 & 0.6 & 60 & 0.6 & 60 & 0.6 && VIS-VIS & 1.1 	
&    60  &  68   &  68   && 1.40 &  22  & 38  &  47  \\
\noalign{\smallskip}
DK Tau .............        		  & 2002 Nov 13 & S13  	& 30 & 1.4 & 30 &1.0 & 30 & 0.9 && IR-N20C80 & 0.9 	
&   64   &  70   &  68    && 1.59 & 15   & 29  & 32  \\
\noalign{\smallskip}
HK Tau .............        		  & 2003 Feb 17 & S13  	& 12 & 5.5 & 12 & 2.9 & 12 & 2.9 && IR-N20C80 & 1.0 	
&   80   &  73   &  73    && 1.75 & 7   & 21  & 26 \\
\noalign{\smallskip}
LkH$\alpha$ 266 .........   	  & 2002 Oct 22 & S27  	& 105 & 0.3 & 105 &	0.3 & 105 & 0.3 && VIS-VIS & 1.7 	
&   90   &  82   &  93    && 1.44 & 5   & 12  & 8 \\
\noalign{\smallskip}
GG Tau .............        		  & 2002 Oct 22 & S27  	& 60 & 1.0 & 30 & 1.0 & 30 & 1.0 && VIS-VIS & 1.7 	
&   67   &  70   &  87   && 1.49 & 12   &  31 & 19     \\
\noalign{\smallskip}
UZ Tau .............        		  & 2002 Nov 14 & S13  	& 45 & 0.6 & 45 & 0.5 & 45 & 0.5 && VIS-VIS & 1.0 	
&   84   &  77   &  77  	&& 1.68 & 8  & 17  &  27       \\
\noalign{\smallskip}
HN Tau .............        		  & 2003 Feb 18 & S13 &\multicolumn{2}{c}{...}&\multicolumn{2}{c}{...}& 15 & 1.5 && VIS-VIS & 0.6 	
&   ...     & ...      &  72   && 1.59 & ...   & ...  & 29 \\
\noalign{\smallskip}
IT Tau ...............      		  & 2003 Feb 19 & S13  	& 15 & 3.8 & 15 & 1.7 & 15 & 1.7 && IR-N20C80 & 1.1 	
&  104  & 77    &  75   && 1.84 & 2   & 17  & 21 \\
\noalign{\smallskip}
L1642-1 ............        		  & 2003 Feb 20 & S13  	& 60 & 0.9 & 60 & 0.6 & 60 & 0.6 && VIS-VIS & 1.1 	
&    78  &  72   & 77  	&& 1.50 & 7   & 23  & 22  \\
\noalign{\smallskip}
RW Aur ............         		  & 2002 Nov 17 & S13  	& 45 & 0.7 & 45 & 0.5 & 45 & 0.5 && IR-N20C80 & 0.8 	
&    66  &  70   & 70   	&& 1.79 & 15   & 34  & 36 \\
\noalign{\smallskip}
CO Ori ..............       		  & 2002 Nov 13 & S13  	& 30 & 2.0 & 30 & 1.0 & 30 & 2.0 && VIS-VIS & 1.0 	
&    68  &  71    & 73   && 1.53 & 12   & 23  & 28 \\
\noalign{\smallskip}
AR Ori .............        		  & 2003 Feb 19 & S13  	& 3 & 20 & 3 & 10 & 3 & 10 && VIS-VIS & 1.1 	
&   151& 122   & 91    && 1.21 & 3   & 8  & 11 \\
\noalign{\smallskip}
LkH$\alpha$ 336 ........    	  & 2003 Feb 19 & S27  	& 6 & 10 & 6 & 9.0 & 6 & 8.5 && VIS-VIS & 1.4 	
&     89 &   87   & 90  	&& 1.20 & 6   & 12  & 12 \\
\noalign{\smallskip}
CGH$\alpha$ 5/6 ........    	  & 2002 Dec 23 &S27  	& 36 & 1.3 & 36 & 1.0 & 36 & 1.0 && VIS-VIS & 0.7 	
&     85 &   84   & 79  	&& 1.08 &  6  & 13  & 21 \\
\noalign{\smallskip}
PH$\alpha$ 14 ............  	  & 2002 Dec 23 &S27  	& 15 & 4.0 & 15 & 2.0 & 15 & 2.0 && VIS-VIS & 0.7 	
&     86 &   80   & 78    && 1.02 & 8   & 16  & 24 \\
\noalign{\smallskip}
PH$\alpha$ 30 ............  	  & 2003 Jan 17 &S27  	& 3 & 28 & 3 & 21 & 3 & 20 && VIS-VIS &  ...   	
&   104 &   84   & 83  	&& 1.02 &  9  &  27 & 32      \\
\noalign{\smallskip}
vBH 16 .............        		  & 2003 Jan 27 &S13  	& 15 & 5.5 & 15 & 4.5 & 15 & 4.5 && IR-N20C80 & 0.9 	
&   122 &  112  & 97   && 1.15 & 4   & 10  & 13 \\
\noalign{\smallskip}
PH$\alpha$ 51 ............. 	  & 2003 Jan 23 &S13  	& 3 & 85 & 3 & 31 & 3 & 30 && VIS-VIS & 1.0 	
&   135 &  107  & 88    && 1.02 & 3   & 7  & 11 \\
\noalign{\smallskip}
HD 76534 .........          	  & 2003 Jan 20 &S13  	& 60 & 0.6 & 60 & 0.6 & 60 & 0.6 && VIS-VIS & 1.6 	
&     69 &    80   & 72    && 1.18 & 12   & 18  & 26 \\
\noalign{\smallskip}
SX Cha .............        		  & 2003 Jan 22 &S13  	&12 & 8.5 & 12 & 2.5 & 12 & 2.5 && IR-N20C80 & 0.9 	
&     93 &    82   & 76    && 1.66 & 7   & 16  & 21 \\
\noalign{\smallskip}
Sz 15 .................     		  & 2003 Jan 22 &S13  	& 12 & 3.9 & 12 & 4.0 & 12 & 3.9 && VIS-VIS & 0.7 	
&     85 &    85   & 83    && 1.67 & 7   & 17  & 16 \\
\noalign{\smallskip}
ESO H$\alpha$ 281 .....      & 2003 Jan 22	 &S13  	& 3 & 51 & 3 & 34 & 3 & 33 && VIS-VIS & 0.8 	
&     99 &    83   & 87    && 1.62 & 5   & 16  & 16	 \\ 
\noalign{\smallskip}
Sz 19$^b$ ................      	  & 2003 Jan 20 &S13  	& 6 & 22 & 6 & 16 & 6 & 16 && VIS-VIS & 1.5 	
&     69 &    76   & 76    && 1.66 & 12   & 22  & 23 \\
\noalign{\smallskip}
VV Cha ............         		  & 2003 Feb 19 &S13  	& 3 & 18 & 3 & 8.0 & 3 & 7.5 && IR-N20C80 & 0.8 	
&   173 &  108   & 100   && 1.66 & 3   & 9  & 9 \\
\noalign{\smallskip}
VW Cha ............         		  & 2003 Feb 20 &S27  	& 75 & 0.8 & 75 & 0.3 & 75 & 0.3 && IR-N20C80 & 1.2 	
&     77 &     76   &   72  && 1.68 &  11  & 25  &  33    \\
\noalign{\smallskip}
Glass I ...............     		  & 2003 Feb 20 &S13  	& 12 & 5.0 & 12 & 3.5 & 12 & 3.0 && IR-N20C80 & 1.3 	
&     88 &     79   &   82  && 1.67 & 5   & 21  & 24 \\
\noalign{\smallskip}
CoD $-$29$\degr$8887 ... & 2003 Jan 20  &S13  	& 90 & 0.3 & 90 & 0.3 & 90 & 0.4 && VIS-VIS  & 1.1 	
&     62 &     69   &   69  && 1.01 & 18   & 31  & 34 \\
\noalign{\smallskip}
Sz 30$^c$ ..................   	& 2003 Jan 20 &S13 &\multicolumn{2}{c}{...}&\multicolumn{2}{c}{...}& 3 & 20 && IR-N20C80 & 1.5 	
&  ...     &   ...       & 184  && 1.67 &  ...  &  ... &  7    \\
\noalign{\smallskip}
Hen 3-600 ..........        	  & 2003 Feb 20 &S13  	& 60 & 0.4 & 60 & 0.4 & 60 & 0.4 && IR-N20C80 & 1.5 	
&  74    &   91     &   72    && 1.03 & 10   & 19  & 26 \\
\noalign{\smallskip}
Sz 41 .................     		  & 2003 Jan 22 &S27  	& 75 & 0.8 & 75 & 0.5 & 75 & 0.5 && VIS-VIS & 0.9 	
&  70    &   74     &   78  && 1.64 &  11  & 27  &  30   \\
\noalign{\smallskip}
CV Cha$^b$ .............        	  & 2003 Jan 20 &S27  	&  6 & 16 & 6 & 11 & 6 & 11 && VIS-VIS & 1.3 	
& 110   &   93     &   75    && 1.65 & 5   & 15  & 22 \\
\noalign{\smallskip}
Sz 48 ..................	  	  & 2004 Apr 06	 & S13 & 
4 & 48 &\multicolumn{2}{c}{...}&\multicolumn{2}{c}{...}&& IR-K & 0.6 & 
184   &   ...   &   ...      && 1.66 & 2   & ...  & ...      \\  
\noalign{\smallskip}
BK Cha ..............		  & 2004 Apr 03	 & S13 & 
4 & 30 &\multicolumn{2}{c}{...}&\multicolumn{2}{c}{...}&& IR-K & 0.5	 &  
78    &   ...   &   ...      && 1.66 & 7   & ...  & ...           \\  
\noalign{\smallskip}
Sz 60 ..................		  & 2004 Apr 05	 & S13 & 
4 & 60 &\multicolumn{2}{c}{...}&\multicolumn{2}{c}{...}&& IR-K & 0.5	 & 
115   &   ...   &   ...      && 1.66 & 5   & ...  & ...         \\  
\noalign{\smallskip}
Sz 62 ..................     		  & 2003 Jan 22 & S27 & 
15 & 3.0 & 15 & 1.6 & 15 & 1.6 && VIS-VIS & 1.0 & 
150   &  105    &   91 && 1.68 & 3   & 7  & 11     \\
\noalign{\smallskip}
Herschel 4636 ....          	  & 2003 Jan 20 & S13 & 
9 & 3.0 & 9 & 4.5 & 9 & 4.5 && VIS-VIS & 1.0 &
    70   &  112    & 104&& 1.08 & 15   & 17  & 17 \\
\noalign{\smallskip}
ESO H$\alpha$ 283 ......     & 2003 Mar 26 & S13 &
3 & 46 & 3 & 35 & 3 & 33 && VIS-VIS & 0.7 &
    76   &    74    &   75  && 1.36 & 13   & 35  & 40     \\  
\noalign{\smallskip}
Sz 65 .................. 		  & 2004 Apr 06	 & S13 & 
36 & 2.2 &\multicolumn{2}{c}{...}&\multicolumn{2}{c}{...}&& VIS-VIS  & 0.8	 &
    60   &   ...   &   ... 	&& 1.36 & 20   & ...  & ...    \\  
\noalign{\smallskip}
Sz 68 ..................		  & 2004 Apr 06	 & S13 & 
88 & 0.6 &\multicolumn{2}{c}{...}&\multicolumn{2}{c}{...}&& VIS-VIS & 0.6	 &
    61   &   ...   &   ...    && 1.34 &  27  & ...  & ...          \\  
\noalign{\smallskip}
HO Lup ..............		  & 2004 Apr 06	 & S13 & 
4 & 13.8 &\multicolumn{2}{c}{...}&\multicolumn{2}{c}{...}&& IR-K & 0.7 &
    80   &   ...   &   ...      && 1.37 & 7   & ...  & ...  \\  
\noalign{\smallskip}
Sz 101 ................		  & 2004 Apr 10	 & S13 & 
4 & 18 &\multicolumn{2}{c}{...}&\multicolumn{2}{c}{...}&& IR-K & 0.7 &
    79   &   ...   &   ...      && 1.03 & 10   & ...  & ...  \\  
\noalign{\smallskip}
Sz 108 ................		  & 2004 Apr 10 & S13 & 
8 & 8.0 &\multicolumn{2}{c}{...}&\multicolumn{2}{c}{...}&& IR-K & 0.6	&
    76   &   ...   &   ...      && 1.03 & 13   & ...  & ... \\  
\noalign{\smallskip}
Sz 120 ................		  & 2004 Apr 10	 & S13 &
180 & 0.6 &\multicolumn{2}{c}{...}&\multicolumn{2}{c}{...}&& VIS-VIS & 0.7 &
    70   &   ...   &   ...        && 1.04 & 18   & ...  & ...  \\  
\noalign{\smallskip}
WSB 3 ...............		  & 2004 May 01 & S13 & 
4 & 16.8 &\multicolumn{2}{c}{...}&\multicolumn{2}{c}{...}&& IR-K & 0.9 &
 133    &   ...   &   ...         && 1.10 & 4   & ...  & ... \\  
\noalign{\smallskip}
WSB 11 .............		  & 2004 Jun 18 & S13 & 
4 & 56 &\multicolumn{2}{c}{...}&\multicolumn{2}{c}{...}&& IR-K & 1.1 &
    93   &   ...   &   ...          && 1.04 & 2   & ...  & ...     \\  
\noalign{\smallskip}
WSB 20 .............		  & 2004 May 01 & S13 & 
12 & 5.0 &\multicolumn{2}{c}{...}&\multicolumn{2}{c}{...}&& IR-K & 0.8 &
    75   &   ...   &   ...          && 1.06 & 12   & ...  & ... \\  
\noalign{\smallskip}
WSB 28 .............		  & 2004 May 01 & S13 & 
4 & 24 &\multicolumn{2}{c}{...}&\multicolumn{2}{c}{...}&& IR-K & 0.8 &
  117   &   ...   &   ...          && 1.04 & 8   & ...  & ...     \\  
\noalign{\smallskip}
SR 24 .................	 	  & 2004 May 01 & S13 & 
16 & 4.0 &\multicolumn{2}{c}{...}&\multicolumn{2}{c}{...}&& IR-K & 1.1 &
    72   &   ...   &   ...     && 1.03 & 20   & ...  & ...     \\  
\noalign{\smallskip}
Elias 2-30 ..........		  & 2004 May 01 & S13 &
40 & 2.0 &\multicolumn{2}{c}{...}&\multicolumn{2}{c}{...}&& IR-K & 1.0 &
    72   &   ...   &   ...      && 1.02 & 13   & ...  & ...\\  
\noalign{\smallskip}
WSB 46 .............		  & 2004 May 01 & S27 &
4 & 13.6 &\multicolumn{2}{c}{...}&\multicolumn{2}{c}{...}&& IR-K & 0.8 &
    148   &   ...   &   ...       && 1.01 & 3   & ...  & ... \\  
\noalign{\smallskip}
Haro 1-14c .........	  	  & 2004 May 01 & S27 & 
52 & 1.0 &\multicolumn{2}{c}{...}&\multicolumn{2}{c}{...}&& VIS-VIS & 1.1 &
    82   &   ...   &   ...        && 1.07 & 7   & ...  & ... \\  
\noalign{\smallskip}
ROXs 43 .............		  & 2004 May 04 & S13 &
100 & 0.6 &\multicolumn{2}{c}{...}&\multicolumn{2}{c}{...}&& VIS-VIS & 0.5 &
    80   &   ...   &   ...      && 1.06 &  36  & ...  & ...          \\  
\noalign{\smallskip}
Elias 2-49 ...........		  & 2004 May 01 & S13 & 
56 & 1.2 &\multicolumn{2}{c}{...}&\multicolumn{2}{c}{...}&& VIS-VIS & 0.9 &
    77   &   ...   &   ...            && 1.03 & 11   & ...  & ... \\  
\noalign{\smallskip}
HBC 652 ............		  & 2004 May 01 & S27 & 
100 & 0.5 &\multicolumn{2}{c}{...}&\multicolumn{2}{c}{...}&& IR-K & 0.9 &
    84   &   ...   &   ...            && 1.03 & 9   & ...  & ...  \\  
\noalign{\smallskip}
LkH$\alpha$ 346 ........... 	  & 2004 May 01 & S27 &
16 & 4.0 &\multicolumn{2}{c}{...}&\multicolumn{2}{c}{...}&& IR-K & 0.9 &
  104    &   ...   &   ...            && 1.02 &  16  & ...  & ...       \\
\noalign{\smallskip}
\hline
\noalign{\smallskip}
\end{tabular}
\end{center}
\label{Tab:obs_log}
\begin{minipage}[position]{17cm}
$^a$\,: The seeing value quoted is a median value of all frames at 550\,nm at zenith.    \\   
$^b$\,: Observed with a neutral density filter.  \\
$^c$\,: Only 2 acquisition frames in Br$\gamma$ with very poor Strehl ratio. \\
\end{minipage}
\end{table*}


\section{The sample}
\label{sect:sample}
\begin{figure}
\centering
\includegraphics[width=8.5cm]{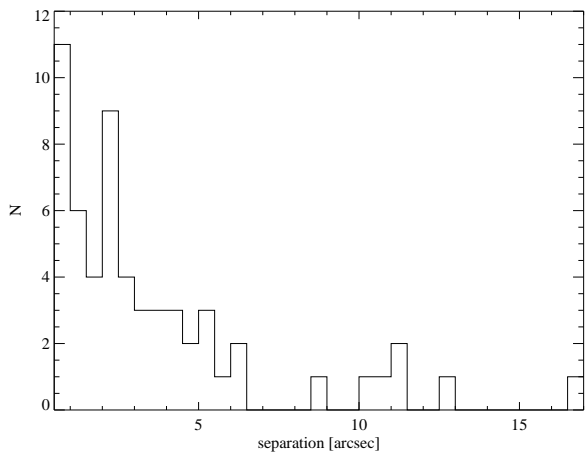}
\caption{Distribution of separations/orbital semimajor axes of the wide binaries of our sample.}
\label{fig:separation_dist_sample}
\end{figure}

We observed a sample composed of 58 wide PMS binaries randomly chosen from the 87 listed in RZ93 (see Table\,\ref{Tab:sample}). 
The separations range from 0$\farcs$5 to 16$\farcs$7, with a median of 2$\farcs$4 and the bulk of the binaries having separations 
$\la$\,6$\arcsec$ (Fig\,\ref{fig:separation_dist_sample}). 
It includes sources from various nearby southern star-forming region such as Chamaeleon, Gum Nebula, Lupus and Ophiuchus, 
while northern targets are mainly from Taurus-Auriga. Some objects from RZ93 and included in our sample are now known to be located 
closer and belong to the TW Hya and MBM12 associations.   
Nearly all sources are T Tauri binary stars, hence composed of young, late-type (mainly K and M-type here) and low mass stars. 
Five sources (HD 76534, Herschel 4636, PH$\alpha$30, Sz 120, Elias 2-49) are of earlier-type, intermediate mass Herbig Ae/Be binary stars. 
The sample we consider here is presumably strongly biased by selection effects which are difficult to quantify because they are the results of a 
complex combination of discovery biases. 
The sample is {\it obviously not} volume-limited since the original sample of RZ93 was mainly drawn from several H$\alpha$ 
surveys which are magnitude-limited. This might have favoured the detection of multiple systems which are brighter than single stars, or in our case, 
high-order multiples which are brighter than binaries (\"{O}pik\,\cite{Opik_1924}). However, we note that our sample is less biased towards bright 
systems in comparison to those of previous multiplicity surveys (see Sect.\,\ref{sect:multiplicity_stat_compa_surveys}). 
Our sample tends also to include mostly actively accreting objects, i.e. Classical T Tauri systems (CTTS).
We should also point out that, since the survey of RZ93, sparse NIR high-angular resolution observations have either confirmed or discovered that ten 
wide binaries of the sample have additional physical companions. These systems are identified in the column 'additional visual companions' in 
Table\,\ref{Tab:sample}. In addition to a likely triple system (Sz\,30), two other systems were reported in RZ93 as possible triple systems 
(LkH$\alpha$\,336 and Sz\,41). Despite all those known and/or suspected multiple systems, not all sources of our sample have been searched for 
additional companions (especially the faintest one) and our present AO survey uncovered additional new candidate companions.

\section{Observations and data reduction}
\label{sect:obs_data_red}
Observations were carried out during two periods. A first set of 37 objects were observed from October 22$^{th}$ 2002 to 
March 26$^{th}$ 2003 while observations of another 21 systems were conducted from April 4$^{th}$ 2004 to 
June 17$^{th}$ 2004. Magnitudes from the literature in V and from 2MASS (Two Micron All-Sky Survey) in K are reported 
in Table\,\ref{Tab:sample} for each combined system or alternatively for the brightest components if resolved. 
NACO was mainly used with the visible wavefront sensor (WFS) since most of the objects resolved by the wavefront camera 
and used as natural guide star for the AO system have bright enough V-magnitude. Only in a few cases the use of the IR-WFS was needed.  
Each object of the first set has been observed through the three narrow-band filters 
Br$\gamma$ (2.166\,$\mu$m, 0.023\,$\mu$m width), H$_2$ (2.122\,$\mu$m, 0.022\,$\mu$m width) and 
[FeII] (1.644\,$\mu$m, 0.018\,$\mu$m width) in order to prevent saturation from the brightest stars of our sample. 
Objects of the second set were observed only through the [FeII] filter.
Three or four dithered exposures were taken in each filter. 
Single exposure integration times vary from typically 0.5\,s to 30\,s which imply a magnitude limit for the median 
exposures of H\,$\sim$\,14 and K\,$\sim$\,13, and for the deepest exposures H\,$\sim$\,16 and K\,$\sim$\,15.5, respectively 
(see Tables\,\ref{Tab:obs_log} and \ref{Tab:undetec_comp}). The field of view (FOV) is either $\sim$\,13$\farcs$6$\times$13$\farcs$6 or 
$\sim$\,27$\farcs$8$\times$27$\farcs$8 for the S13 and S27 cameras respectively, with some enlargement which 
can be obtained with the dithering depending on the position of the target in each frame. 
The combination of natural guide star magnitude and seeing lead to AO-corrections with typical Strehl ratios of 
$\sim$\,30\% and $\sim$\,10\% in Br$\gamma$ and [FeII], respectively, which provides mainly diffraction-limited cores. 
The resolutions of the images, measured as the FWHM of a Gaussian fit to the radial profile of the brightest 
component, have a median value of 80.2\,mas, 79.7\,mas and 77.8\,mas in [FeII], H$_2$ and Br$\gamma$, respectively 
(see Table\,\ref{Tab:obs_log}). 

Data reduction was performed in IRAF\footnote{IRAF is distributed by the National Optical Astronomy Observatory, which is 
operated by the Association of Universities for Research in Astronomy, Inc., under contract to the National Science Foundation.} 
in the usual way\,: sky subtraction, flat-fielding, bad-pixels and cosmics corrections. Sky background frames obtained from median 
averaging of the dithered frames were subtracted from the individuals frames.  All frames were then flat-fielded 
and corrected for bad-pixels using calibration files provided by ESO. Cosmic rays have been removed using the 
LACOS\footnote{LACOS webpage\,: http://www.astro.yale.edu/dokkum/lacosmic/} package (Van Dokkum \cite{VanDokkum_2001}) and finally the 
frames were registered and combined together. 

The detection of companions was first performed by eye. We then searched for eventual close and faint companions using psf-subtraction. 
Specifically, each component image of each system was subtracted by the other scaled component images of that system and the residuals analyzed.
The efficiency of this method is substantially improved by the fact that the components were imaged simultaneously and that their separation are 
usually within the isoplanatic angle, i.e. the AO-correction is very similar. No companion undetected by eye was found, however a few components 
exhibit residuals which are presumably to be attributed to faint extended emission of scattered light from disks or envelopes (an example is 
presented in Appendix\,\ref{sect:comments_ind_objects}).   

\begin{figure*}
\centering
\begin{tabular}{c}
\includegraphics[width=17cm]{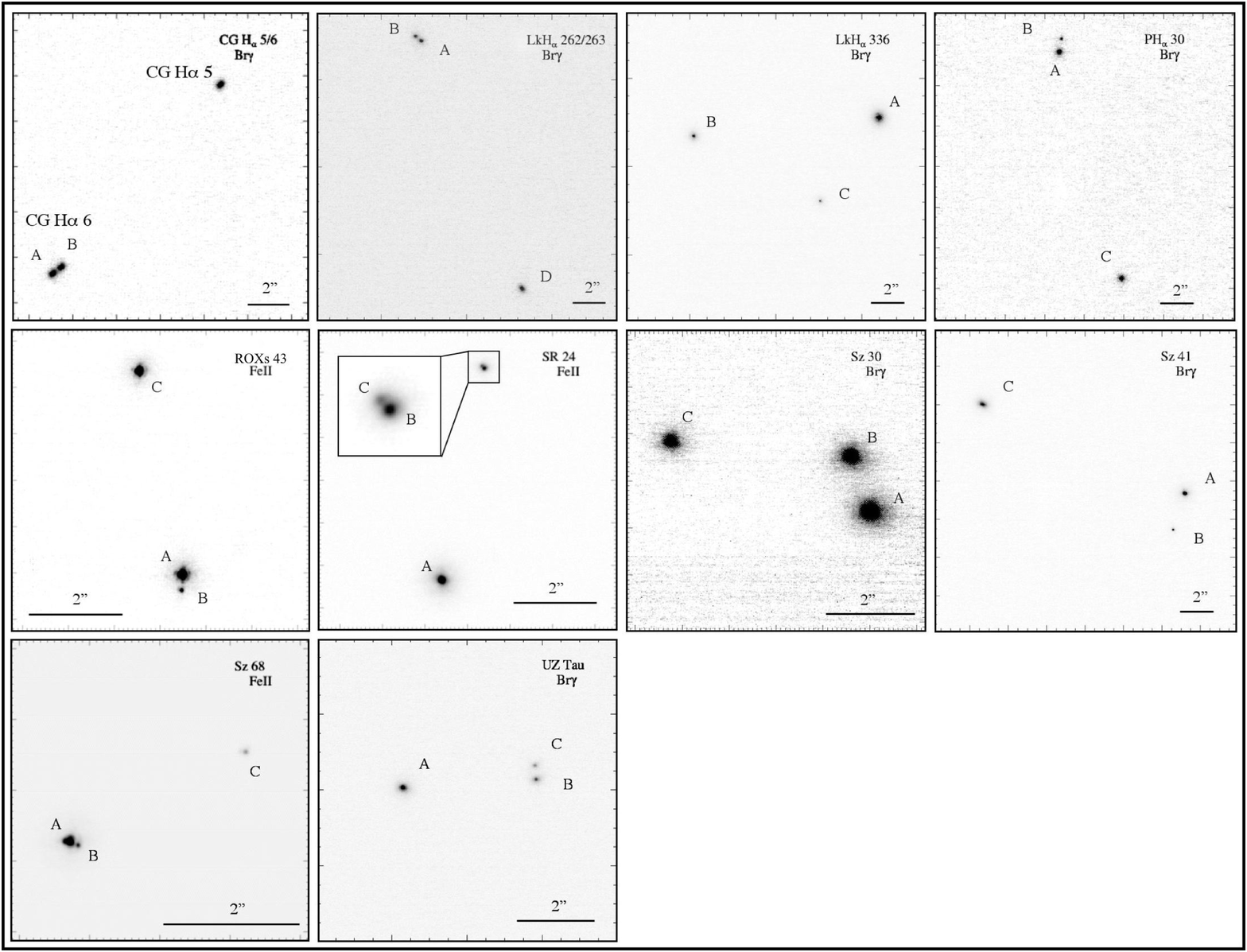}\\
\vspace{0.2cm}\\
\includegraphics[width=17cm]{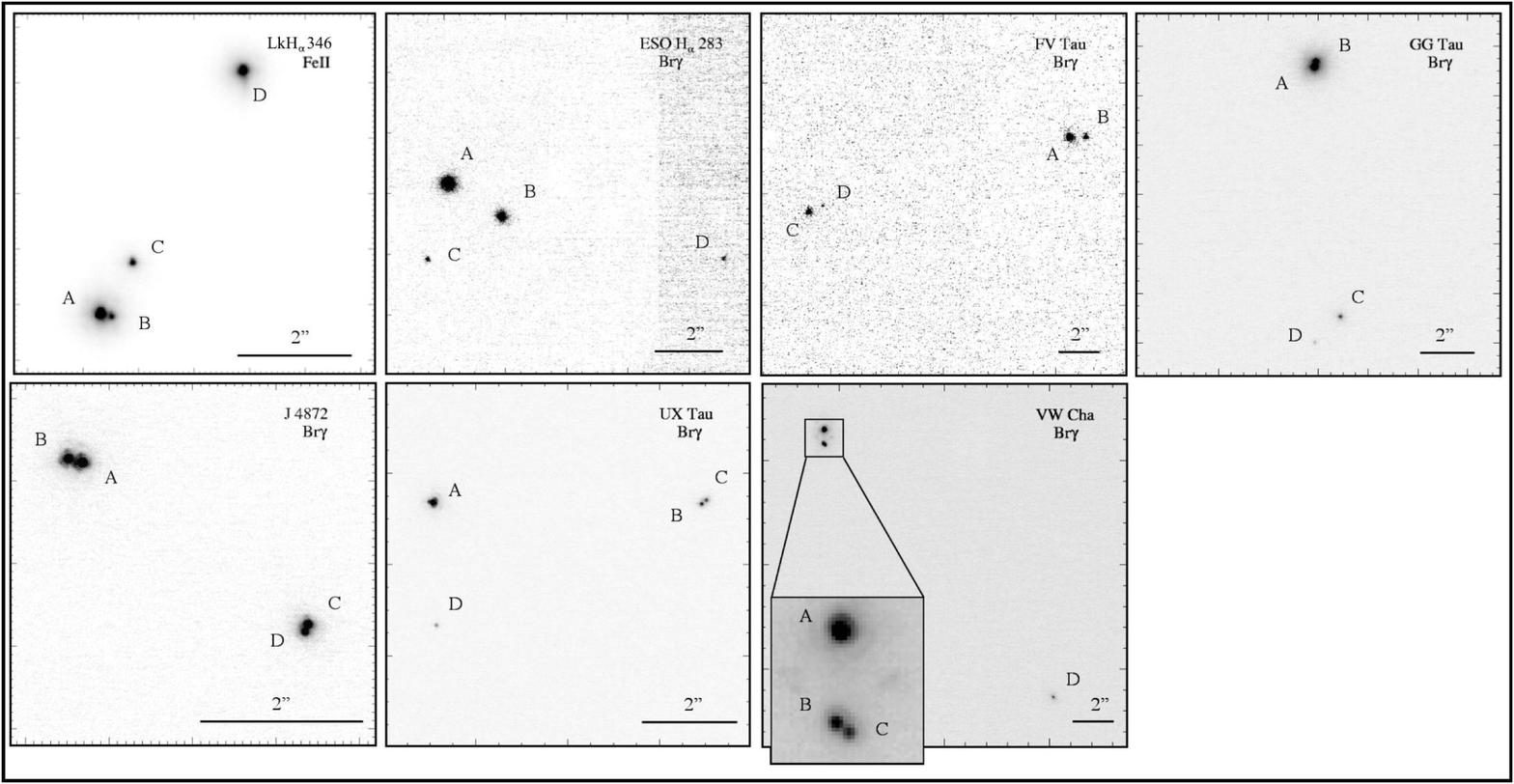}
\end{tabular}
\caption{Already known and candidate triple (upper panel) and quadruple (lower panel) systems detected in our VLT/NACO survey. 
These systems are the only apparent high-order multiple systems found in our sample, physical systems will be identified in Sect.\,\ref{sect:chance_proj}. North is up, east is left.}
\label{fig:triples_quadruples}
\end{figure*}

Astrometry and relative photometry of the high-order multiple systems were performed using the IRAF package DAOPHOT. 
The relative photometry was based on the results of PSF-fitting using the task ALLSTARS (Stetson \cite{Stetson_1987}) averaged with the results 
obtained with aperture photometry techniques, with radius of typically 3 pixels. The PSF used was usually the main component of each system. 
Relative positions were determined with the PSF-fitting. To estimate uncertainties, we performed the same analysis on both 
the individual frames and the combined frames for all wavelengths and calculate their standard deviation.  
We assumed that the systematics introduced by the PSF-fitting procedure are of the order of 0$\farcs$001 and 0.01 in the derived separation and flux ratio, respectively. The final error in relative positions are estimated by combining quadratically the rms variations of the results from the different data sets 
with the uncertainty in the plate scale (S13\,: 13.26$\pm$0.03\,mas, Masciadri et al.\,\cite{Masciadri_2003}, S27\,: 27.01$\pm$0.05\,mas, Chauvin et al.\,\cite{Chauvin_etal2005}) 
and detector orientation ($\pm$0.5$^{\circ}$, Masciadri et al.\,\cite{Masciadri_2003}).
The astrometric and photometric results are presented in Table\,\ref{Tab:systems_parameters} for all possible triple/quadruple 
systems of our sample.


\section{Results}
\label{sect:results}
%


\subsection{New candidate companions}
As a result of our survey, among the 58 observed systems, 10 were found to be apparent triples and 7 to be apparent quadruples 
(see Fig.\,\ref{fig:triples_quadruples} and Table\,\ref{Tab:systems_parameters}). The remaining systems remain binary within our resolution 
and detection limits (see Table\,\ref{Tab:undetec_comp}). No new edge-on disks other than the two previously known in our sample 
(HK\,Tau\,B and LkH$\alpha$\,263\,C) were detected. However, since LkH$\alpha$\,263\,C was not detected it is probable that our survey might not have been 
deep enough to detect those (see Appendix\,\ref{sect:edge_on_disks}).
Throughout this paper, the single components of all systems are named with upper 
case letters, starting from A for the brightest component in Br$\gamma$ and [FeII], for respectively the first and second data set. 
This nomenclature\,\footnote{It should be noted that the Multiple Star Catalogue (MSC, Tokovinin \cite{Tokovinin_1997}) uses a 
more sophisticated nomenclature.} is adopted in Fig.\,\ref{fig:triples_quadruples}. 
Our designation is only practical when dealing with multiple systems, and does not reflect primary, secondary, tertiary etc. which 
are based on the component masses.

\begin{table*}
\scriptsize
\caption{Measured parameters of all candidate triple/quadruple systems resolved in our survey, with their apparent multiplicity in the projected separation 
range 0\farcs07-12$^{\prime\prime}$. Also quoted in last column is the probable and assumed multiplicity of this systems in this separation range based 
on the chance projection analysis of Sect.\,\ref{sect:chance_proj}.}
\begin{center}
\renewcommand{\arraystretch}{0.7}
\setlength\tabcolsep{5pt}
\begin{tabular}{lcc@{\hspace{3mm}}r@{\,$\pm$\,}l@{\hspace{2mm}}r@{\,$\pm$\,}l@{\hspace{3mm}}r@{\,$\pm$\,}l@{\hspace{2mm}}r@{\,$\pm$\,}l@{\hspace{2mm}}r@{\,$\pm$\,}l@{\hspace{2mm}}r@{\hspace{1mm}}r@{\,$\pm$\,}l@{\hspace{1mm}}r@{\,$\pm$\,}l@{\hspace{1mm}}r@{\hspace{1mm}}r@{\,$\pm$\,}l@{\hspace{1mm}}r@{\,$\pm$\,}l@{\hspace{1mm}}c@{\hspace{2mm}}c}
\hline\noalign{\smallskip}
System                       		&  app. & Pair        &  \multicolumn{2}{c}{Separation}   &    \multicolumn{2}{c}{P.A.}     
& \multicolumn{6}{c}{Flux ratio}   & & \multicolumn{4}{c}{1st component pair}  & & \multicolumn{4}{c}{2nd component pair}  & Ref & Prob. \\
\cline{8-13}  \cline{15-18} \cline{20-23} \\
& mult. &  &   \multicolumn{2}{c}{[arcsec]}        &    \multicolumn{2}{c}{[$^{\circ}$]}  
& \multicolumn{2}{c}{[FeII]}   & \multicolumn{2}{c}{H$_2$} & \multicolumn{2}{c}{Br$\gamma$}  
& & \multicolumn{2}{c}{H} & \multicolumn{2}{c}{K} & & \multicolumn{2}{c}{H} & \multicolumn{2}{c}{K} & (mult.) & mult.\\
\noalign{\smallskip}
\hline
\noalign{\smallskip}
ESO H$\alpha$ 283....   
& Q & A-B & 2.058 & 0.002 & 238.5 & 0.5 & 0.61 &  0.01 & 0.42 & 0.01 & 0.42 & 0.01 && 11.64 & 0.06 & 10.69 & 0.03 && 12.18 & 0.07 & 11.64 & 0.05 & 1 & B \\
&     & A-C & 2.571 & 0.003 &165.1 & 0.5 & 0.04 &  0.01 & 0.06 & 0.01 & 0.06 & 0.01 &&\multicolumn{2}{c}{}&\multicolumn{2}{c}{}&& 15.10 & 0.17 &  13.79 & 0.07 &  \\
&     & A-D & 9.326 & 0.007 &254.6 & 0.5 & 0.03 &  0.01 & 0.03 & 0.01 & 0.03 & 0.01 &&\multicolumn{2}{c}{}&\multicolumn{2}{c}{}&& 14.77 & 0.15 & 14.09 & 0.10 & \\
\noalign{\smallskip}

PH$\alpha$ 30............    
& T  & A-B & 0.661 & 0.002 & 350.3 & 0.5 & 0.25 & 0.01 & 0.25 & 0.01 & 0.24 & 0.01 &&12.70 & 0.03 & 12.39 & 0.03 && 14.19 & 0.06 & 13.97 & 0.05 & 1 & B \\
&     & A-C & 11.58 & 0.01   & 195.7 & 0.5 & 0.99 & 0.02 & 1.02 & 0.02 & 0.95 & 0.01 &&\multicolumn{2}{c}{}&\multicolumn{2}{c}{}&& 12.81 & 0.02 & 12.60 & 0.03 & \\
\noalign{\smallskip}

CGH$\alpha$ 5/6........    
& T & A-B & 0.482    & 0.001 & 306.8 & 0.5 & 0.59 & 0.01 & 0.91 & 0.01 & 0.92 & 0.01 && 10.04 & 0.02 & 9.82 & 0.03 && 10.61 & 0.03 & 9.91 & 0.03 & 1 & T \\
&     & A-C & 11.289 & 0.007 & 317.4 & 0.5 & 0.31 & 0.03 & 0.54 & 0.03 & 0.60 & 0.01 &&\multicolumn{2}{c}{}&\multicolumn{2}{c}{}&& 10.82 & 0.03 & 10.02 & 0.02 & \\
\noalign{\smallskip}

VW Cha...........                  
& Q &  A-B & 0.653 & 0.001 & 177.2 & 0.5 & 0.39 & 0.01 & 0.25 & 0.01 & 0.24 & 0.01 && 8.17 & 0.07 & 7.32 & 0.04 && 9.19 & 0.10 & 8.87 & 0.07 & 3, 9, 10 & Q \\
&     &  B-C & 0.113 & 0.001 & 232.9 & 0.5 & 0.62 & 0.05 & 0.61 & 0.03 & 0.62 & 0.04 &&\multicolumn{2}{c}{}&\multicolumn{2}{c}{}&& 9.71 & 0.19 & 9.40 & 0.13 & \\
&     &  A-D & 16.78 & 0.01   & 220.4 & 0.5 & 0.03 & 0.01 & 0.04 & 0.01 & 0.05 & 0.01 &&\multicolumn{2}{c}{}&\multicolumn{2}{c}{}&& 10.86 & 0.02 & 9.89 & 0.02 & \\
\noalign{\smallskip}

Sz 30.................  
& T & A-B & 1.237 & 0.004 & 19.4 & 0.5 &\multicolumn{2}{c}{...}&\multicolumn{2}{c}{...}& 0.77 & 0.01 
&&\multicolumn{2}{c}{...}& 9.58 & 0.05 &&\multicolumn{2}{c}{...}& 9.86	& 0.05 & 2 & T \\
&    & B-C & 3.854 & 0.004 & 85.3 & 0.5 &\multicolumn{2}{c}{...}& \multicolumn{2}{c}{...}& 0.86 & 0.05 
&&\multicolumn{2}{c}{}&\multicolumn{2}{c}{}&&\multicolumn{2}{c}{...}& 9.98 & 0.03 & \\
&    & A-C & 4.504 & 0.002 & 70.8 & 0.5 &\multicolumn{2}{c}{...}& \multicolumn{2}{c}{...}& 0.66 & 0.03 
&&\multicolumn{2}{c}{}&\multicolumn{2}{c}{}&&\multicolumn{2}{c}{}&\multicolumn{2}{c}{}& \\
\noalign{\smallskip}

Sz 41.................  
& T &  A-B & 1.974 & 0.002 & 162.4 & 0.5 & 0.13 & 0.01 & 0.11 & 0.01 & 0.12 & 0.01 && 8.66 & 0.06 & 8.12 & 0.03 && 10.87 & 0.07 & 10.45 & 0.04 & 2, 15 & B$^b$ \\
&    &  A-C & 11.46 & 0.01 & 66.1   & 0.5 & 0.23 & 0.02 & 0.26 & 0.02 & 0.22 & 0.02 &&\multicolumn{2}{c}{}&\multicolumn{2}{c}{}&& 8.72 & 0.05 & 8.57 & 0.02 &  \\
\noalign{\smallskip}

LkH$\alpha$ 336.......... 
& T & A-B & 10.777 & 0.008 & 95.5  & 0.5 & 0.26 & 0.01 & 0.27 & 0.01 & 0.27 & 0.01 && 9.72 & 0.03 & 9.15 & 0.03 && 12.00 & 0.05 & 11.47 & 0.03 & 2 & T \\
&     & B-C & 8.199  & 0.007 & 243.4 & 0.5 & 0.46 & 0.01 & 0.39 & 0.01 & 0.41 & 0.01 &&\multicolumn{2}{c}{}&\multicolumn{2}{c}{}&& 11.41 & 0.03 & 10.73 & 0.02 & \\
&     & A-C & 5.799  & 0.004 & 144.2 & 0.5 & 0.12 & 0.01 & 0.10 & 0.01 & 0.11 & 0.01 &&\multicolumn{2}{c}{}&\multicolumn{2}{c}{}&&\multicolumn{2}{c}{}&\multicolumn{2}{c}{}& \\
\noalign{\smallskip}

J 4872...............        		
& Q & A-B & 0.173 & 0.001 & 74.8  & 0.5   & 0.86 & 0.01 & 0.86 & 0.01 & 0.85 & 0.01 && 9.53 & 0.08 & 9.63 & 0.05 && 9.69 & 0.09 & 9.80 & 0.05 & 4, 5, 1 & Q \\
&     & C-D & 0.102 & 0.001 & 157.9 & 1.3 & 0.76 & 0.04 & 0.71 & 0.01 & 0.73 & 0.01 && 10.42 & 0.17 & 10.39 & 0.06 && 10.71 & 0.14 & 10.74 & 0.07 & \\   
&     & A-C & 3.339 & 0.003 & 234.1 & 0.5 & 0.44 & 0.04 & 0.49 & 0.01 & 0.50 & 0.01 &&\multicolumn{2}{c}{}&\multicolumn{2}{c}{}&&\multicolumn{2}{c}{}&\multicolumn{2}{c}{}& \\   
\noalign{\smallskip}

UX Tau.............          		
& Q & A-B & 5.856 & 0.003 & 269.7 & 0.5 & 0.12 & 0.01 & 0.11 & 0.01 & 0.13 & 0.01 && 8.03 & 0.02 & 7.61 & 0.02 && 9.58 & 0.05 & 9.56 & 0.02 & 2, 4 & Q \\
&     & B-C & 0.136 & 0.001 & 309.0 & 0.5 & 0.80 & 0.01 & 0.81 & 0.02 & 0.81 & 0.02 &&\multicolumn{2}{c}{}&\multicolumn{2}{c}{}&& 9.83 & 0.03 & 9.80 & 0.04 & \\
&     & A-D & 2.692 & 0.002 & 181.6 & 0.5 & 0.07 & 0.01 & 0.05 & 0.01 & 0.05 & 0.01 &&\multicolumn{2}{c}{}&\multicolumn{2}{c}{}&& 10.94 & 0.03 & 10.79 & 0.04 & \\
\noalign{\smallskip}

UZ Tau.............          		
& T & A-B & 3.560 & 0.006 & 273.5 & 0.5 & 0.22 & 0.04 & 0.22 & 0.02 & 0.24 & 0.01 &&\multicolumn{2}{c}{8.12}& 7.35 & 0.03 && 8.46 & 0.04 & 7.93 & 0.03 & 6, 7, 8 & T \\
&    & B-C & 0.365 & 0.006 & 4.3      & 3.0 & 0.52 & 0.03 & 0.51 & 0.01 & 0.52 & 0.01 &&\multicolumn{2}{c}{}&\multicolumn{2}{c}{}&& 9.18 & 0.11 & 8.64 & 0.05 & \\
\noalign{\smallskip}

FV Tau.............          		
& Q & A-B & 0.780    & 0.004 & 273.2 & 0.6 & 0.16 & 0.01 &  0.24 & 0.01 & 0.29 & 0.01 && 8.49 & 0.04 & 7.72 & 0.03 && 10.48 & 0.11 & 9.07 & 0.08 & 6 & Q \\
&     & C-D & 0.693   & 0.003 & 294.6 & 0.7 & 0.11 & 0.02 & 0.15 & 0.02 & 0.18 & 0.02 && 9.60 & 0.04 & 9.05 & 0.04 && 12.03 & 0.20 & 10.91 & 0.13 & \\
&     & A-C & 12.081 & 0.009 & 106.1 & 0.5 & 0.71 & 0.01 & 0.56 & 0.01 & 0.51 & 0.01 &&\multicolumn{2}{c}{}&\multicolumn{2}{c}{}&&\multicolumn{2}{c}{}&\multicolumn{2}{c}{}& \\
\noalign{\smallskip}

GG Tau.............          		
& Q & A-B & 0.249   & 0.001  & 346.0 & 0.5 & 0.58 & 0.02 & 0.57 & 0.01 & 0.58 & 0.01 && 8.31 & 0.04 & 7.86 & 0.02 && 8.91 & 0.07 & 8.46 & 0.04 & 11 & Q \\
&     & C-D & 1.460   & 0.003 & 135.3 & 0.5 & 0.21 & 0.01 & 0.20 & 0.01 & 0.21 & 0.01 && 10.60 & 0.03 & 10.18 & 0.03 && 12.30 & 0.06 & 11.87 & 0.07 & \\
&     & A-C & 10.100 & 0.007 & 186.0 & 0.5 & 0.08 & 0.01 & 0.08 & 0.01 & 0.08 & 0.01 &&\multicolumn{2}{c}{}&\multicolumn{2}{c}{}&&\multicolumn{2}{c}{}&\multicolumn{2}{c}{}& \\
\noalign{\smallskip}

LkH$\alpha$ 262/263...    
& Q$^a$ & A-B & 0.414 & 0.001 &  52.0 & 0.5 & 1.12 & 0.02 & 0.87 & 0.01 & 0.88 & 0.01 && 10.66 & 0.03 & 10.18 & 0.02 && 10.53 & 0.04 & 10.32 & 0.03 & 12, 13 & Q \\
&    	       & A-D & 15.38 & 0.01 & 202.5  & 0.5 & 0.90 & 0.06 & 1.05 & 0.03 & 0.99 & 0.05 &&\multicolumn{2}{c}{}&\multicolumn{2}{c}{}&& 10.29 & 0.02 & 9.64 & 0.02 & \\
\noalign{\smallskip}

LkH$\alpha$ 346 ........... 	
& Q & A-B & 0.204 & 0.002 & 256.5 & 1.0 & 0.24  & 0.03 &\multicolumn{2}{c}{...}&\multicolumn{2}{c}{...}  
&& 9.23 & 0.07 & \multicolumn{2}{c}{...} && 10.8 & 0.2 & \multicolumn{2}{c}{...} & 1 & Q \\
&     & A-C & 1.086 & 0.001 & 328.2 & 0.5 & 0.26  & 0.01 &\multicolumn{2}{c}{...}&\multicolumn{2}{c}{...}  
&&\multicolumn{2}{c}{}&\multicolumn{2}{c}{}&& 10.70 & 0.08 &  \multicolumn{2}{c}{...} & \\
&     & A-D & 5.068 & 0.006 & 329.7 & 0.5 & 0.60  & 0.01 &\multicolumn{2}{c}{...}&\multicolumn{2}{c}{...}  
&&\multicolumn{2}{c}{}&\multicolumn{2}{c}{}&& 8.99 & 0.03 & \multicolumn{2}{c}{...} & \\
\noalign{\smallskip}

ROXs 43 .............    		
& T & A-B & 0.334 & 0.001 & 168.9 & 0.5 & 0.05 & 0.01 &\multicolumn{2}{c}{...}&\multicolumn{2}{c}{...} 
&& 7.22 & 0.06	& \multicolumn{2}{c}{...} && 10.54 & 0.07 & \multicolumn{2}{c}{...} & 1 & T \\
&    & A-C & 4.487 & 0.001 & 12.1   & 0.5 & 0.64 & 0.01 &\multicolumn{2}{c}{...}&\multicolumn{2}{c}{...}  
&&\multicolumn{2}{c}{}&\multicolumn{2}{c}{}&& 7.48 & 0.08 & \multicolumn{2}{c}{...} & \\
\noalign{\smallskip}

SR 24 .................   		
& T & A-B & 5.068 & 0.001 & 348.5 & 0.5 & 0.17 & 0.01 &\multicolumn{2}{c}{...}&\multicolumn{2}{c}{...} 
&& 8.17 & 0.04 & \multicolumn{2}{c}{...} && 9.1 & 0.2	 & \multicolumn{2}{c}{...} & 14 & T \\
&    & B-C & 0.081 & 0.001 & 45.6   & 0.5 & 0.5   &  0.1   &\multicolumn{2}{c}{...}&\multicolumn{2}{c}{...}  
&&\multicolumn{2}{c}{}&\multicolumn{2}{c}{}&& 9.8 & 0.4	& \multicolumn{2}{c}{...} & \\
\noalign{\smallskip}

Sz 68 ..................    		
& T & A-B & 0.126 & 0.001 & 246.7 & 0.5 & 0.16 & 0.01 &\multicolumn{2}{c}{...}&\multicolumn{2}{c}{...} 
&& 7.09 & 0.03	& \multicolumn{2}{c}{...} && 9.05 & 0.05 & \multicolumn{2}{c}{...} & 9 & T \\
&    & A-C & 2.808 & 0.002 & 296.8 & 0.5 & 0.07 & 0.01 &\multicolumn{2}{c}{...}&\multicolumn{2}{c}{...} 
&&\multicolumn{2}{c}{}&\multicolumn{2}{c}{}&& 10.05 & 0.05 & \multicolumn{2}{c}{...} & \\

\noalign{\smallskip}
\hline
\noalign{\smallskip}
\end{tabular}
\end{center}
\label{Tab:systems_parameters}
\begin{minipage}[position]{18cm}
$^a$\,: includes the edge-on disk LkH$\alpha$ 262/263C (Jayawardhana et al.\,\cite{Jayawardhana_etal2002}, Chauvin et al.\,\cite{Chauvin_etal2002}) 
             undetected in our survey. \\  
$^b$\,: Sz\,41\,C was unambiguously identified as a background giant (Walter\,\cite{Walter_1992}). \\  
Notes\,: multiplicity abbreviations are B=Binary, T=Triple and Q=Quadruple.\\
Multiplicity references\,:  
(1) This work. 
(2) RZ93. 
(3) Brandeker et al.\,\cite{Brandeker_etal2001}. 
(4) Duch\^{e}ne\,\cite{Duchene_1999}.  
(5) White \& Ghez\,\cite{White_Ghez_2001}. 
(6) Simon et al.\,\cite{Simon_etal1992}. 
(7) Leinert et al.\,\cite{Leinert_etal1993}. 
(8) Ghez et al.\,\cite{Ghez_etal1993}. 
(9) Ghez et al.\,\cite{Ghez_etal1997a}.
(10) Brandner et al.\,\cite{Brandner_etal1996}.
(11) Leinert et al.\,\cite{Leinert_etal1991}.
(12) Jayawardhana et al.\,\cite{Jayawardhana_etal2002}.
(13) Chauvin et al.\,\cite{Chauvin_etal2002}.
(14) Simon et al.\,\cite{Simon_etal1995}.
(15) Walter \,\cite{Walter_1992}.
\end{minipage}
\end{table*}

\begin{table*}
\scriptsize
\caption{Upper limits for the relative brightness of undetected companions to both components of the binaries in our survey, 
measured at 0$\farcs$07, 0$\farcs$15 and 0$\farcs$5, as well as the limiting magnitudes at large separations.}
\begin{center}
\renewcommand{\arraystretch}{0.7}
\setlength\tabcolsep{5pt}
\begin{tabular}{l@{\hspace{2mm}}r@{\,$\pm$\,}l@{\hspace{2mm}}r@{\,$\pm$\,}l@{\hspace{2mm}}c@{\hspace{2mm}}cc@{\hspace{2mm}}c@{\hspace{2mm}}c@{\hspace{2mm}}cc@{\hspace{2mm}}c@{\hspace{2mm}}c@{\hspace{2mm}}cc@{\hspace{2mm}}c@{\hspace{2mm}}c@{\hspace{2mm}}cc@{\hspace{2mm}}c@{\hspace{2mm}}c@{\hspace{2mm}}cc@{\hspace{1mm}}rr}
\hline\noalign{\smallskip}
 	&\multicolumn{2}{c}{}&\multicolumn{2}{c}{}& &  & & \multicolumn{8}{c}{Upper limits in [FeII]} & \multicolumn{7}{c}{Upper limits in Br$\gamma$}  && \multicolumn{2}{c}{Limiting}\\
\cline{9-15}  \cline{17-23}  \\
System &   \multicolumn{2}{c}{Separation}  &    \multicolumn{2}{c}{P.A.$^a$}     &   \multicolumn{2}{l}{Flux ratio}$^a$  & & \multicolumn{4}{c}{comp. A}   & \multicolumn{4}{c}{comp. B}  & \multicolumn{4}{c}{comp. A}   & \multicolumn{3}{c}{comp. B} && \multicolumn{2}{c}{Magnitudes} \\
\cline{6-7} \cline{9-11}  \cline{13-15}  \cline{17-19}   \cline{21-23} \cline{25-26} \\                         
                         &    \multicolumn{2}{c}{[arcsec]}       &    \multicolumn{2}{c}{[$^{\circ}$]}  &    [FeII]  &   Br$\gamma$  &  & 0$\farcs$07 & 0$\farcs$15 & 0$\farcs$5 & & 0$\farcs$07 & 0$\farcs$15 & 0$\farcs$5 & & 0$\farcs$07 & 0$\farcs$15 & 0$\farcs$5 & & 0$\farcs$07 & 0$\farcs$15 & 0$\farcs$5 && H & K  \\
\noalign{\smallskip}
\hline
\noalign{\smallskip}

AR Ori ............. &  2.002 & 0.001 & 239.4 & 0.5 & 0.74 & 0.55 
&& 0.19 & 0.18 & 0.08 && 0.18 & 0.17 & 0.13 
&& 0.17 & 0.12 & 0.05 && 0.17 & 0.12 & 0.11 
&& 13.8 & 13.6 \\
CoD $-$29$\degr$8887 . &  0.515 & 0.003 &  28.2 & 0.5 & 0.52 & 0.51 
&& 0.04 & 0.02 & 0.007 && 0.04 & 0.02 & 0.02 
&& 0.03 & 0.01 & 0.005 && 0.03 & 0.01 & 0.01
&& 12.9 & 12.8 \\
CO Ori .............. &  2.058 & 0.004 & 274.7 & 0.5 & 0.08 & 0.07 
&& 0.39 & 0.13 & 0.02 && 0.39 & 0.31 & 0.25 
&& 0.15 & 0.11 & 0.007 && 0.15 & 0.14 & 0.11
&& 12.3 & 12.5 \\
CV Cha ............. & 11.203 & 0.005 &  98.7 & 0.5 & 0.12 & 0.14 
&& 0.76 & 0.23 & 0.04 && 0.79 & 0.63 & 0.51 
&& 0.26 & 0.14 & 0.02 && 0.28 & 0.30 & 0.23 
&& 11.1 & 11.2 \\
DK Tau ............. &  2.360 & 0.001 & 119.0 & 0.5 & 0.33 & 0.28 
&& 0.06 & 0.05 & 0.01 && 0.06 & 0.04 & 0.03 
&& 0.04 & 0.04 & 0.01 && 0.05 & 0.04 & 0.03 
&& 13.4 & 12.9 \\
ESO H$\alpha$ 281 ..... &  1.748 & 0.001 & 172.6 & 0.5 & 0.21 & 0.22 
&& 0.09 & 0.07 & 0.02 && 0.09 & 0.07 & 0.06 
&& 0.09 & 0.05 & 0.01 && 0.09 & 0.05 & 0.05
&& 15.3 & 15.1 \\
Glass I$^b$ ............. &  2.430 & 0.002 & 285.1 & 0.5 & 0.27 & 0.23 
&& 0.10 & 0.07 & 0.02 && 0.09 & 0.06 & 0.05 
&& 0.05 & 0.02 & 0.004 && 0.04 & 0.02 & 0.02 
&& 13.0 & 13.4 \\
HD 76534 ......... &  2.073 & 0.002 & 303.6 & 0.5 & 0.36 & 0.33 
&& 0.09 & 0.08 & 0.02 && 0.09 & 0.08 & 0.07 
&& 0.07 & 0.07 & 0.02 && 0.07 & 0.07 & 0.07
&& 12.4 & 12.4 \\
Hen 3-600 ......... &  1.481 & 0.003 & 213.4 & 0.5 & 0.69 & 0.68 
&& 0.05 & 0.04 & 0.02 && 0.05 & 0.04 & 0.03 
&& 0.03 & 0.03 & 0.01 && 0.03 & 0.03 & 0.02
&& 11.8 & 12.2 \\
Herschel 4636 ... &  3.754 & 0.004 & 213.8 & 0.5 & 0.23 & 0.11 
&& 0.10 & 0.05 & 0.005 && 0.10 & 0.05 & 0.03 
&& 0.42 & 0.08 & 0.004 && 0.42 & 0.09 & 0.02
&& 14.0 & 13.6 \\
HK Tau ............. & 2.318 & 0.006 & 171.3 & 0.5 & ... & ... 
&& 0.70 & 0.08 & 0.03 &&...&...&...
&& 0.61 & 0.04 & 0.02 &&...&...& ...
&& 13.2 & 13.2 \\
HN Tau ............. &  3.142 & 0.001 & 219.7 & 0.5 &...& 0.12 
&&  ... &  ... &   ... &&  ... &  ... &  ... 
&& 0.31 & 0.09 & 0.02 && 0.41 & 0.33 & 0.33
&& ... & 13.0 \\
IT Tau ............... &  2.416 & 0.008 & 225.1 & 0.5 & 0.28 & 0.27 
&& 0.40 & 0.22 & 0.07 && 0.39 & 0.36 & 0.30 
&& 0.14 & 0.12 & 0.02 && 0.15 & 0.14 & 0.13
&& 12.0 & 12.3 \\
L1642-1 ............ &  2.744 & 0.001 & 348.7 & 0.5 & 0.36 & 0.23 
&& 0.17 & 0.15 & 0.03 && 0.17 & 0.15 & 0.13 
&& 0.11 & 0.09 & 0.02 && 0.11 & 0.09 & 0.09 
&& 12.5 & 12.5 \\
LkH$\alpha$ 266 ......... &  3.224 & 0.003 & 177.7 & 0.5 & 0.79 & 0.95 
&& 0.10 & 0.05 & 0.02 && 0.10 & 0.05 & 0.03 
&& 0.07 & 0.04 & 0.03 && 0.07 & 0.04 & 0.03
&& 14.0 & 13.9 \\
PH$\alpha$ 14 ............. &  0.650 & 0.006 & 168.3 & 0.5 & 0.85 & 0.93 
&& 0.05 & 0.05 & 0.02 && 0.05 & 0.04 & 0.02 
&& 0.06 & 0.05 & 0.02 && 0.06 & 0.05 & 0.02
&& 16.1 & 15.7 \\
PH$\alpha$ 51 ............. &  0.627 & 0.003 & 184.0 & 0.5 & 0.59 & 0.46 
&& 0.19 & 0.18 & 0.07 && 0.19 & 0.18 & 0.13 
&& 0.17 & 0.16 & 0.05 && 0.18 & 0.16 & 0.13
&& 15.2 & 14.8 \\
RW Aur ............ &  1.447 & 0.003 & 255.0 & 0.5 & 0.20 & 0.21 
&& 0.14 & 0.08 & 0.01 && 0.16 & 0.14 & 0.10 
&& 0.08 & 0.06 & 0.007 && 0.08 & 0.06 & 0.04
&& 12.9 & 13.1 \\
SX Cha ............. &  2.211 & 0.003 & 288.7 & 0.5 & 0.25 & 0.22 
&& 0.12 & 0.09 & 0.01 && 0.13 & 0.09 & 0.05 
&& 0.13 & 0.11 & 0.02 &  & 0.13 & 0.11 & 0.10 
&& 15.3 & 13.5 \\
Sz 15 ................. & 10.506 & 0.008 &  27.5 & 0.5 & 0.45 & 0.60 
&& 0.49 & 0.20 & 0.03 && 0.53 & 0.31 & 0.12 
&& 0.44 & 0.14 & 0.02 && 0.50 & 0.29 & 0.09
&& 14.9 & 14.9 \\
Sz 19 ................. & \multicolumn{2}{c}{...} & \multicolumn{2}{c}{...} & ... & ... 
&& 0.43 & 0.07 & 0.04 &&...&...&...
&& 0.56 & 0.07 & 0.04 &&...&...& ...
&& 10.5 & 9.7 \\
Sz 62 ................. &  1.116 & 0.005 & 267.3 & 0.5 & 0.60 & 0.59 
&& 0.09 & 0.09 & 0.03&& 0.10 & 0.08 & 0.06 
&& 0.08 & 0.07 & 0.03 && 0.08 & 0.08 & 0.06 
&& 14.2 & 13.7 \\
vBH 16 ............. &  4.509 & 0.003 & 329.8 & 0.5 & 0.59 & 0.60 
&& 0.20 & 0.18 & 0.09 && 0.21 & 0.18 & 0.13 
&& 0.14 & 0.10 & 0.04 && 0.14 & 0.10 & 0.06
&& 12.5 & 12.8 \\
VV Cha ............ &  0.790 & 0.001 &  13.4 & 0.5 & 0.72 & 0.63 
&& 0.22 & 0.19 & 0.10 && 0.21 & 0.19 & 0.14 
&& 0.25 & 0.23 & 0.10 && 0.24 & 0.23 & 0.17
&& 13.1 & 12.6 \\
BK Cha$^b$ ........... &  0.766 & 0.001 & 153.6 & 0.5 & 0.55 & ... 
&& 0.04 & 0.02 & 0.01 && 0.05 & 0.02 & 0.01 
&& ... & ... & ... &  & ... & ... & ...
&& 15.4 & ... \\
Elias 2-30 ......... &  6.317 & 0.001 & 175.3 & 0.5 & 0.06 & ... 
&& 0.28 & 0.11 & 0.005 && 0.28 & 0.13 & 0.10 
&& ... & ... & ... &  & ... & ... & ...
&& 13.7 & ... \\
Elias 2-49 ......... &  1.097 & 0.001 & 224.9 & 0.5 & 0.17 & ... 
&& 0.04 & 0.02 & 0.005 && 0.04 & 0.02 & 0.01 
&& ... & ... & ... &  & ... & ... & ...
&& 13.7 & ... \\
Haro 1-14c ....... & 13.047 & 0.001 & 122.6 & 0.5 & 0.42 & ... \
&& 0.28 & 0.03 & 0.003 && 0.92 & 0.27 & 0.06 
&& ... & ... & ... &  & ... & ... & ... 
&& 14.4 & ... \\
HBC 652 .......... &  8.695 & 0.001 & 321.7 & 0.5 & 0.13 & ... 
&& 0.23 & 0.07 & 0.005 && 0.18 & 0.07 & 0.02 
&& ... & ... & ... &  & ... & ... & ...
&& 14.7 & ... \\
HO Lup ............ &  1.536 & 0.001 &  34.9 & 0.5 & 0.24 & ... 
&& 0.09 & 0.06 & 0.02 && 0.09 & 0.06 & 0.05 
&& ... & ... & ... &  & ... & ... & ...
&& 14.3 & ... \\
Sz 101 .............. &  0.771 & 0.001 & 302.9 & 0.5 & 0.54 & ... 
&& 0.04 & 0.02 & 0.01 && 0.05 & 0.02 & 0.02 
&& ... & ... & ... &  & ... & ... & ...
&& 15.3 & ... \\
Sz 108 .............. &  4.029 & 0.001 &  24.5 & 0.5 & 0.14 & ... 
&& 0.07 & 0.04 & 0.009 && 0.07 & 0.04 & 0.04 
&& ... & ... & ... &  & ... & ... & ...
&& 14.8 & ... \\
Sz 120 .............. &  2.600 & 0.001 & 140.8 & 0.5 & 0.09 & ... 
&& 0.03 & 0.01 & 0.002 && 0.03 & 0.01 & 0.01 
&& ... & ... & ... &  & ... & ... & ...
&& 14.1 & ... \\
Sz 48 ................ &  1.480 & 0.001 & 232.5 & 0.5 & 0.58 & ... 
&& 0.12 & 0.10 & 0.06 && 0.12 & 0.10 & 0.08 
&& ... & ... & ... &  & ... & ... & ... 
&& 14.2 & ... \\
Sz 60$^b$ .............. &  3.347 & 0.001 &  99.7 & 0.5 & 0.80 & ... 
&& 0.07 & 0.05 & 0.02 && 0.07 & 0.05 & 0.02 
&& ... & ... & ... &  & ... & ... & ... 
&& 14.6 & ... \\
Sz 65 ................ &  6.371 & 0.001 &  97.8 & 0.5 & 0.31 & ... 
&& 0.06 & 0.03 & 0.005 &  & 0.06 & 0.03 & 0.02 
&& ... & ... & ... &  & ... & ... & ...
&& 14.5 & ... \\
WSB 11 ........... &  0.472 & 0.001 & 316.8 & 0.5 & 0.89 & ... 
&& 0.11 & 0.09 & 0.04 && 0.11 & 0.09 & 0.06 
&& ... & ... & ... &  & ... & ... & ... 
&& 15.2 & ... \\
WSB 20 ........... &  0.779 & 0.002 &  31.6 & 0.5 & 0.46 & ... 
&& 0.02 & 0.01 & 0.006 && 0.02 & 0.01 & 0.01 
&& ... & ... & ... &  & ... & ... & ... 
&& 14.5 & ... \\
WSB 28 ........... &  5.119 & 0.001 & 358.3 & 0.5 & 0.07 & ... 
&& 0.27 & 0.26 & 0.03 && 0.27 & 0.27 & 0.23 
&& ... & ... & ... &  & ... & ... & ...
&& 14.0 & ... \\
WSB 3 ............. &  0.600 & 0.004 & 160.2 & 0.5 & 0.65 & ... 
&& 0.08 & 0.07 & 0.03 && 0.08 & 0.07 & 0.06 
&& ... & ... & ... &  & ... & ... & ...
&& 14.0 & ... \\
WSB 46 ........... & 10.209 & 0.001 & 301.6 & 0.5 & 0.50 & ... 
&& 0.24 & 0.16 & 0.03 && 0.24 & 0.15 & 0.04 
&& ... & ... & ... &  & ... & ... & ...
&& 14.6 & ... \\

\noalign{\smallskip}
\hline
\noalign{\smallskip}
\end{tabular}
\end{center}
\label{Tab:undetec_comp}
\begin{minipage}[position]{18cm}
$^a$\,: Assumes as for the triples/quadruples that the primary is the brightest component in Br$\gamma$ and [FeII], 
	    for respectively the first and second data set. \\  
$^b$\,: Components inverted with respect to RZ93, given the above assumption.\\
Notes\,: Separations and P.A.s are computed from aperture photometry and the uncertainties combine the rms variations over the individual frames and the uncertainties 
due to the plate scale and detector orientation.\\	    
\end{minipage}
\end{table*}

We mainly searched for unresolved pairs in each individual component of the wide binaries of the sample. 
Nevertheless some of the targets of the sample were so poorly known and observed that it has been possible 
to detect new distant companion candidates. Indeed there is, in some cases, an obvious lack of NIR follow-up 
observations since the survey of RZ93 which was performed with Gunn-z filter ($\lambda \sim $0.9$\mu$m). 

Three of the triple candidate (CGH$\alpha$\,5/6, PH$\alpha$\,30 and ROXs\,43) and two of the quadruple candidate 
systems (ESO H$\alpha$\,283 and LkH$\alpha$\,346) are new detections. 
CGH$\alpha$\,6 was known from~RZ93 to form a binary with the 11$\farcs$1 north-western 
companion CGH$\alpha$\,5. Here we resolved for the first time CGH$\alpha$\,6 as a close 0$\farcs$5 binary, making it 
a hierarchical triple. PH$\alpha$\,30 is a 0$\farcs$67 binary with a wide companion found in our survey 
at 11$\farcs$9 separation. ESO H$\alpha$\,283 appears to be a possible quadruple system composed of the original 
2$\farcs$08 binary detected by~~RZ93 plus a third (C) and a faint fourth (D) component located 2$\farcs$57 SE and 
9$\farcs$3 SW of the primary, respectively.  The LkH$\alpha$\,346 system contains two additional candidate companions. The 
first one is located $\sim$0$\farcs$2 from the primary of the system (spectral type K7, aka LkH$\alpha$\,346\,NW), while the second 
one is lying approximately in between the primary and the secondary (spectral type M5, aka LkH$\alpha$\,346\,SE). With flux ratios of 
$\sim$0.3 with respect to the M5 component, both might well be substellars, if bound. This system is further discussed in 
Appendix\,\ref{sect:comments_ind_objects}. 
ROXs\,43\,A (NTTS 162819-2423S) is a 89-days single-lined spectroscopic PMS binary in the $\rho$ Oph cloud 
(Mathieu et al.\,\cite{Mathieu_etal1989}), with eccentricity e=0.4 and {\it a}.sin(i)=0.1 AU. It is part of a hierarchical quadruple system, 
ROXs\,43\,A lying 4$\farcs$8 south from ROXs\,43\,B (NTTS 162819-2423N), a 0$\farcs$016 binary (Simon et al.\,\cite{Simon_etal1995}). 
We found a candidate fifth component in this system 0$\farcs$3 from ROXs\,43\,A, corresponding to 42\,AU at 140\,pc, with flux ratio 
$\sim$0.05 in H-band. Interestingly, this candidate companion was not detected in a recent speckle survey of $\rho$ Oph (Ratzka et al.\,\cite{Ratzka_etal2005}). 
Part of the near-IR excess of ROXs\,43\,A could originate from this candidate companion. Jensen \& Mathieu 
(\cite{Jensen_Mathieu_1997}) fitted a circumbinary disk with gap to the SED of ROXs 43A. If the candidate companion is physically 
related, the issue of the spatial distribution of circumstellar/cicumbinary material would have to be re-addressed.   
Additionally, J\,4872\,B was already known to be a close pair (White \& Ghez\,\cite{White_Ghez_2001}), forming a quadruple with the 
other pair J\,4872\,A, but no measurements were published before ours.


\subsection{Limits for undetected companions and completeness}
\label{sect:sens_limit_completeness}

Table\,\ref{Tab:undetec_comp} shows the upper limits for undetected companions to the only binaries of our sample at separations 0$\farcs$07, 0$\farcs$15 and 0$\farcs$5 in [FeII] and Br$\gamma$, as well as the estimated limiting magnitudes at large separations in H and K. The detection limits were computed using the 
following method. At each radial distance and position angle from a star, the standard deviation of the flux was calculated over a circular region of radius 70\,mas, i.e. equivalent to the mean size of the PSF cores. The detection limit as a function of separation to a star is the average of the 5$\sigma$ flux over all 
position angles excepted those lying in the direction of the companion. We repeated the procedure with the PSF-subtracted images, and defined the total detection 
limit as the minimum of the two detection limits. The sensitivity is increased by the PSF-subtraction at separations smaller than typically 0$\farcs$1-0$\farcs$3. 
Our procedure was checked by adding artificial companions at a few separations and position angles, which are scaled versions of the target star PSF, and looking for the maximum flux ratio at which the companion is detected by visual inspection. A statisfying agreement was found between the limiting flux ratio given by the two methods.

Concerning the limiting magnitude at large separations, these were calculated by combining 2MASS H- and K-magnitudes of resolved or unresolved systems and our 
limiting flux ratios in [FeII] and Br$\gamma$. We checked our estimated magnitude using undetected or marginally detected stars in our frames which are in 
common with 2MASS and found a relatively good agreement. The discrepancies are up to $\sim$1\,mag and can be explained either by possible photometric variability between 2MASS and our observation epochs or by the fact that we assumed that the relative flux in H and K is given by that in [FeII] and Br$\gamma$. 
The more remarkable cases are further discussed in Sect.\,\ref{sect:chance_proj}. 
 
\begin{figure}
\centering
\begin{tabular}{c}
\includegraphics[width=8.7cm]{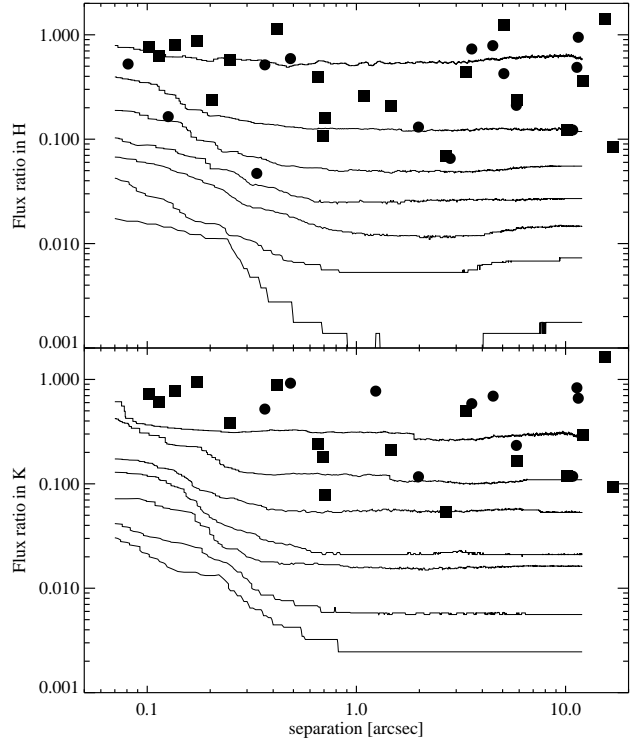}\\
\end{tabular}
\caption{Results of our survey in a plot of flux ratio or magnitude difference in H and K vs. pair separation. The circle symbol refers to pairs in triple 
systems while square symbols are for pairs in quadruple systems. The lines represent the completeness of our observations as derived from the 
sensitivity limit for undetected companions of the only binaries of the sample (see Sect.\,\ref{sect:sens_limit_completeness}). 
The lines are, from top to bottom, 99\%, 90\%, 70\%, 50\%, 30\%, 10\% and 1\% completeness.}
\label{fig:diff_H_K_vs_sep}
\end{figure}

For any given component of the binaries we thus have the limiting flux ratio for undetected companion as a function of separation. We can therefore produce 
a completeness map for each of these sources, i.e. a map giving the probability to detect a companion as a function of separation and magnitude difference 
for each source. The underlying assumption is that all companions above the sensitivity limits are detected, i.e. there is a sharp limit in sensitivity. 
This might not be completely true especially for AO imaging data because of the anisotropy of the correction, but this simplification should still provide 
meaningful results (see Tokovinin et al.\,\cite{Tokovinin_etal2006} and K{\"o}hler et al.\,\cite{Koehler_etal2006} for further discussions). 
The total completeness map of all the sources is simply the 
average of all individual completeness maps. We will further assume in the following that the total completeness map created this way is statistically equivalent 
to that obtained if we would have included the triple and quadruple systems. Fig.\,\ref{fig:diff_H_K_vs_sep} shows the total completeness map in H and K with 
the detected companions of the triple/quadruple systems overploted. The majority of the companions fall above the 90\% completeness level and all are above 
the 50\% level, giving us confidence in these limits.

Assuming that both the distribution of separation and the distribution of flux ratio of companions are flat, which is obviously a very rough assumption, 
we estimate that the completeness within the separation range 0$\farcs$07-10$\arcsec$ and above a flux ratio of 0.001 is of the order of 95\%. 
Given the statistical errors involved in this study, which are a direct consequence of the sample size, the correction for incompleteness is considered 
not significant enough to be applied. It should be noted that systems observed with the S13 camera were not probed for companions at large separations 
(says $\la$\,7-8$\arcsec$, depending of the position of the system on the detector). Even though the brighter companions at such separations would show 
up in 2MASS, the faintest would likely be missed.

\begin{figure}
\centering
\includegraphics[width=8.5cm]{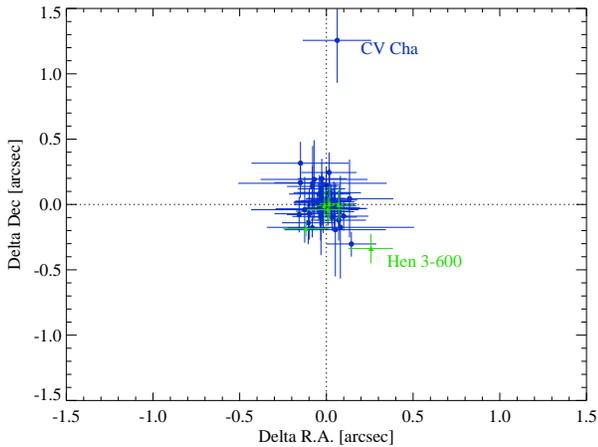}
\caption{Relative motion in RA and Dec between the RZ93 positions and those measured in this study (NACO-RZ93). 
Green triangles are pairs with projected separations $<$\,150\,AU.}
\label{fig:cpm}
\end{figure}

\subsection{Microjets and Herbig-Haro objects}
\label{subsect: HH-jets}
Since the exposure times for all sources was scaled in order to avoid saturation, the sensitivity to extended emission is limited. 
We could however be sensitive to bright Herbig-Haro (HH) objects and/or HH-jets close to the star by using PSF-subtraction techniques 
in H$_2$ and [FeII]. This is of particular interest in view of the known correlation of outflows and multiple stars (Reipurth \cite{Reipurth_2000}), 
even associated with relatively old sources like classical T Tauri stars (e.g. Mundt \& Eisl\"{o}ffel \cite{Mundt_Eisloeffel_1998}). 
Similarly, dynamical interaction in multiple stars systems following an ejection of one component by close triple encounter are also suspected 
to be at the origin of the FU Orionis phenomenon (Reipurth \& Aspin \cite{Reipurth_Aspin_2004}). 

Our clear non detection of any prominent outflows/jets could however be explained by a lack of sensitivity since some sources known to harbor jets 
were not detected. The bipolar jet originating from RW Aur A (HH 229) is a good example in this respect. First identified spectroscopically 
by Hirth et al. (\cite{Hirth_etal1994}), its inner part (0$\farcs$1-10") has later been imaged as a bipolar micro-jet in the optical forbidden 
emission-lines [OI] and [SII] (Dougados et al.\,\cite{Dougados_etal2000}), but was already known to have a large extension, up to 100" 
(Mundt \& Eisl\"{o}ffel \cite{Mundt_Eisloeffel_1998}), and could be even more extended if HH 835, some 5$\arcmin$ distant, 
is associated with the outflow (McGroarty \& Ray \cite{McGroarty_Ray_2004}). Recently, the microjet has been detected in [FeII] 
(Davis et al. \cite{Davis_etal2002}, Pyo et al. \cite{Pyo_etal2005}), and both high and low-velocity components identified. 
Our detection limit is H=12.9 which translates to $\sim$\,9.4 mag.arcsec$^{-2}$, which is probably 1-2mag brighter than typical [FeII] fluxes 
measured from HH-objects (e.g. Gredel \cite{Gredel_1994}). Similarly, probably for the same reason, the recently discovered microjet in UZ Tau E, 
imaged in [OI] using HST/STIS slitless spectroscopy (Hartigan et al. \cite{Hartigan_etal2004}), was not detected in our images in [FeII] and H2. 

However, we note that we might have detected some outflow emission from another known HH-jet, namely HH 186, whose driving source is Sz\,68 
(Heyer \& Graham \cite{Heyer_Graham_1989}). This is a highly collimated jet of forbidden [SII] emission extending 34$\arcsec$ at a position angle of 
135$\degr$, and composed of three emission peaks at 21$\arcsec$, 28$\arcsec$ and 34$\arcsec$. These knots are outside our field of view and we did 
not detect any jet emission on this side of Sz\,68\,A. But, we detected a faint nebulosity at PA$\sim$295$\degr$ and separation $\sim$1$\farcs$27 
which could be the possible counter jet quoted in Heyer \& Graham (\cite{Heyer_Graham_1989}). Deeper imaging is necessary in order to confirm this result.

\subsection{Relative motions}
\label{subsect:relative_motion}
Fig.\,\ref{fig:cpm} shows the variations of relative positions between our measurements and the values reported in RZ93. 
Uncertainties of the RZ93 astrometry were considered quite conservatively and include relative position measurement errors of $\pm$0$\farcs$13 (1\,pixel) 
as well as plate scale and detector orientation uncertainties of 0$\farcs$0013 (1\,\% of plate scale) and $\pm$1.0$^{\circ}$, respectively. 
While most of the pairs for which this comparison was possible shows no significant relative motion, a few cases retained our attention. 
The wide pair CV\,Cha ($\sim$ 10$\arcsec$ separation) is a clear outlier and the two components of this visual pair, CV\,Cha and CW\,Cha, 
are most probably not physical. Hen\,3-600 (TWA\,3) is another binary which shows a significant relative motion and deserves further investigations 
(see Appendix\,\ref{sect:comments_ind_objects}) 
Another interesting point is that the seemingly non-hierachical systems LkH$\alpha$\,346 (AB and AC pair ), Sz\,30 (AB and AC pair), 
and UX\,Tau (AB pair) does not show any significant relative motion within the $\sim$\,13\,yrs time span.

In some cases, it was possible to compare the inferred relative motion with measured proper-motions (e.g. Ducourant et al.\,\cite{Ducourant_etal2005}). 
In addition, among the four systems with resolved proper-motions in the catalog of Ducourant et al. (\cite{Ducourant_etal2005}), namely VW\,Cha/Sz\,23, 
CV\,Cha/CW\,Cha, FV\,Tau/FV\,Tau\c, LkH$\alpha$\,263/LkH$\alpha$\,262, only CV\,Cha/CW\,Cha shows a significant (more than 3 sigma) relative proper-motion.     
Although the relative proper motion of the CV\,Cha system is mainly in declination (also confirmed by Tycho-2 catalog), which is consistent with the relative motion 
we see here and would tend to favor the chance projection scenario, its amplitude would lead to a relative motion not exceeding $\sim$0$\farcs$3 in 13\,yrs. 
We therefore suggest either an underestimate of the uncertainties on the proper-motions or an error in the astrometry reported in RZ93. 

\subsection{Chance projections}
\label{sect:chance_proj}

\begin{table}
\scriptsize
\caption{Probability of unrelated companions to the triple/quadruple systems as estimated from 2MASS.}
\begin{center}
\renewcommand{\arraystretch}{0.5}
\setlength\tabcolsep{7pt}
\begin{tabular}{lccccc}
\hline\noalign{\smallskip}
System                       		& Comp. 		&	$\Sigma$ (K$<$K$_{comp}$) 		& P$_{unrelated}$ \\
                             		& 			&	[arcsec$^{-2}$]   				& \\
\noalign{\smallskip}
\hline
\noalign{\smallskip}
ESO H$\alpha$ 283....      &   	C		&	1.36\,$10^{-3}$    				&  2.9\,$10^{-2}$ \\
					&   	D		&	1.70\,$10^{-3}$    				&  0.37 \\
PH$\alpha$ 30............    	&   	C		&	2.19\,$10^{-4}$    				&  8.8\,$10^{-2}$ \\
VW Cha...........            	&   	D		&	1.27\,$10^{-5}$    				&  5.4\,$10^{-3}$ \\
CGH$\alpha$ 5/6........      	&   	B		&	7.10\,$10^{-6}$    				&  5.6\,$10^{-6}$ \\
					&   	C		&	7.10\,$10^{-6}$    				&  2.8\,$10^{-3}$ \\
Sz 41.................       		&   	C		&	4.94\,$10^{-6}$    				&  2.1\,$10^{-3}$ \\
Sz 30.................       		&   	C		&	1.45\,$10^{-5}$    				&  9.2\,$10^{-4}$ \\
UZ Tau.............          		&   	B		&	3.09\,$10^{-7}$    				&  1.3\,$10^{-5}$ \\
J 4872...............        		&   	C		&	5.25\,$10^{-6}$    				&  1.8\,$10^{-4}$ \\
UX Tau.............          		&   	D		&	7.10\,$10^{-6}$    				&  1.6\,$10^{-4}$ \\
LkH$\alpha$ 336..........    	&   	C		&	1.67\,$10^{-5}$    				&  1.8\,$10^{-3}$ \\
FV Tau.............          		&   	C		&	2.78\,$10^{-6}$    				&  1.3\,$10^{-3}$ \\
GG Tau.............          		&   	C		&	8.02\,$10^{-6}$    				&  2.6\,$10^{-3}$ \\
LkH$\alpha$ 262/263...    &   	D		&	2.16\,$10^{-6}$    				&  1.6\,$10^{-3}$ \\
LkH$\alpha$ 346 ..........    &       C		&	1.54\,$10^{-4}$					&  5.9\,$10^{-4}$ \\
					&       D		&	3.70\,$10^{-5}$					&  3.0\,$10^{-3}$ \\
ROXs 43 ............. 	   	&       B		&	1.88\,$10^{-5}$					&  5.3\,$10^{-6}$ \\
		   			&       C		&	2.16\,$10^{-6}$					&  1.4\,$10^{-4}$ \\
SR 24 .................   		&       C		&	1.36\,$10^{-5}$					&  1.1\,$10^{-3}$ \\
Sz 68 ..................    		&       B		&	4.94\,$10^{-6}$					&  2.4\,$10^{-7}$ \\
	   				&       C		&	1.27\,$10^{-5}$					&  3.1\,$10^{-4}$ \\
\noalign{\smallskip}
\hline
\noalign{\smallskip}
\end{tabular}
\end{center}
\label{Tab:Proba_unrelated}
\end{table}

\begin{figure*}
\centering
\begin{tabular}{cc}
\includegraphics[width=8.7cm]{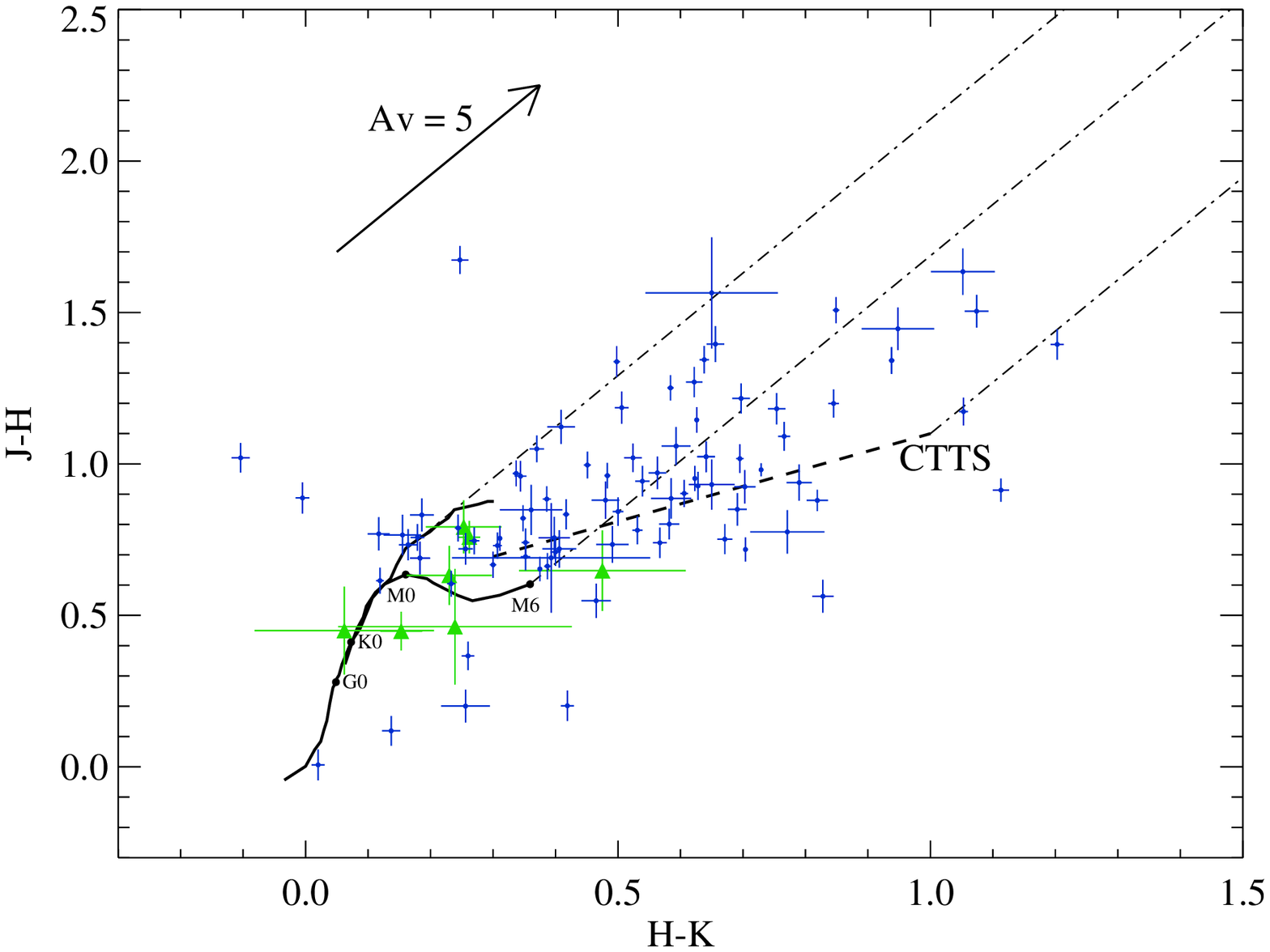}
\includegraphics[width=8.7cm]{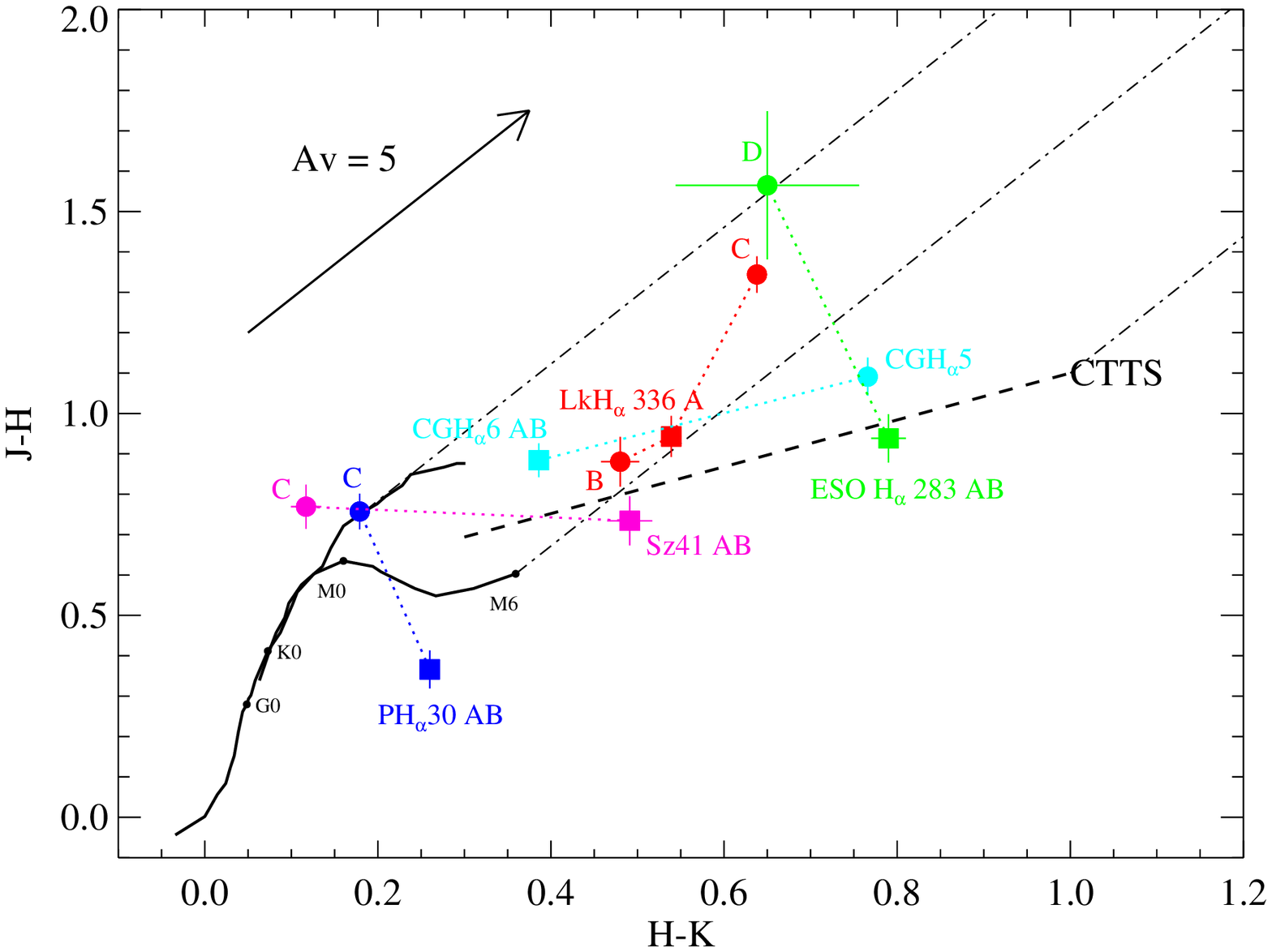} \\
\includegraphics[width=8.7cm]{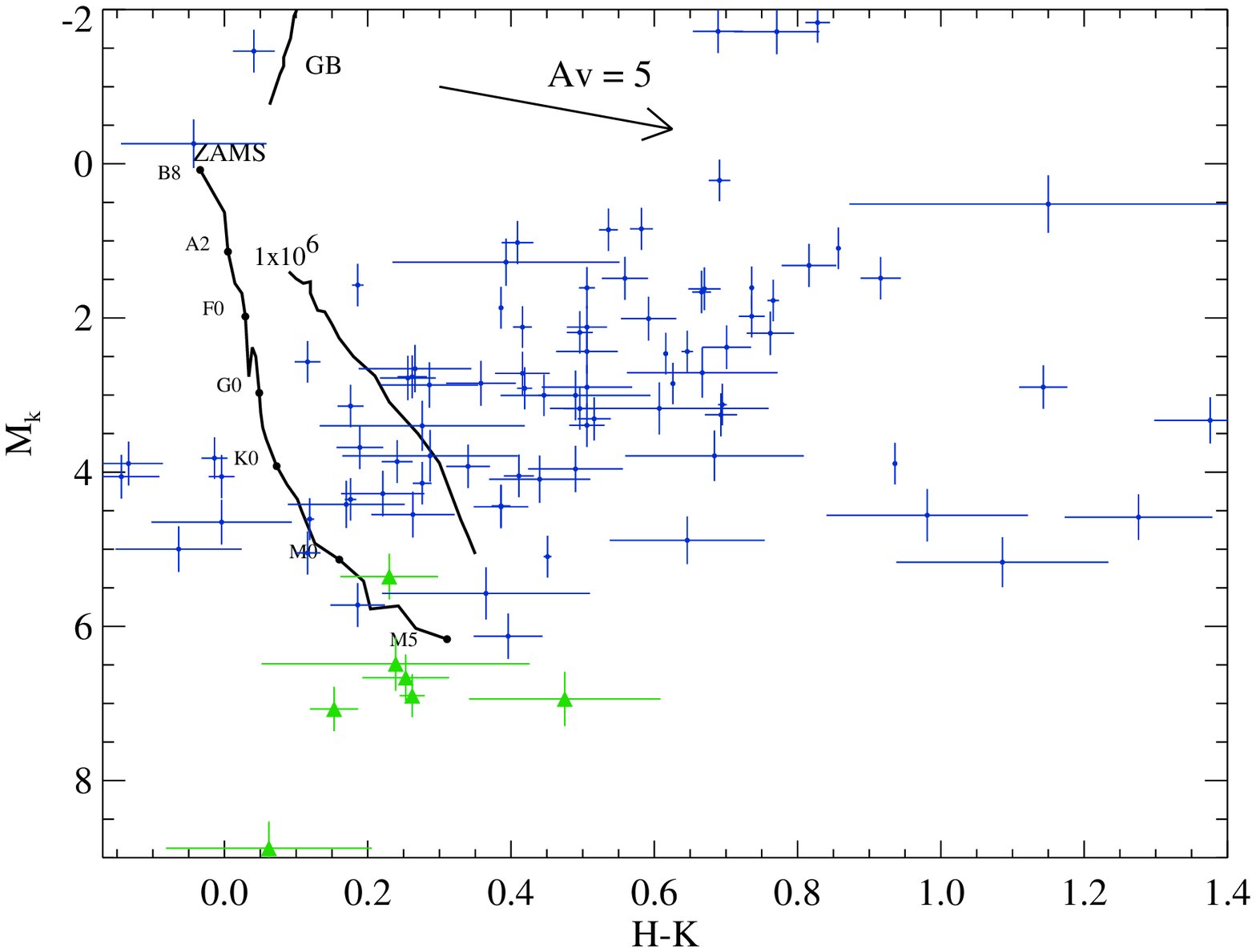}
\includegraphics[width=8.7cm]{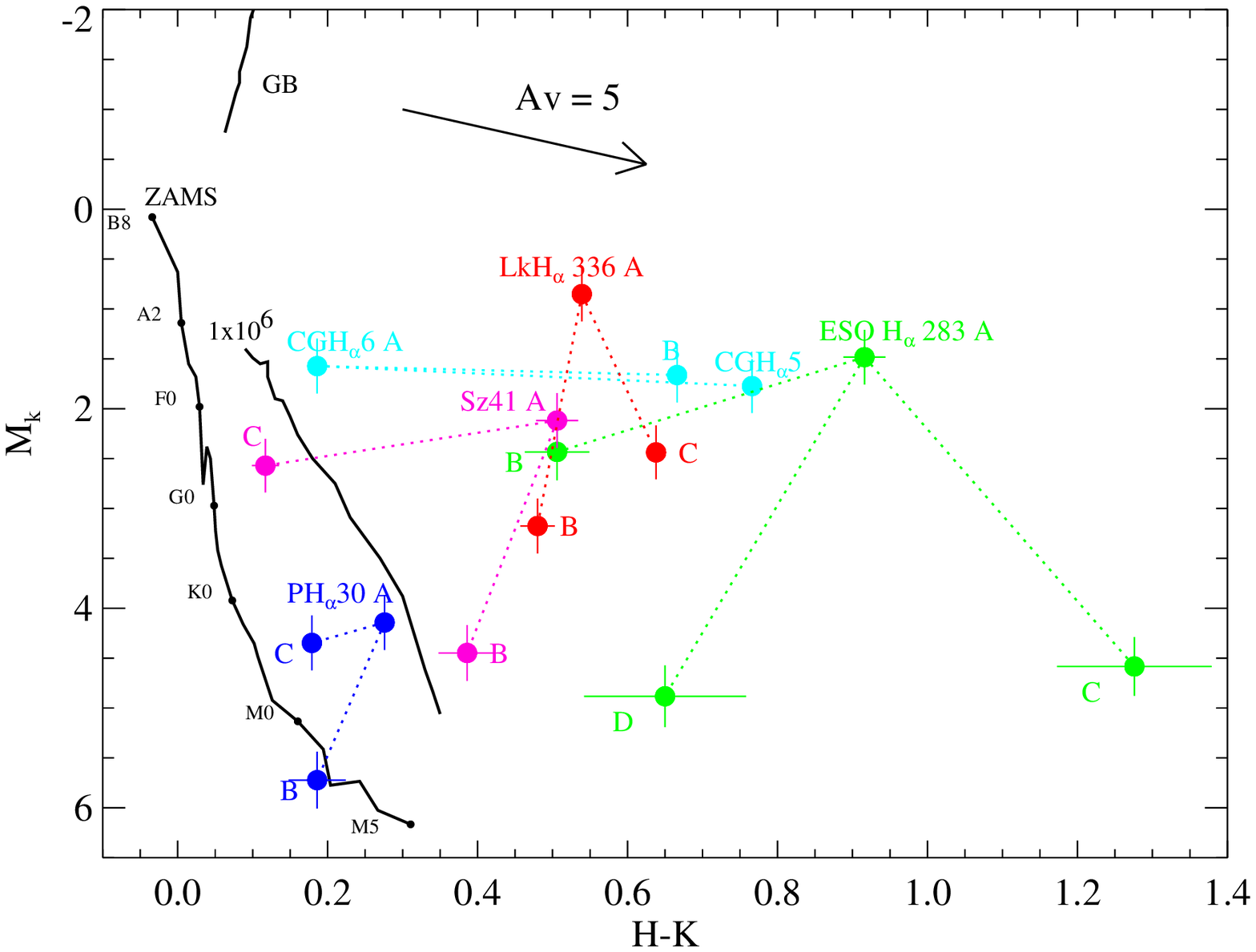} \\
\end{tabular}
\caption{Upper left\,: J-H/H-K color-color diagram for all sources with reliable 2MASS magnitudes, i.e. detected in all three bands, in the FOV of our targets. 
  Blue small dot symbols denote sources detected in our survey while green triangle symbols are those non or marginally detected. 
  Upper right\,: J-H/H-K color-color diagram for the new (PH$\alpha$\,30, CGH$\alpha$\,5/6, 
  ESO\,H$\alpha$\,283) and unconfirmed (LkH$\alpha$\,336 and Sz\,41) systems. 
  Filled squares represent component(s) included in the $\sim$2\,arcsec beam of 2MASS along with the main component, 
  filled circles are possible companions linked to the latter by dotted lines. 
  The loci of unreddened dwarf and giant stars are shown as the solid lines (Bessel \& Brett\,\cite{Bessel_Brett1988}), 
  as well as the locus of unreddened classical T Tauri stars (CTTS) as the dashed line (Meyer et al.\,\cite{Meyer_etal1997}). 
  A reddening vector for a standard reddening law is shown (E(J-H) = 0.11 A$_V$, E(H-K) = 0.065 A$_V$, Meyer et al.\cite{Meyer_etal1997}). 
  Lower left\,: H-K/M$_k$ color-magnitude diagram for {\it all resolved components} in the FOV of our targets for which we have a H-K color. 
  As above, blue small dot symbols denote sources detected in our survey while green triangle symbols are those non-detected.
  The distribution of component is relative to the ZAMS, to the beginning of the giant branch (GB), and to a 1\,Myr isochrone for PMS stars from 
  0.08 to 1.4\,M$_{\odot}$ (with [M/H]=0, Y=0.275 and L$_{mix}$ = H$_p$, Baraffe et al.\,\cite{Baraffe_etal1998}). 
  Lower right\,: H-K/M$_k$ color-magnitude diagram for {\it all resolved components} of the new and unconfirmed systems above.
  Absolute magnitudes M$_K$ are computed using the adopted distances shown in Table\,\ref{Tab:sample} and assuming 10\% relative error. 
  The reddening vector assumes A$_K$ = 0.11 A$_V$.
  All data are in the CIT system.
  }
\label{fig:col_col_col_mag_diags}
\end{figure*}

In order to infer what are the systems whose components are gravitationally bound and those which are only the 
result of chance projection, we used two approaches. 

The first one is a statistical approach which consists in estimating the probability that the companions we found are physically bound to 
their primary. It implies to determine first the local surface density of background/foreground sources in each field. 
For that purpose we compiled the number of 2MASS objects at least as bright in K-band as 
the candidate companion(s) in a 30x30\,arcmin field surrounding each primary. This leads to the average surface density of objects 
brighter than this limiting magnitude, noted $\Sigma$ (K$<$K$_{comp})$. Assuming a random uniform distribution of unrelated objects 
across the field, the resulting probability $P(\Sigma, \Theta)$ of at least an unrelated source to be located within a certain angular distance $\Theta$ 
from a particular target is given by\,: 

\begin{equation}
 P(\Sigma, \Theta) = 1 - e^{-\pi \Sigma \Theta^2}.
\label{P_unrelated}
\end{equation}

\noindent Since we did not attempt to distinguish between cloud members and background/foreground 
stars, the resulting probability is therefore also accounting for unrelated cloud members. The last column of Table\,\ref{Tab:Proba_unrelated} 
gives the resulting probabilities for a companion to be unrelated to the primary of a system. 
Most of the companions detected in our survey have probabilities to be projected unrelated stars well below the percent level. 
This means that they are very likely bound to their systems, although considering probabilities to individual sources is known to be 
prone to error (see e.g. Brandner et al. \cite{Brandner_etal2000} for a discussion). Three candidate companions (ESO H$\alpha$\,283 C, 
ESO H$\alpha$\,283 D, and PH$\alpha$\,30 C) show however a non-negligible probability of being the result of chance projections with 
probabilities of respectively 2.9\%, 37\% and 8.8\%. 

The second approach is an attempt to determine the nature of the new or so far unconfirmed components through 
the use of both a color-color J-H/H-K diagram and a H-K/M$_k$ color-magnitude diagram. While it was possible to composed the latter with 
all resolved components of the first data set, for which we have H-K color measurements, the former is only composed of the well separated 
components for which 2MASS J, H and K magnitudes are available. 
Thus, 2MASS combined magnitudes and/or single-component magnitudes for the well-resolved ($\geq$\,2\,arcsec) companions are used 
for the color-color diagram. This means that through this latter we are probing the only well-separated candidate companions in our survey. 
This is nevertheless still useful since these candidate companions are usually those displaying the larger probabilities to be unrelated 
background sources (Table\,\ref{Tab:Proba_unrelated}). 

The five systems PH$\alpha$\,30, CGH$\alpha$\,5/6, ESO H$\alpha$\,283, LkH$\alpha$\,336 and Sz\,41 are plotted 
in the upper-right plot of Fig.\,\ref{fig:col_col_col_mag_diags} with the combined color of the brightest components as derived by 2MASS as filled squares. 
It turns out that PH$\alpha$\,30\,C and Sz\,41\,C might well be giant background stars with very low line-of-sight 
extinction in view of this diagram. ESO H$\alpha$\,283\,D can also be interpreted as a background giant seen under 5-6\,mag of 
visual extinction. However, since WTTS are also often located near the giant locus, as it can be noted in the upper-left plot of 
Fig.\,\ref{fig:col_col_col_mag_diags} where all systems have been plotted, one cannot conclude on the sole basis of this 
diagram, but it provides further evidence for PH$\alpha$\,30\,C and ESO H$\alpha$\,283\,D to be chance projections. 

The color-magnitude diagram presented in the lower-right plot of Fig.\,\ref{fig:col_col_col_mag_diags} is composed of 
{\it all resolved components} of the five sytems in our frames, while the color-magnitude diagram represented in the lower-left plot 
of the same figure stands for all systems with known H-K colour. Magnitudes for the single components are derived using H and K 2MASS combined magnitudes 
and relative photometry in [FeII] and Br$\gamma$, respectively.
This plot shows that, given the known distance to Sz 41, PH$\alpha$\,30 and ESO H$\alpha$\,283 
(respectively 160, 450 and 700\,pc), the hypothesis that Sz\,41\,C, PH$\alpha$\,30\,C, ESO H$\alpha$\,283\,C and  D 
are actually background giants implies distances of at least $\sim$480\,pc, $\sim$4,5\,kpc and $\sim$7\,kpc, respectively, 
which are consistent with giant absolute brightness. It has been actually shown that Sz\,41\,C is indeed a background giant 
(see Appendix\,\ref{sect:comments_ind_objects}).

In summary, it is clear from these diagrams that some candidate companions (especially the faintest objects) are likely 
extincted background stars. Others are consistent with T Tauri stars suffering differential extinction, or showing perhaps 
different accretion disk properties and orientations. 
Although spectroscopy and common proper-motion are necessary in order to unambiguously identify any chance 
projection, we conclude from the above analysis that PH$\alpha$\,30\,C, ESO\,H$\alpha$\,283\,C and 
ESO\,H$\alpha$\,283\,D are consistent with being projected background stars. As for Sz\,41\,C, we will not consider further 
these companion candidates in our analysis.

The compilation of all 2MASS sources with magnitudes in all JHK bands present in our FOV reveals some objects which are 
marginally detected ($\la$\,5-sigma) or undetected in our frames. These sources are reported in the plots of Fig.\,\ref{fig:col_col_col_mag_diags} 
with triangle symbols. 
These 7 sources are in the FOV of CV\,Cha (2 objects), Sz\,62, PH$\alpha$\,14, PH$\alpha$\,30, CGH$\alpha$\,5/6 and GG\,Tau. 
These sources are mostly consistent with background main-sequence stars or giants in view of their locations on these diagrams. 
In addition, it is most probable that they would have been identified as cloud members by previous H$_{\alpha}$ surveys, even if 
we cannot totally exclude that some could still be very-low mass ($\la$\,0.3\,M$_{\sun}$) young ($\la$10\,Myr) non-accreting stars, 
therefore difficult to be identified by such surveys. 
A more complete census  of the low-mass content of clouds as the one provided by e.g. Luhman (\cite{Luhman_2004}) for Cha\,I, in 
association with resolved X-ray studies would certainly shed light on the nature of these sources.
Finally, it is noteworthy to point out that some systems present unsual bluer colors and are located to the right of the main-sequence in both diagrams. 
These sources corresponds to J\,4872\,ABCD (only resolved in the color-magnitude diagram), UX\,Tau\,BC (idem) and Sz\,60\,A. 
While further investigations would be necessary, especially resolved spectroscopy and JHK photometry, we suspect that the proximity with the other components 
in those systems (with separations $\la$\,3$\farcs$5) might have biased the 2MASS magnitudes.


\subsection{Nature of the companions}
\label{sect:comp_Spt_mass_age}

\begin{figure}
\centering
\includegraphics[width=9cm]{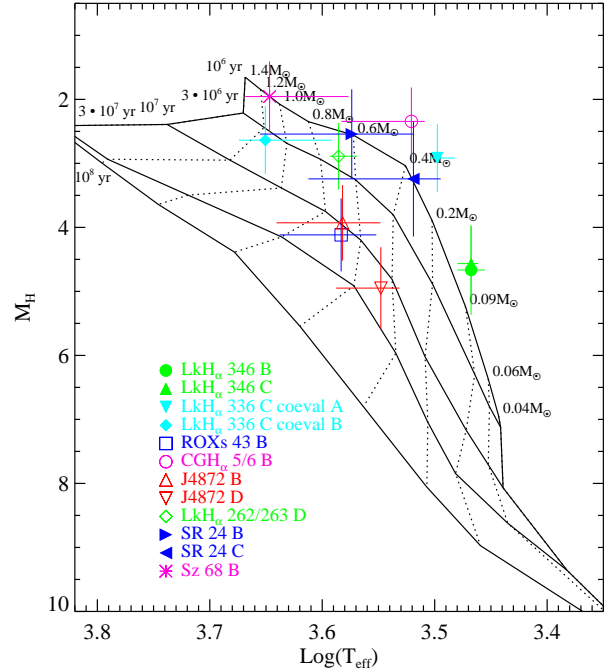}
\caption{Locus of new and poorly studied companions in the (T$_{\mathrm{eff}}$, M$_{\mathrm{H}}$) plane together with the PMS evolutionary tracks and isochrones of 
Baraffe et al. (\cite{Baraffe_etal1998}). The components are placed in the diagram using their M$_{\mathrm{H}}$ and assuming coevality with the closest component of the system.}
\label{fig:HR_diag}
\end{figure}

We derived an estimate of the spectral types and masses of the new, presumably physical, companions as well as those that are poorly 
studied (and lack of spectral type and/or mass estimates) from their H-band magnitude and the assumption of coevality with the other stars 
in the systems. These values are reported in Table\,\ref{Tab:systems_accretion}.

With the H-band magnitude from Table\,\ref{Tab:systems_parameters}  and spectral type from the literature we placed the known and studied 
members of each system on the (T$_{\mathrm{eff}}$, M$_{\mathrm{H}}$) plane of the H-R diagram with PMS evolutionary tracks and isochrones 
overlaid and derived an age for the systems. 
The values of T$_{\mathrm{eff}}$ for these known and studied components of the systems are derived from the system's spectral type quoted in literature 
(reported in Table\,\ref{Tab:sample}) and the temperature scale from Sherry et al. (\cite{Sherry_etal2004}). 
When T$_{\mathrm{eff}}$ values are quoted in literature they are within half a subclass the one derived with the adopted temperature scale. 
Uncertainties of typically one subclass in spectral type are assumed, apart for some cases for which uncertainties could be much larger 
(e.g. CGH$_{\alpha}$\,5/6\,B). Uncertainties in absolute H-band magnitude assume a distance uncertainty of 10\%, and values of A$_V$ from 
the literature were used to deredden the H-band magnitudes.
In Fig.\,\ref{fig:HR_diag} we show the {\it only} new multiple system components as well as poorly studied components plotted on the H-R diagram 
together with the PMS tracks and isochrones of Baraffe et al. (\cite{Baraffe_etal1998}) with $\alpha$=1.9 for M$>$0.6\,M$_{\sun}$. A similar diagram 
not shown here was formed for Sz\,30\,C for which only K-magnitudes are available. Uncertainties in the inferred masses and spectral types of these 
new and/or poorly studied components are set by changing the absolute H-band magnitude by its uncertainty and searching for the intersection with 
the isochrone corresponding to the inferred age. 
An example of the procedure in the case of LkH$\alpha$\,346 is presented in Appendix\,\ref{sect:comments_ind_objects}. 
Due to the fact that H-band flux might not be completely 
photospheric (Greene \& Meyer \cite{Greene_Meyer_1995}), the spectral types and a fortiori the masses derived here should only be considered 
as crude estimates. This is especially the case for Sz\,30\,C for which we had to use K-band fluxes. We would therefore expect an underestimate of 
the ages of the primaries which reflects on earlier spectral-types and overestimated masses for the assumed coeval companions. 
Future follow-up spectroscopy of these objects will give much more accurate T$_{\mathrm{eff}}$ values allowing to test coevality of these system components. 
In this respect, if the two new components of the LkH$\alpha$\,346 system are confirmed to be physical, then this system would be an important 
test case for the PMS evolutionary models, spanning a relatively large range of masses and possibly extending to the substellar regime, in the same fashion as 
GG\,Tau (White et al.\,\cite{White_etal1999}). 


\subsection{Multiplicity statistics}
\label{sect:multiplicity_stat}

Among the 58 wide binaries surveyed, five are Herbig Ae/Be binary stars (HD\,76534, Herschel\,4636, PH$\alpha$\,30, Sz\,120, Elias\,2-49) 
and one is likely to be a foreground (older) object (Sz\,15, see Appendix\,\ref{sect:comments_ind_objects}). 
Excluding these systems from our statistics, we have 52 T Tauri star binary systems 
with mostly K and M-type primary, apart CO\,Ori (F8e), Sz\,19 (G2), CV\,Cha (G8), Elias\,2-30 (G2.5) and ROXs\,43 (G0). 
Taking into account the edge-on disk LkH$\alpha$\,262/263\,C we end up with 38 binaries, 7 triples (CGH$\alpha$\,5/6, LkH$\alpha$\,336, 
ROXs\,43, SR\,24, Sz\,30, Sz\,68, UZ\,Tau) and 7 quadruples (LkH$\alpha$\,262/263, LkH$\alpha$\,346, FV\,Tau, GG\,Tau, J\,4872, 
UX\,Tau, VW\,Cha). 
We did not attempt to correct for incompleteness (see discussion in Sect.\,\ref{sect:sens_limit_completeness}). 
In order to characterize the multiplicity, we here define two quantities. First,  a quantity we call {\it degree of multiplicity per wide binary} 
(or a multiplicity frequency per wide binary, MF/wB)\,:

\begin{equation}
 MF/wB = \frac{T+Q+...}{wB+T+Q+...},
\label{MF/wB}
\end{equation}

\noindent where wB represents the number of wide binaries (with projected component separations typically $\ga$\,0$\farcs$5, see Sect.\,\ref{sect:sample}), 
T the number of triples and Q the number of quadruples. This quantity here equals 26.9\,$\pm$7.2\%. 
The errors have been estimated by taking the square root of the number of triples and quadruples. 

Second, a so-called  {\it companion star frequency per wide binary} (CSF/wB) which is defined as \,:

\begin{equation}
 CSF/wB = \frac{2T+3Q+...}{wB+T+Q+...}.
\label{CSF/wB}
\end{equation} 

\noindent Our study gives 0.67\,$\pm$0.11 for this quantity. Note that the latter quantity is sensitive to a decrease 
of high-order multiplicity by, e.g., dynamical decay while the former quantity is not. In general, while MF/wB is only a measure 
of how common high-order systems are, CSF/wB also give us information about the relative fractions of high-order multiplicity 
via the number of companions.

In the rest of the study, we will consider a distance-limited sample in order to ensure a similar range of linear 
projected separations probed, thus providing meaningful statistical results. 
We limit ourselves to only the wide T Tauri binaries of the sample at distance 140-190pc (41 systems). This includes only 
the systems for which the separation between companions are in the range 10/14 AU - 1700/2300 AU, i.e. with 
projected separations between 0\farcs07 and 12$^{\prime\prime}$. In that case, MF/wB=26.8\,$\pm$8.1\%  
and CSF/wB=0.68\,$\pm$0.13 (30 binaries, 5 triples, 6 quadruples).


\section{Discussion}
\label{sect:Discussion}

\subsection{Multiplicity as a function of mass}
\label{sect:mass_depend_mult}
A breakdown of the distance-limited sample by primary mass, i.e. by spectral type of the primary, 
leads to MF/wB=38\,$\pm$14\%  and CSF/wB=1.1\,$\pm$0.2 for K types (13 binaries, 2 triples, 6 quadruples), 
MF/wB=13\,$\pm$9\%  and CSF/wB=0.3\,$\pm$0.1 for M types  (13 binaries, 2 triples). 
The trend is consistent with the known decrease of multiplicity with decreasing primary mass (Sterzik \& Durisen \cite{Sterzik_Durisen2003}).

\subsection{Comparison between clouds}
\label{sect:clouds_comparison}
The question that arises naturally is the one about the multiplicity frequency in different clouds, although 
the latter is of lower statistical significance than that of the total sample. The cloud with the highest value is 
Tau-Aur (5 triples-quadruples/11 wide binaries, MF/wB=45\,$\pm$20\%, CSF/wB=1.3\,$\pm$0.3), followed 
by Cha\,I (2/10, MF/wB=20\,$\pm$14\%, CSF/wB=0.5\,$\pm$0.2) and Ophiuchus (2/10, MF/wB=20\,$\pm$14\%, 
CSF/wB=0.4\,$\pm$0.2). 
Some star formation models suggest that cloud intrinsic properties might imprint different binary or multiplicity properties 
(Sterzik \& Durisen \cite{Sterzik_Durisen1998}, Sterzik et al. \cite{Sterzik_etal2003}), but no model is able to quantify the 
differences observed.


\begin{table*}
\scriptsize
\caption{Comparison of the multiplicity frequency per binary (MF/wB and MF/B) and the companion star frequency per binary (CSF/wB and CSF/B) 
of our work with those derived from previous multiplicity surveys among T Tauri stars in a similar separation range $\sim$14-1700 AU$^a$, and with 
recent numerical simulations (bottom part). Also indicated are the number of wide (separation $>$\,0$\farcs$5) 
binaries (wB), binaries (B), triples (T), quadruples (Q) and higher multiples ($>$4).}
\begin{center}
\renewcommand{\arraystretch}{0.5}
\setlength\tabcolsep{7pt}
\begin{tabular}{llcccccr@{\,$\pm$\,}lr@{\,$\pm$\,}lr@{\,$\pm$\,}lr@{\,$\pm$\,}l}
\hline\noalign{\smallskip}
Reference                      & cloud   	& wB & B & T & Q & $>$4 &  \multicolumn{2}{c}{MF/wB} 	&  \multicolumn{2}{c}{CSF/wB} &  \multicolumn{2}{c}{MF/B} 	&  \multicolumn{2}{c}{CSF/B}\\
\noalign{\smallskip}
\hline
\noalign{\smallskip}
\noalign{\smallskip}
This work           			&   several   				&  30 & 42.3 & 5 & 6 & - &  26.8 & 8.1\% &   0.68 &	0.13  &    20.6   &  6.2\%  &   0.53   &   0.10 \\
\noalign{\smallskip}
\noalign{\smallskip}
\noalign{\smallskip}
\noalign{\smallskip}
						\multicolumn{15}{c}{Other Multiplicity Surveys}	 \\	
\noalign{\smallskip}
\hline
\noalign{\smallskip}
L93           		&   Tau-Aur    				&  24 & 39 & 3 & 2 & - &   17.2 & 7.7\% &   0.41	&	0.12  &    11.4   &  5.1\%  &   0.27   &   0.08 \\
KL98  		&   Tau-Aur    				&  21 & 30 & 6 & 1 & - &   25.0 & 9.4\% &   0.54	&	0.14  &    18.9   &  7.2\%  &   0.41   &   0.11  \\
G97            	&   Cha/Lup/CrA    			&  19 & 21 & 2 & -  & - &   9.5   & 6.7\% &   0.19	&	0.10  &    8.7     &  6.1\%  &   0.17   &   0.09 \\

K00     		&   Sco-Cen		   		& 24  & 37 & 6 & - & - &  20.0	&	8.2\% 		&   0.40	&	0.12  &    14.0   &  5.7\%  &   0.28   &   0.08\\
K01    		&   Cha					&  5   & 16 & 1 & - & - &  16.7  	&	16.7\%		&   0.33	&	0.24 &    5.9   &  5.9\%  &   0.12   &   0.08\\
R05    		&   Oph					&  22 & 45 & 5 & - & - &  18.5  	&	8.3\%		&   0.37	&	0.12 &    10.0   &  4.5\%  &   0.20   &   0.06\\
\noalign{\smallskip}
Tot(Other surveys)		&				&  115 & 188 & 23 & 3 & - &  18.4  	&	3.6\%	&   0.39	&	0.05 &    12.2   &  2.4\%  &   0.26   &   0.03 \\
Tot(All surveys)			&				&  132 & 217.3 & 23 & 7 & - &  18.5  	&	3.4\%	&   0.41	&	0.05 &    12.1   &  2.2\%  &   0.27   &   0.03 \\
\noalign{\smallskip}
\noalign{\smallskip}
\noalign{\smallskip}
						\multicolumn{15}{c}{Numerical Simulations}	\\	
\noalign{\smallskip}
\hline
\noalign{\smallskip}
SD03	&   	& -  & 0.21 & 0.08 & 0.03 &  & \multicolumn{2}{c}{...} &  \multicolumn{2}{c}{...} &   \multicolumn{2}{c}{34\%}	&   \multicolumn{2}{c}{0.78} \\
D03		&   	& -  & 0.154 & 0.108 & 0.045 &  & \multicolumn{2}{c}{...} &  \multicolumn{2}{c}{...} &  \multicolumn{2}{c}{49.8\%}&   \multicolumn{2}{c}{1.14} \\
D04		&   	& -  & 11 & 2 & 2 & 3 & \multicolumn{2}{c}{...} &  \multicolumn{2}{c}{...} &  38.9  	&	14.7\%		&   1.28	&	0.27 \\
G04		&   	& -  &  7  & 6  & 2 & 1 & \multicolumn{2}{c}{...} &  \multicolumn{2}{c}{...} &  56.3  	&	18.8\%		&   1.38	&	0.29 \\
\noalign{\smallskip}
\hline
\noalign{\smallskip}
\end{tabular}
\end{center}
\label{Tab:mult_other_surveys}
\begin{minipage}[position]{17cm}
$^a$\,: This range is assuming that all systems are at 140\,pc. Additionally the upper limit is actually set by confusion with background/foreground stars, and varies 
from survey to survey. The surveys of Leinert et al. (\cite{Leinert_etal1993}), K\"{o}hler \& Leinert (\cite{Koehler_Leinert1998}), and Ghez et al. (\cite{Ghez_etal1997a}) 
have set their upper limit to 12$\arcsec$ ($\sim$\,1700\,AU at 140\,pc), while the surveys of K\"{o}hler et al. (\cite{Koehler_etal2000b}), K\"{o}hler (\cite{Koehler_2001}) 
and Ratzka et al. (\cite{Ratzka_etal2005}) have upper limits of 6$\arcsec$.\\   
References\,: 
L93\,: Leinert et al. \cite{Leinert_etal1993}, 
KL98\,: K\"{o}hler \& Leinert \cite{Koehler_Leinert1998}, 
G97\,:  Ghez et al. \cite{Ghez_etal1997a}, 
K00\,: K\"{o}hler et al. \cite{Koehler_etal2000b}, 
K01\,: K\"{o}hler \cite{Koehler_2001}, 
R05\,: Ratzka et al. \cite{Ratzka_etal2005}, 
SD03\,: Sterzik \& Durisen \cite{Sterzik_Durisen2003},
D03\,: Delgado-Donate et al. \cite{Delgado_Donate_etal2003}, 
D04\,: Delgado-Donate et al. \cite{Delgado_Donate_etal2004}, 
G04\,: Goodwin et al. \cite{Goodwin_etal2004}.
\end{minipage}
\end{table*}

\subsection{Comparison with previous multiplicity surveys}
\label{sect:multiplicity_stat_compa_surveys}
We compare our result with the proportion of triples/quadruples found in previous NIR multiplicity surveys which have focussed on the 
brightest sources (generally K\,$\la$\,10) with broadly similar separation range, sensitivity\footnote{However, our AO survey is probably slightly more 
sensitive than previous speckle surveys.}, and statistical uncertainties. 
These are the studies by Leinert et al. (\cite{Leinert_etal1993}) and K\"{o}hler \& Leinert (\cite{Koehler_Leinert1998}) in Tau-Aur, 
Ghez et al. (\cite{Ghez_etal1997a}) in Chamaeleon, Lupus and CrA, K\"{o}hler et al. (\cite{Koehler_etal2000b}) 
in the Scorpius-Centaurus OB association, K\"{o}hler (\cite{Koehler_2001}) in Chamaeleon, and  Ratzka et al. 
(\cite{Ratzka_etal2005}) in Ophiuchus. 
We primarily base this comparison on the multiplicity frequency per wide binary (MF/wB), as defined above (Eq.\ref{MF/wB}), and 
the companion star frequency per wide binary (CSF/wB, Eq.\ref{CSF/wB}). Since those surveys used samples drawn from lists of apparently single stars, 
and not wide binaries like our, they include all resolved binaries with separations down to the resolution limit. They therefore give multiplicity fractions 
per binary, i.e. MF/B and CSF/B, where B includes binaries with separations $\la$0$\farcs$5. We thus have to count in these surveys 
only binaries with separations larger than 0$\farcs$5 (i.e. $\sim$70 AU at 140\,pc). This will give us an estimate of MF/wB and CSF/wB for these surveys, 
and allow a direct comparison with our own survey results. Note that in all these surveys we consider a companion number uncorrected for either 
completeness or chance projection, as these corrections are difficult to assess properly and moreover roughly compensating each other. While 
completeness depends on the sensitivity of these surveys which is rather uniform, chance projection is most of the time treated by assuming a 'safe' 
maximum separation which depends on the background/foreground surface density of stars.  
On our side, we would have to slightly restrict the separation range to the resolution achieved by those surveys (i.e. typically 
$\sim$0$\farcs$1-12$^{\prime\prime}$ which is $\sim$14-1700\,AU at 140\,pc), but this is not significant since only 
one companion was detected below this limit (SR\,24\,C at separation 0$\farcs$08). Table\,\ref{Tab:mult_other_surveys} summarizes 
the result of this comparison. It shows that in first order our newly derived multiplicity agrees with the previous surveys, 
within the uncertainties. 

In a second step, we will attempt to estimate the fraction of binaries which were not included in our survey, i.e. those with a separation range 
$\sim$0$\farcs$1-0$\farcs$5. This will allow us to convert our MF/wB to MF/B (as well as CSF/wB to CSF/B) and to perform a direct comparison 
with the results from the other surveys. For that purpose, we derived the fraction of binaries of those surveys with separations between 
$\sim$0$\farcs$1 and 0$\farcs$5. We only consider those surveys for which the maximum separation was set to 12$\arcsec$, and 
find a fraction of close binaries $\alpha$=28.9\,$\pm$5.7\%. The difference with the fraction derived from all surveys, including those with upper 
separations of 6$\arcsec$, is not statistically different since in that case $\alpha$=38.8\,$\pm$4.5\%. The corrected number of binaries with 
separations $>$0$\farcs$1 for our survey is then B=wB/(1-$\alpha$)=30/(1-0.29)=42.3. Thus, MF/B = 20.6\,$\pm$6.2\% and CSF/B = 0.53\,$\pm$0.10 
leading to the same conclusion as above. 

Noticeably, the CSF/B (and CSF/wB) is significantly higher in our survey than for all the other surveys. If we average the result for all the other surveys the difference is at 
the 2$\sigma$ level. This overabundance of companions is a consequence of a higher fraction of quadruples over triples in our survey with respect to all previous surveys. 
The fraction of systems with multiplicity higher or equal than four to the number of systems with multiplicity higher or equal than three 
(called $f_4$ in the Sect.\,\ref{sect:MS_mult}) is 55.0\,$\pm$22.2\% for our survey, while it is only 13.0\,$\pm$7.5\% on average for all other surveys.


\subsection{Comparison with theory}
\label{sect:multiplicity_stat_compa_theory}
There is a probable overabundance of high-order multiples produced by the current simulations of star formation with 
respect to current observations. Direct comparison is not possible since theoretical multiplicity frequencies include  
binaries with separations down to typically $\sim$\,3-5\,AU. Wider high-order companions with separations $\ga$\,2000\,AU are quite 
rare, so that the upper bound of the separation range should be less of a concern. 
We will assume here that the corrections to be applied in order to obtain MF/B in the same separation range as the one probed here 
are minor, i.e. we will neglect the correction for systems in the separation range $\sim$\,3-5\,AU-10\,AU. 
In some cases, this might not be the case. For example in the simulations of Goodwin et al. (\cite{Goodwin_etal2004}) the fraction of close 
binaries with separations $\la$\,10\,AU would be up to half of all binaries formed (S. Goodwin 2006, private communication). We therefore have to 
keep in mind that the theoretically derived MF/B should be considered as lower-limits. In the following, we summarize the theoretical studies used for the comparison.

Sterzik \& Durisen (\cite{Sterzik_Durisen2003}) performed few-body cluster decay simulations. 
Although that study neglected the effect of remnant molecular gas and disk accretion and treated only the process of 
dynamical evolution of young small N-body clusters, it yields highly significant and robust statistics since a large number 
of realizations (10 000) has been computed. A degree of multiplicity of MF/B=34\% (CSF/B = 0.78) was found. 
Delgado-Donate et al. (\cite{Delgado_Donate_etal2003}) modeled the 
dynamical decay of a large number (a hundred) of small-N (N=5) star-forming clusters including the effects of 
competitive accretion and dynamical evolution through Smoothed Particles Hydrodynamics (SPH) simulations with a $\sim$1AU spatial 
resolution, and found a rather high multiplicity frequency close to MF/B=50\%.
A similar high frequency of multiple systems was the outcome of two other recent and more sophisticated SPH simulations. 
Delgado-Donate et al. (\cite{Delgado_Donate_etal2004}) simulated the fragmentation of 10 small-scale 
turbulent molecular clouds and their subsequent dynamical evolution, including this time the effect of accretion disks 
into the evolution of multiples. Goodwin et al. (\cite{Goodwin_etal2004}) followed the collapse and fragmentation of 20 
dense star-forming cores with a low-level of turbulence. In both cases a high frequency of high-order multiples was obtained
(Table\,\ref{Tab:mult_other_surveys}).


\subsection{Comparison with Main-Sequence multiplicity}
\label{sect:MS_mult}

Most of our targets will evolve towards G and K-types on the Main-Sequence, therefore we will compare our results with the 
field star surveys focussing on those spectral types. Duquennoy \& Mayor (\cite{Duquennoy_Mayor1991}, hereafter DM91) gives an 
estimate of the high-order multiplicity among solar-type Main-Sequence dwarfs. 
However selection effects may apply, as already realized in DM91, and more recent surveys and compilations (e.g. Tokovinin \& Smekhov \cite{Tokovinin_Smekhov_2002}) argue for a larger fraction of high-order multiples among MS dwarfs. 
In the following, we will compare the multiplicity fractions from the MSC catalog (Tokovinin \cite{Tokovinin_1997}) with ours.  
MSC gives their results using a {\it multiplicity fraction $f_n$} defined as (Batten\,\cite{Batten_1973})\,: 

\begin{equation}
 f_n = \frac{N_n}{N_{n-1}}.
\label{f_n}
\end{equation} 

\noindent where $N_{n}$ is the number of systems of multiplicity {\it at least} $n$. Batten (\cite{Batten_1973}) estimated $f_n$ to be of the order of 25\%, independently 
of the order of multiplicity. Tokovinin (\cite{Tokovinin_2006}) estimated from MSC a fraction of triple systems (or higher) relative to binaries (or higher) $f_3$=19\,$\pm$4\% and a fraction of (at least) quadruples to (at least) triples $f_4$=22\,$\pm$2\%. A similar value of $f_3$=20\,$\pm$6\% was found by Tokovinin \& Smekhov (\cite{Tokovinin_Smekhov_2002}) in a sample 
of dwarfs with spectral types later than F5 extracted from MSC. This latter value is very similar to the former one, suggesting that the high-mass evolved stars included 
in MSC should have little effects on the multiplicity fractions. In comparison, a value of $f_3$=11\,$\pm$4\% can be estimated from the sample of DM91. 
In our case, it happens that $f_3$=MF/B, so that $f_3$=20.6\,$\pm$6.2\% and $f_4$=55.0\,$\pm$22.2\%. 
While the fraction of triples to binaries is in agreement with what is found for the main-sequence, there is a marginally significant overabundance of quadruples 
with respect to triples in our survey when compared to MSC values. A clear caveat of this comparison is that multiplicity fractions for main-sequence stars are derived 
for much larger linear separation range than for our survey. We attempt to address this issue in the next section.


\subsection{Total multiplicity}
\label{sect:total_mult}

We searched in the literature for spectroscopic binaries (SBs) included in our sample. Systematic searches for spectrsocopic companions 
of T Tauri stars comprise the radial-velocity surveys of Mathieu et al. (\cite{Mathieu_etal1989}), Reipurth et al. (\cite{Reipurth_etal2002}), 
Torres et al. (\cite{Torres_etal2003}), and 
Melo (\cite{Melo_2003}). According to them, there are 8 SBs, of which 5 need confirmation, in our sample (see Table\,\ref{Tab:sample}).  
Since not all the components of our sample have been searched for spectroscopic companions, this number is likely to be a lower limit. 
We can therefore tentatively estimate the increase of multiplicity that would result if we could include SBs by taking into account all those potential 8 SBs. 
This would lead to multiplicities MF/wB=39.0\,$\pm$9.8\%  and CSF/wB=1.02\,$\pm$0.16 for companions with separation down to a few tens 
of AU (25 binaries, 8 triples, 6 quadruples, 2 quintuples)\,\footnote{This includes the 0$\farcs$016 companion to ROXs\,43\,C (Simon et al.\,\cite{Simon_etal1995}).}. 
This also corresponds to multiplicity fractions (Eq.\,\ref{f_n}) $f_3$=39\,$\pm$10\%, which is an upper limit here (see below), $f_4$=50\,$\pm$18\%, and 
$f_5$=25\,$\pm$18\%, i.e. the ratio of quadruples to triples is still higher than that of the main-sequence 
($f_5$ is within the uncertainties of that of MSC which is 20\,$\pm$4\%). 

We try to better estimate $f_3$ by tentatively correcting for the binaries with separations lower than 0$\farcs$5 as before, 
but not for those systems with separation between a few tenths of AU to $\sim$\,10\,AU, i.e. almost pure SBs, as these systems might not be so ubiquitous 
(Sterzik et al.\,\cite{Sterzik_etal2005}, Tokovinin et al.\,\cite{Tokovinin_etal2006}). This gives MF/B=$f_3$=31.3\,$\pm$7.8\%  and CSF/B=0.82\,$\pm$0.13. 
Therfore, the ratio of triples to binaries appears also higher for young systems than for the main-sequence, although at a marginally significant statistical level and with the 
caution that this result might be in part biased by the assumption we made about the paucity of tight binaries. 
Noticeably, a relatively good match is found with the numerical models of dynamical evolution of young small N-body clusters 
(see Sect.\,\ref{sect:multiplicity_stat_compa_theory} and Table\,\ref{Tab:mult_other_surveys}).


\subsection{Hierarchical configurations}
\label{sect:hierarchy}

Among the quadruple systems, three systems are composed of two pairs separated by a distance larger in projection than the pair separations, 
and four systems are composed of a pair with a wide companion and an even wider companion. An almost equal number of systems between 
these two hierarchical configurations is also what is found among Main-Sequence systems in the MSC catalog (Tokovinin\,\cite{Tokovinin_2001}).


\begin{table}
\scriptsize
\caption{Systems dynamical stability. Application of the stability criteria from Eggleton  \& Kiseleva (1995) (their Y$^{min}_{0}$, cf Eq. 2) 
and from Mardling \& Aarseth (2001) ($R^{min}_{crit} = \frac{R^{out}_{p}}{a_{in}}$ their Eq. 90) for a given partition (1+2)+3, q$_{in}$=$m_1/m_2$ and q$_{out}$=$(m_1+m_2)/m_3$. R$_{mes}$ is the ratio of the outer to the inner 
projected separation of the triple partition considered$^a$.}
\begin{center}
\renewcommand{\arraystretch}{0.7}
\setlength\tabcolsep{7pt}
\begin{tabular}{lcccccc}
\hline\noalign{\smallskip}
System                       		& Partition$^b$ &	q$_{in}$ 	&q$_{out}$$^c$&	Y$^{min}_{0}$$^c$	&	R$^{min}_{crit}$$^c$&	R$_{mes}$	\\
                             		& 			&			& 			&					&					&				\\
\noalign{\smallskip}
\hline
\noalign{\smallskip}
VW Cha...........            	&   (B+C)+A	&	1.14		&	0.75		&		6.16			&		3.92			&		5.8	   	 \\
CGH$\alpha$ 5/6........      	&   (A+B)+C	&	1.84		&	0.93		&		5.89			&		3.75			&		23.4	   	 \\
Sz 30.................       		&   (A+B)+C	&	1.60		&	1.34		&		5.45			&		3.49			&		3.1		  \\
UZ Tau.............          		&   (B+C)+A	&	1.08		&	0.60		&		5.25			&		3.38			&		9.8	   	 \\
J 4872...............        		&   (A+B)+CD	&	1.10		&	1.39		&		5.43			&		3.48			&		19.3	   	 \\
UX Tau.............          		&   (A+D)+BC	&	6.81		&	2.40		&		4.81			&		3.22			&		2.2	  	 \\
LkH$\alpha$ 336..........    	&   (A+C)+B	&	2.83		&	1.33		&		5.46			&		3.5			&		1.41		  \\
FV Tau.............          		&    (C+D)+AB	&	3.15		&	0.27		&		7.87			&		5.20			&		17.4	  	 \\
GG Tau.............          		&    (C+D)+AB	&	2.85		&	0.11		&		9.80			&		...$^d$		&		6.92	  	 \\
LkH$\alpha$ 262/263...    &    (A+B)+C	&	0.95		&	1.17		&		5.62			&		3.58			&		9.9	   	 \\
LkH$\alpha$ 346 ......... 	&    (A+B)+CD	&	3.11		&	0.42		&		7.08			&		4.57			&		5.3	      	 \\
ROXs 43 ............    		&    (A+B)+C	&	1.13		&	1.16		&		5.63			&		3.59			&		13.4	        	\\
SR 24 .................    		&    (B+C)+A	&	1.79		&	0.63		&		6.40			&		4.09			&		62.6	   	 \\
Sz 68 ..................    		&    (A+B)+C	&	3.77		&	25.3		&		3.26			&		2.84			&		22.3	       	\\
\noalign{\smallskip}
\hline
\noalign{\smallskip}
\end{tabular}
\end{center}
\label{Tab:syst_stability}
\begin{minipage}[position]{9cm}
  $^a$\,:  In case of ambiguity the lowest value is adopted. \\
  $^b$\,: When several configurations exist, the more unstable is reported. \\
 $^c$\,:   When several mass estimates exist, these columns report the values corresponding to the greater values of Y$^{min}_{0}$ and R$^{min}_{crit}$.\\
 $^d$\,: Not applicable since $m_3/(m_1+m_2)>5$.
 \end{minipage}
\end{table}

\subsection{System dynamical stability}
\label{sect:syst_stability}

In order to quantitatively investigate the system dynamical stability, we applied to our sample the semi-empirical criteria developed by Eggleton  \& Kiseleva (1995) 
and Mardling \& Aarseth (2001). 
The result is shown in Table\,\ref{Tab:syst_stability} where R$_{mes}$ should be at least larger than Y$^{min}_{0}$ or R$^{min}_{crit}$ to ensure stability. 
Assumptions are that the periastron of the outer orbit as well as the semi-major axis and the apastron of the inner orbit are 
approximated by their projected separations, i.e. the excentricity of the outer orbit is assumed to be zero and the system is supposed to be seen face-on. 
The component masses used are taken from the literature when available and quoted in Table\,\ref{Tab:systems_accretion}, or roughly estimated by us 
using flux ratios and the assumption of coevality (Sect.\,\ref{sect:comp_Spt_mass_age}). Under the above assumptions, 3 systems 
(LkH$\alpha$\,336, Sz\,30 and UX\,Tau) appear to be unstable by both criteria while three others (LkH$\alpha$ 346, VW\,Cha and GG\,Tau) 
appear unstable by the Eggleton  \& Kiseleva (1995) criterion only. In that respect, it is interesting to point out that recently, Beust \& Dutrey (\cite{Beust_Dutrey_2006}) 
have shown through a detailed analysis of the dynamics of the GG\,Tau system that only some special orbital configuration would allow GG\,Tau CD to be stable, 
although this result depends strongly on the adopted size of the circumbinary disk around GG\,Tau\,AB. 

Even without the values reported in Table\,\ref{Tab:syst_stability}, one can immediately note from Fig.\,\ref{fig:triples_quadruples} that the 
three systems LkH$\alpha$\,336, Sz\,30 and UX\,Tau are apparently non-hierarchical. 
This corresponds to a frequency 3/14=21$\pm$12\%. 
Note that K\"{o}hler \& Leinert (\cite{Koehler_Leinert1998}) found 1 apparent non-hierarchical triple among 7 triples, while K\"{o}hler et al (\cite{Koehler_etal2000b}) and Ratzka et al. (\cite{Ratzka_etal2005}) found1/6 and 1/5, respectively. This yields a fraction of apparent non-hierarchical triples among triples of $\sim$15-20\%, 
and in combination with our survey, a total fraction of 6/32=19$\pm$8\%.

Ambartsumian (\cite{Ambartsumian_1954}) provided a detailed statistical analysis of how many Trapezium systems are in fact pseudo-Trapezium 
systems, i.e. projected hierarchical systems. 
With the assumption of circular orbits, the probability for a hierarchical triple system to be seen in a pseudo-Trapezium configuration is (his Eq. 10)\,: 

\begin{equation}
U = \frac{lg2-\frac{1}{2}-\frac{1}{24}\frac{k_0^2}{k_1^2}}{lgk_1-lgk_0}.
\label{Amba_proba}
\end{equation} 

\noindent where $k_0 < AC/AB < k_1$, AC and AB are the maximum and minimum projected separations, respectively, $k_0$ and $k_1$ their ratios 
in the case of a Trapezium and hierarchical configuration, respectively.  
In our multiple system sample, we have $k_0 \sim$\,2 and $k_1\sim$\,10, thus U=0.05. Following Ambartsumian (\cite{Ambartsumian_1954}), we include 
the quadruple systems by considering an equal number of the two hierarchical configurations discussed in Sect.\,\ref{sect:hierarchy} and a fraction of quadruples 
among systems with multiplicity higher than 2 of one half. This leads to a probability P=0.07 that apparent non-hierarchical systems are in fact projected 
hierarchical systems and this is in statistical agreement with the value we found.

However, if we suppose that the relative orbital inclinations could be rather small, as the study of Sterzik \& Tokovinin (\cite{Sterzik_Tokovinin_2002}) 
among field triple systems seems to indicate, then the fraction of projected hierarchical systems among apparent non-hierarchical systems would 
be much smaller and 1 or 2 of our 3 systems 
could be really non-hierarchical. These systems would then be either very young, or in re-arrangement after a recent close triple approach following or 
not the ejection of a companion. In all cases, these systems are not expected to be long-term stable and may later decay. For example, 
Delgado-Donate et al. (\cite{Delgado_Donate_etal2003}) found that the non-hierarchical systems formed in their simulations decay in less than $\sim$0.5\,Myr. 
UX\,Tau is the only systems with an estimated age for A, D and the combined BC system (see Table\,\ref{Tab:systems_accretion}), and it seems not 
consistent with a very young ($\la$\,1\,Myr) system. On the other hand, this could indeed be the case for LkH$\alpha$\,336.


\subsection{Disk evolution}
\label{sect:disk_evolution}

\begin{table*}
\scriptsize
\caption{Known T Tauri types (CTTS or WTTS) for all single components or otherwise pairs of the triple/quadruple systems.
Also included are spectral types, A$_V$, masses, and ages of the components from the literature, as well as our own estimates for the new and poorly studied 
components when possible (see Sect.\,\ref{sect:comp_Spt_mass_age}).}
\begin{center}
\renewcommand{\arraystretch}{0.7}
\setlength\tabcolsep{7pt}
\begin{tabular}{l@{\hspace{2mm}}c@{\hspace{3mm}}l@{\hspace{0.5mm}}rlccr@{\hspace{2mm}}rl@{\hspace{2mm}}c@{\hspace{3mm}}c
@{\hspace{1mm}}r@{\,$\pm$\,}l@{\hspace{1mm}}r@{\,$\pm$\,}l@{\hspace{1mm}}r}
\hline\noalign{\smallskip}
System   &   Comp.  	& other name 	&    HBC 	&   SpT & A$_V$ &   Ref  & EW(H${_\alpha}$)  & $\Delta$(K$-$L) & Type & Ref  		     & Ref  	    &  \multicolumn{2}{c}{Mass}  &  \multicolumn{2}{c}{Age}   & Ref   \\
                &    or pair   &           		&         	&	      &	            &  (SpT) &    [\AA]    			&  [mag]   		      &           & (H${_\alpha}$)  & (K$-$L)&  \multicolumn{2}{c}{[M$_{\sun}$]}  &  \multicolumn{2}{c}{[Myr]} &   	\\
\noalign{\smallskip}
\hline
\noalign{\smallskip}
CGH$\alpha$ 5/6 ........      	  
&    A-B 	& CGH$\alpha$ 6	&	&  K7        	& ... & 24  	&      5.0  &  ...  	&   W &  24   &    & \multicolumn{2}{c}{...} &  \multicolumn{2}{c}{...} &  \\
&     A 	& 			 	&   	&  K7  	&  & 24  	&  ...	       &  ...   	&  ...   &  	    &     &\multicolumn{2}{c}{0.79\,$^{+0.09}_{-0.03}$}&\multicolumn{2}{c}{$<$\,1}& 1\\
&     B 	& 			 	&   	&  M0-4  	&  & 1  	&  ...	       &  ...   	&  ...   &  	    &     &\multicolumn{2}{c}{0.43\,$^{+0.26}_{-0.14}$}&\multicolumn{2}{c}{$<$\,1}& 1\\
&     C 	& CGH$\alpha$ 5 	&   	&  K2-5  	& ... & 24  	&  126.9  &  ...   	&   C  &  24   &    &\multicolumn{2}{c}{1.31\,$^{+0.09}_{-0.3}$}&\multicolumn{2}{c}{5\,$^{+8}_{-4}$}& 1\\
\noalign{\smallskip}
VW Cha ...........            	  
&     A              	& 	& 575    & K5-7   & 1.37 &  2  	&      7.9-116  &  ...	&   C 	&   2-3  &    &\multicolumn{2}{c}{1.00}&\multicolumn{2}{c}{0.4} & 21 \\
&     B-C           	&	&            &   K7     & 1.23 &  2  	&      0.48   & 	...	&   W   	&   2     &     &\multicolumn{2}{c}{0.40-0.35} & \multicolumn{2}{c}{0.4}& 21 \\
&     D 		&	Sz 23  &&   M2.5 &  6.4  &  8  	&      ...       	 & 	...	&   C$^a$ &           &     &\multicolumn{2}{c}{0.20} & \multicolumn{2}{c}{0.3} & 8 \\
\noalign{\smallskip}
Sz 30 .................       		  
&     A        &	&   	     &  M0.5  	& 0.58  &  2  		&    11.0      & 	...	&   C    &   2   &    &\multicolumn{2}{c}{0.48}&\multicolumn{2}{c}{2} & 2 \\
&     B        &	&   	     &  M2     	& 0.19  &  2  		&      2.6      &       ...	&   W   &   2   &    &\multicolumn{2}{c}{0.30}&\multicolumn{2}{c}{2} & 2 \\
&     C$^{c1}$        &	&   &  M3-5 	&   & 1   	&      ...       	 &       ...	&   ...    &        &    &\multicolumn{2}{c}{0.23\,$^{+0.14}_{-0.10}$}&\multicolumn{2}{c}{4\,$^{+6}_{-3}$}&1\\
&     C$^{c2}$        &	&   &  K8-M3 	&   & 1   	&      ...       	 &       ...	&   ...    &        &    &\multicolumn{2}{c}{0.58\,$^{+0.23}_{-0.18}$}&\multicolumn{2}{c}{17\,$^{+47}_{-13}$}&1\\
\noalign{\smallskip}
LkH$\alpha$ 336 ..........    	  
&     A     &LkH$\alpha$ 336	&  190   &  K7    & 1.03 &  4,9  &      24.0-35.0   & 	...	&   C   	&    4,9   &   &\multicolumn{2}{c}{0.51} &\multicolumn{2}{c}{0.5} &  9 \\
&     B     &LkH$\alpha$ 336/c&  516  &  M0.5     & 2.62 &  4,9  &        5.1-8.0   & 	...	&   W   	&    4,9   &   &\multicolumn{2}{c}{0.52} &\multicolumn{2}{c}{2.2} &  9  \\
&     C$^{c1}$     &	&  	    &  M5-6	 & &  1  &        ...   & 	...	&   ...   	&         &   	  &\multicolumn{2}{c}{0.18\,$^{+0.10}_{-0.05}$}&\multicolumn{2}{c}{$<$\,1}&1 \\
&     C$^{c2}$     &	&  	    &  K3-9	 & &  1  &        ...   & 	...	&   ...   	&         &   	  &\multicolumn{2}{c}{1.18\,$^{+0.22}_{-0.42}$}&\multicolumn{2}{c}{4\,$^{+5}_{-3}$}&1 \\
\noalign{\smallskip}
J 4872 ...............        		  
&     A-B              &	&   	     &  K9   & ... &  5  		&        0.8   & 	...	&   W   &    5  &   &\multicolumn{2}{c}{...} &\multicolumn{2}{c}{...} &   \\
&     C-D             &	&   	     &  M1  & ... &  5  		&        4.2   & 	...	&   W   &    5  &   &\multicolumn{2}{c}{...} &\multicolumn{2}{c}{...} &  \\
&     A             	  &	&   	     &  K9  & &  5  		&        ...	 & 	...	&   ...	   &        &   &\multicolumn{2}{c}{0.78\,$^{+0.03}_{-0.06}$}&\multicolumn{2}{c}{10\,$^{+15}_{-6}$}&1\\
&     B             	  &	&   	     &  K5-M2 & &  1  		&        ...	 & 	...	&   ...	   &        &   &\multicolumn{2}{c}{0.71\,$^{+0.31}_{-0.23}$}&\multicolumn{2}{c}{10\,$^{+15}_{-6}$}&1\\
&     C             	  &	&   	     &  M1  	    & &  5  		&        ...	 & 	...	&   ...	   &        &   &\multicolumn{2}{c}{0.59\,$^{+0.14}_{-0.11}$}&\multicolumn{2}{c}{17\,$^{+18}_{-11}$}&1\\
&     D             	  &	&   	     &  M0-3    & &  1  		&        ... 	 & 	...	&   ...	   &        &   &\multicolumn{2}{c}{0.48\,$^{+0.23}_{-0.13}$}&\multicolumn{2}{c}{17\,$^{+18}_{-11}$}&1\\
\noalign{\smallskip}
UX Tau .............          		  
&     A                  &	UX Tau E		& 43      &  K4-5  &0.26&  5,6  	&     3.9-9.5    		&     0.65	&   C  	&  4-6  		&  6  	& 1.09 & 1.4 & 4.5 & 1.5 & 6 \\
&     B-C              &	UX Tau W		& 42      &  M2  	  &0.26&  5  	&     4.0-4.5    		&     -0.08	&   W   	& 4-6   		&  6  	&  0.52 & 1.4 & 5.9 & 1.9 & 6  \\
&     D                  &				&            &  M3-5  &0.57&  5,6  	&       8.5         		&     0.15	&   W   	&   6     		&  6 	&  0.16 & 1.3 & 1.6 & 4.1 & 6 \\
\noalign{\smallskip}
UZ Tau .............          		  
&     B-C              &	UZ Tau W		& 53      &  M3  	  & &  4,9  	&      80.0    		&       ...	&         &   7     	&         &  \multicolumn{2}{c}{...}  &\multicolumn{2}{c}{...} &   \\
&     B                  &				&            &  M2  	  & 0.55 &  6,20     	&      54.0       		&       0.68	&   C   &   20    	&   6   & 0.56 &1.3 & 2.6 & 1.8 & 6 \\
&     C                  &				&            &  M2-3  & 1.75 &  6,20     	&      97.0     		&       0.38	&   C   &   20    	&   6   & 0.52 & 1.3 & 3.2 & 1.8 & 6 \\
&     A                  &	UZ Tau E		&  52     &  M1  	   & 0.33 &  4,9,6  	&      74-82       		&       ...	&   C   &   6,7 	&    	  & 0.65 & 1.2 & 5.1 & 2.3 & 6 \\
\noalign{\smallskip}
FV Tau .............          		  
&     A     	            &				&  386  & K5   	    &5.40&  4,9,20 	&       6.0-15.0    	&    0.21	&   C    	&   6,20    &  6  & 1.12 & 1.3 & 3.7 & 1.5 & 6 \\
&     B	            &				&           & K6-M2   &5.40&  6,20   	&    41.0-63.0   		&    0.94	&   C    	&   6,20    &  6  & 0.89 & 1.3 & 6.3 & 1.7 & 6 \\
&     C	            &	FV Tau/c		&  387  & M2.5-3  &3.25&  4,9,20 	&    17.0-21.0     	&    0.18	&   W$^d$&   6,20    &  6   & 0.41 & 1.3 & 1.8 & 2.3 & 6 \\
&     D	            &	FV Tau/c		&           & M3.5-5  &7.00&  6,20  	&  224.0-800       	&    0.86	&   C$^e$ &   6,20    &  6  & 0.13 & 1.2 & 19 & 2.8 & 6 \\
\noalign{\smallskip}
GG Tau .............          		  
&     A &	GG Tau Aa	& 54     &  K7-M0      &0.30& 10,20   	&   $ >$42-57.0    	&    0.71	&  C  &   6,20 &   6   & 0.76 & 0.09 & 1.0 & 2.2 & 6 \\
&     B &	GG Tau Ab	&           &  M0.5-2     &0.45& 10,20   	&    16.0-21.0  		&    0.62	&  C  &   6,20 &   6   & 0.68 & 0.02 & 1.1 & 1.8 & 6 \\
&     C &	GG Tau Ba	&           &  M5.5        &0.00&  11  	&    21.0-22.0 		&    0.52	&  C  &   6,5   &   6    & 0.12 & 0.03 & 1.8 & 2.5 & 6 \\
&     D &	GG Tau Bb	&   	     &  M7.5  	&0.00&  11  	&    32.0-19.0  		&    0.43	&  C  &   6,5   &   6    & 0.042 & 0.019 &\multicolumn{2}{c}{$<$1} & 6 \\
\noalign{\smallskip}
LkH$\alpha$ 262/263...      
&    A-B && 9  	     & M3-4 &0.00& 13,14,23 & 30.0 &  0.32 &  :C  & 4,14 & 13  & \multicolumn{2}{l}{$\sim$0.45-0.45} &\multicolumn{2}{c}{$\sim$2}& 12 \\
&     A  &	&	     & M3	&&  13,14,23 & ...      & ...	&   ...  &    &    & \multicolumn{2}{c}{0.40\,$^{+0.15}_{-0.13}$} &\multicolumn{2}{c}{2\,$^{+4}_{-1}$}& 1 \\
&     B  &	&	     & M3	&&  13,14,23 & ...      & ...	&   ...  &    &    & \multicolumn{2}{c}{0.42\,$^{+0.17}_{-0.12}$} &\multicolumn{2}{c}{2\,$^{+4}_{-1}$}& 1 \\
&     C  &	&	     & M0	&& 12 		   & ...      & ...	&   C  &    &    & \multicolumn{2}{l}{$\sim$0.7} & \multicolumn{2}{c}{$\sim$2} & 12 \\
&     D  & LkH$\alpha$ 262 & 8  & M0   &1.17& 4,13,14 	&     31.0   &  0.62 &   C & 4,14 & 13 &  \multicolumn{2}{c}{0.69\,$^{+0.08}_{-0.09}$}& \multicolumn{2}{c}{2\,$^{+4}_{-1}$}& 1 \\
\noalign{\smallskip}

LkH$\alpha$ 346 ...........     	  
&     A-B-C 	  & LkH$\alpha$ 346 SE &   275    & M5  & 0.65 & 4  &  18.0  &  ... &  W  & 4  &  & \multicolumn{2}{c}{...} & \multicolumn{2}{c}{...} & \\
&     A 	  	  &  				&	&  M5    &&     1	&      ...   	   &  ...  &  ...  &  &  &  \multicolumn{2}{c}{0.28\,$^{+0.16}_{-0.12}$} &  \multicolumn{2}{c}{$<$\,1} & 1  \\
&     B 	  	  &  				&	&  M6-7 &&     1	&      ...	   &  ...  &  ...  &   &  &  \multicolumn{2}{c}{0.09\,$^{+0.03}_{-0.02}$} & \multicolumn{2}{c}{$<$\,1} & 1  \\
&     C 	  	  &  				&	&  M6-7 &&     1	&      ...	   &  ...  &  ...  &   &   &  \multicolumn{2}{c}{0.09\,$^{+0.03}_{-0.01}$} & \multicolumn{2}{c}{$<$\,1} & 1  \\
&     D		  & LkH$\alpha$ 346 NW &   275    & K7 & 1.16 & 4 & 28.0      &  ...  &  C  & 4 &   & \multicolumn{2}{c}{0.8\,$^{+0.16}_{-0.04}$} & \multicolumn{2}{c}{2\,$^{+5}_{-1}$} & 1  \\
\noalign{\smallskip}

ROXs 43 ..............         	  
&     A-B   & NTTS 162819-2423S &  &  G0         &2.3& 18,22       &    0.4-5.6 &  0.56-0.98 &  C   & 18 & 18-25  &  \multicolumn{2}{l}{$\sim$1.5$^b$} & \multicolumn{2}{c}{$\sim$7$^b$} & 22 \\
&     B       & 	&       		         &  K5-M2	&&  1         &   ... &  ...  &  ...    &       &        & \multicolumn{2}{c}{0.71\,$^{+0.25}_{-0.21}$} &  \multicolumn{2}{c}{$\sim$7$^b$} & 1 \\
&     C       & NTTS 162819-2423N & &  K3-K5	&1.3&  18,22 &  0.9	&  0.09-0.35   &  W  &  18 & 18-25 &  \multicolumn{2}{c}{$\sim$1.3$^b$} & \multicolumn{2}{c}{$\sim$3$^b$} & 22 \\
\noalign{\smallskip}
SR 24 .................          	  
&     A      &  SR 24S  &  262  &  K2	   &4.49& 4    & 76.0  &    0.99   &  C &  4 &  17  & \multicolumn{2}{c}{$>$ 1.4}& \multicolumn{2}{c}{1\,$^{+7}$} & 1\\
&     B-C  &  SR 24N &  262  &  M0.5	   &3.07& 4    & 24.0  &    0.69   &  C &  4 &  17  &  \multicolumn{2}{c}{ ...} & \multicolumn{2}{c}{ ...}  \\
&     B &   &      			    &  K4-M4 && 1    &   ...     &    1.38   &  C &     &   25  &\multicolumn{2}{c}{0.61\,$^{+0.6}_{-0.27}$} & \multicolumn{2}{c}{1\,$^{+7}$} & 1 \\
&     C &   &   			    &  K7-M5 && 1    &   ...     &    1.07   &  C &      &  25  &\multicolumn{2}{c}{0.34\,$^{+0.46}_{-0.18}$} & \multicolumn{2}{c}{1\,$^{+7}$} & 1 \\
\noalign{\smallskip}
Sz 68 ..................         	  
&     A-B  &   &  248 &   K2		&4.00&  3,16,19   &  6.8 &    0.26    &  C   & 3  &   16  & 2.5 & 0.1 & \multicolumn{2}{c}{$<$ 0.1} & 16  \\
&     B      &   &          &   K5-M1	&&  1 	            &   ...   &    ...	     &  ...   &     &          &\multicolumn{2}{c}{1.10\,$^{+0.40}_{-0.37}$}& \multicolumn{2}{c}{$<$ 0.1} & 1  \\
&     C      &   &          &   M6		&0.00&  16		   &    ...  &    0.12    &  W  &     &   16  & \multicolumn{2}{c}{0.1\,$^{+0.1}_{-0.02}$}& 0.2 & 0.1 & 16  \\
\noalign{\smallskip}
\hline
\noalign{\smallskip}
\end{tabular}
\end{center}
\label{Tab:systems_accretion}
\begin{minipage}[position]{18cm}
  Note\,: HBC = catalog entry number in Herbig \& Bell\,\cite{Herbig_Bell_1988}. \\
  $^a$\,: The T Tauri type classification of Sz 23 is from G\`{o}mez \& Mardones\,\cite{Gomez_Mardones_2003}.  \\
  $^b$\,: Estimated by eye from the plot of Bouvier \& Appenzeller\,\cite{Bouvier_Appenzeller_1992}.\\
  $^{c1}$\,: Coeval with component A. $^{c2}$\,: Coeval with component B.\\
  $^d$\,: Passive disk (McCabe et al.\,\cite{McCabe_etal2006}). \\
  $^e$\,; IRC candidate (Woitas et al.\,\cite{Woitas_etal2001}, see also McCabe et al.\,\cite{McCabe_etal2006}). \\
  References\,: 
  (1) This work.
  (2) Brandner \& Zinnecker\,\cite{Brandner_Zinnecker_1997}. 
  (3) Appenzeller et al.\,\cite{Appenzeller_1983} - Rydgren\,\cite{Rydgren_1980}. 
  (4) Cohen \& Kuhi\,\cite{Cohen_Kuhi_1979}. 
  (5) Duch\^{e}ne et al.\,\cite{Duchene_etal1999}. 
  (6) White \& Ghez\,\cite{White_Ghez_2001}. 
  (7) Cohen \& Kuhi\,\cite{Cohen_Kuhi_1979} - Edwards et al.\,\cite{Edwards_1987}. 
  (8) G\`{o}mez \& Mardones\,\cite{Gomez_Mardones_2003}. 
  (9) Hartigan et al.\,\cite{Hartigan_etal1994}. 
  (10) White et al.\,\cite{White_etal1999}. 
  (11) Luhman\,\cite{Luhman_1999}. 
  (12) Jayawardhana et al.\,\cite{Jayawardhana_etal2002}. 
  (13) Jayawardhana et al.\,\cite{Jayawardhana_etal2001}. 
  (14) Herbst et al.\,\cite{Herbst_etal2004}. 
  (15) Perryman et al. 1997. 
  (16) Prato et al.\,\cite{Prato_etal2003}. 
  (17) Rydgren 1976. 
  (18) Walter et al. 1994. 
  (19) Hughes et al. 1994.
  (20) Hartigan \& Kenyon\,\cite{Hartigan_Kenyon2003}.
  (21) Brandeker et al.\,\cite{Brandeker_etal2001}.
  (22) Bouvier \& Appenzeller\,\cite{Bouvier_Appenzeller_1992}.
  (23) Luhman\,\cite{Luhman_2001}.
  (24) Reipurth \& Pettersson\,\cite{Reipurth_Pettersson1993}. 
  (25) McCabe et al.\,\cite{McCabe_etal2006}.

 \end{minipage}
\end{table*}

The study of disk evolution in young multiple system allows to investigate samples of stars of various masses that are {\it a priori} coeval. 
It is however necessary to first understand the influence of multiplicity on disk evolution, i.e. in which conditions the disks can be considered to 
evolve in isolation in multiple systems. This has been the subject of several early studies among T Tauri binaries (see Monin et al.\,\cite{Monin_etal_2006_PPV} 
and references therein for a recent review). 
Only a few attempts to investigate disk evolution of higher order multiple 
systems have been achieved recently (White et al.\,\cite{White_etal2002}). This is however of particular interest since both close and 
wider pairs are usually present in such hierarchical systems and all components of a system are supposed to be coeval. 
In the following, we analyze the frequency of disks in our sample of triples/quadruples through the use of various 
disk and/or accretion diagnostics and compare to what is known for isolated binaries.

We performed a compilation of T Tauri types for both individual components and close pairs of the triple/quadruple 
systems from the available literature (Table\,\ref{Tab:systems_accretion}). Two criteria were used in order to 
assign these types\,: first, the equivalent width of the H$\alpha$ line with a threshold between WTTS and CTTS 
at about 10\,\AA, depending on the spectral type (Martin\,\cite{Martin_1998}, Hartigan \& Kenyon\,\cite{Hartigan_Kenyon2003}); 
second, the $\Delta$(K-L) color excess which was shown to correlate with both strong [OI] and H$\alpha$ emission lines (Edwards et al.\,\cite{Edwards_etal1993}). 
In the latter case $\Delta$(K-L)$>$0.4 indicates the presence of an optically thick accretion disk (Edwards et al.\,\cite{Edwards_etal1993}). 
The $\Delta$(K-L) color excess was computed from K-L measurements and A$_V$ available in literature (for which the reference 
is reported in Table\,\ref{Tab:systems_accretion}), spectral types either from the literature or estimated by us and intrinsic colors of 
Kenyon \& Hartmann (\cite{Kenyon_Hartmann_1995}). 
For some systems individual component T Tauri types are missing and only the global type of the usually tightest pair is available. 
However, one can consider that a tight pair presenting spectral WTTS properties is in fact composed of two WTTS, while no 
conclusion is possible in the case of a tight pair for which the composite spectrum shows CTTS properties. 
We recently obtained spatially-resolved spectroscopy for some of these tight systems and the results will be presented elsewhere.   
 
While the number of systems in Table\,\ref{Tab:systems_accretion} does not allow us to draw any firm conclusions, some trends can 
nevertheless be identified. Among the 9 systems with CTTS/WTTS information known and/or deduced for each component, 
one is presumably composed of only WTTS (J\,4872). 
Since this system does not contain information about differential disk evolution, we will focus on the remaining 8 systems (CG\,H$\alpha$\,5/6, VW\,Cha, 
UX\,Tau, UZ\,Tau, FV\,Tau, GG\,Tau, LkH$\alpha$\,346, SR\,24). Half of these\,\footnote{FV\,Tau, as discussed below in the text, is not a mixed system.} 
are systems of mixed type (i.e. at least one component with a different type). The fraction of mixed systems is even higher if one includes 
the 4 systems that are already mixed even without the full knowledge of all components' T Tauri types (i.e. Sz\,30, LkH$\alpha$\,336, ROXs\,43, Sz\,68). 
We end up with a fraction of mixed systems of $\sim$\,65\,\% (8/12), similar or even higher to what is found among binaries ($\sim$\,40\,\%, 
Monin et al.\,\cite{Monin_etal_2006_PPV}). 
The much higher fraction of Taurus systems among non-mixed systems (3/4, compared to 1/8 for mixed systems) supports the suggestion of 
a regional dependance of the fraction of mixed systems, or at least a lower fraction in Taurus as in the case of binaries (Monin et al.\,\cite{Monin_etal_2006_PPV}).
On the other hand, we do not find any difference in the distribution of estimated age in mixed- and non-mixed systems which could explain 
the existence of these two populations, i.e. the age of non-mixed systems is not significantly younger than that of mixed systems. 

Disk diagnostics in all close pairs (separation $\la$\,100\,AU) are consistent with what is expected from tidal truncation theory (Armitage et al.\,\cite{Armitage_1999}). 
All close CTTS pairs (GG\,Tau\,AB, UZ\,Tau\,BC, SR\,24\,BC, FV\,Tau\,AB) exhibit estimated mass ratios q$\ga$\,0.5. SR\,24\,BC with q$\sim$\,0.5 has the lowest value, 
and this particular case may need disk replenishment from a common envelope with material falling preferentially onto the secondary's disk 
(Bate \& Bonnell\,\cite{Bate_Bonnell1997}) in order to prevent the more truncated of the secondary disk to be dissipated prior that of the primary. 
In general, all four close CTTS pairs may well need disk replenishment in order for the highly truncated disks to sustain active accretion. 
While both GG\,Tau\,AB and SR\,24\,BC are known to be surrounded by circumbinary disks for replenishment (Dutrey et al.\,\cite{Dutrey_etal1994}, 
Andrews \& Williams\,\cite{Andrews_Williams2005}), the cases of FV\,Tau\,AB and UZ\,Tau\,BC seem less obvious since no circumbinary disk has 
been detected so far. A massive circumbinary disk is known to be surrounding the spectroscopic pair UZ\,Tau\,A located some 500\,AU from UZ\,Tau\,BC 
but it is very unlikely responsible for the resplenishment of the BC pair. It is interesting to note that only gas emission in the CO line was detected 
around the pair SR\,24\,BC (Andrews \& Williams\,\cite{Andrews_Williams2005}) which suggests that other circumbinary disks or common reservoir 
could have been missed by previous continuum millimeter surveys. The case of the FV\,Tau\,CD pair deserves special comment since it seems to be, 
together with VW\,Cha\,A-BC, a close pair of mixed type. While the latter is consistent with exhausted highly truncated secondary disks, FV\,Tau\,CD shows a 
WTTS-CTTS type despite an estimated q$\sim$\,0.25 which would imply a more tidally truncated secondary disk. However, one should note that this apparent contradiction 
can be solved if one considers that component C is in fact a passive disk (i.e. a non-accreting disk, e.g. McCabe et al.\,\cite{McCabe_etal2006}) and component D 
an IRC candidate (Woitas et al.\,\cite{Woitas_etal2001}, see also McCabe et al.\,\cite{McCabe_etal2006}). 
Therefore, while the origin of the IRC phenomenon is still under debate, we can speculate that the latter is likely to be more massive, i.e. the primary. 
This implies that the close pair FV\,Tau\,CD which has two disks would have a star mass ratio presumably closer to unity, hence in agreement with 
synchronized disk evolution. 

The fact that 4 mixed systems include both close {WTTS} pairs and well-separated components with disks seems to suggest that the highly truncated disks 
of these close pairs were not fed by a circumbinary envelope and hence became rapidily exhausted. Alternatively a reservoir could have originally existed in 
such multiple systems and be subsequently disrupted by dynamical interactions with the other component(s) of these systems.


\section{Summary}
\label{sect:summary}

In this paper, we reported on our survey for high-order multiplicity among wide visual Pre-Main Sequence (PMS) binaries conducted 
with NACO at the VLT. The main conclusions of our study are summarized as follows\,:

\noindent (1) Among the 58 PMS wide binaries surveyed, which comprises 52 T Tauri systems from various star-forming regions, we found 7 
triple systems (2 new) and 7 quadruple systems (1 new). The new close companions are most likely physically bound based 
on their probability of chance projection and on their position on a color-color diagram. Some systems might still be not physical and 
future spectroscopy and common-proper motion measurement will be able to give an answer soon. The corresponding degree of multiplicity 
among wide binaries is MF/wB=26.9\,$\pm$7.2\% in the projected separation range 0\farcs07-12$^{\prime\prime}$, with the largest contribution 
from the Taurus cloud. Considering a restricted sample composed of systems at distance 140-190\,pc, we obtained only a slightly different value 
of MF/wB=26.8\,$\pm$8.1\%, in the separation range 10/14 AU - 1700/2300 AU (30 binaries, 5 triples, 6 quadruples).

\noindent (2) A general trend of decreasing multiplicity with lower primary mass is found, consistent with the known decrease of multiplicity with primary mass 
mainly found among binaries (Sterzik \& Durisen \cite{Sterzik_Durisen2003}). 

\noindent (3) Comparison with previous multiplicity surveys focusing on the brightest sources with similar resolution and (perhaps slightly lower) sensitivity 
shows that there is a good agreement between our newly derived multiplicity frequency per wide binary and those derived from these surveys, in the common 
separation range $\sim$14-1700 AU, although a significant overabundance of quadruple systems compared to triple systems is apparent. 
Considering that 4 out of the 6 quadruples of our distance-limited sample are from Taurus and that sources of that region have a frequency of 
quadruples to triples $f_4$=39\,$\pm$17\% when combining the results from previous surveys with ours, indicates that this excess of quadruples is 
mainly due to Taurus. 

\noindent (4) Tentatively including the spectroscopic pairs to our restricted sample and comparing the multiplicity fractions to those 
measured from solar-type main-sequence stars in the solar neighborhood leads to the conclusion that both the ratio of triples to binaries ($f_3$) and the 
ratio of quadruples to triples ($f_4$) seems to be in excess among young stars. While the former is not statistically significant, and depends 
on the assumption of paucity of spectroscopic binaries with no third component, the latter may be statistically significant.

\noindent (5) Our multiplicity frequency per binary seems lower than current predictions from numerical simulations of multiple star formation, 
especially SPH simulations. Noticeably, however, a relatively good match is found with the numerical models of the dynamical evolution of 
young small N-body clusters. 

\noindent (6) Some systems might not be dynamically stable according to the criteria of Eggleton  \& Kiseleva (\cite{Eggleton_Kiseleva_1995}) and 
Mardling \& Aarseth (\cite{Mardling_Aarseth_2001}). 
However, the three apparent non-hierarchical triple systems could be real if one assumes some correlation in the orbital inclinations.  In one case, 
the age of the system could be young enough to explain this configuration.

\noindent (7) We performed a compilation of T Tauri types for both individual components and pairs of the triple/quadruple systems from the 
available literature in order to study the relative disk evolutions in these systems. With the caution of small number statistics we identified the following trends. 
The fraction of mixed systems is high, up to $\sim$\,65\,\%, similar to 
what is found for binaries. Likewise, the fraction of mixed systems in Taurus is low. Disk replenishment by a circumbinary envelope seems necessary 
in order to explain the actively accreting close pairs, while the coexistence of close disk-less pairs with other accreting components in some systems 
argue in favor of no replenishment or a disrupted envelope.  As for what is found for binaries, the disks seems to evolve in isolation at large separations 
and be affected by tidal truncation at small separations.   

\appendix

\section{Comments on individual objects}
\label{sect:comments_ind_objects}

\begin{figure}
\centering
\begin{tabular}{ccc}
\includegraphics[width=2.5cm]{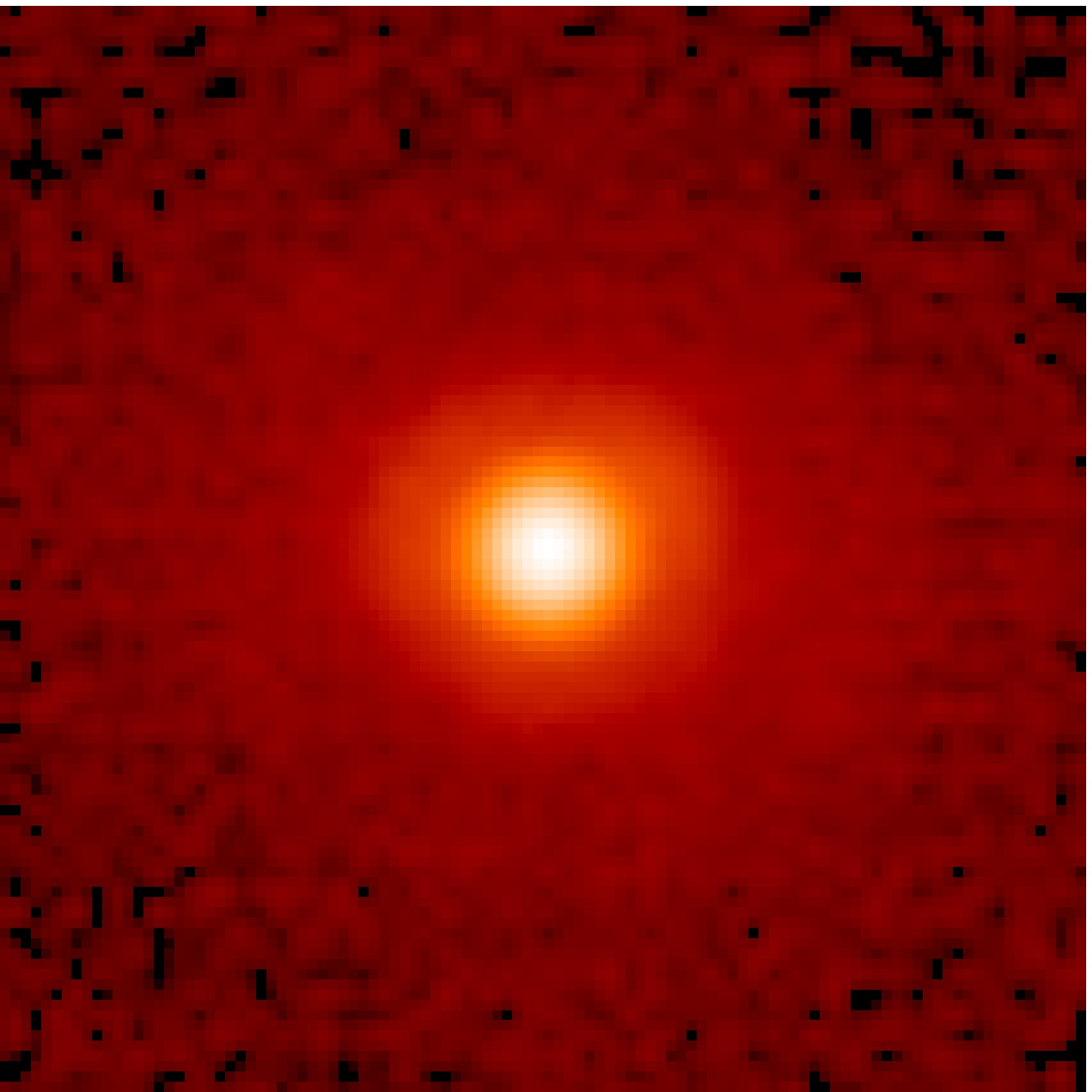}&
\includegraphics[width=2.5cm]{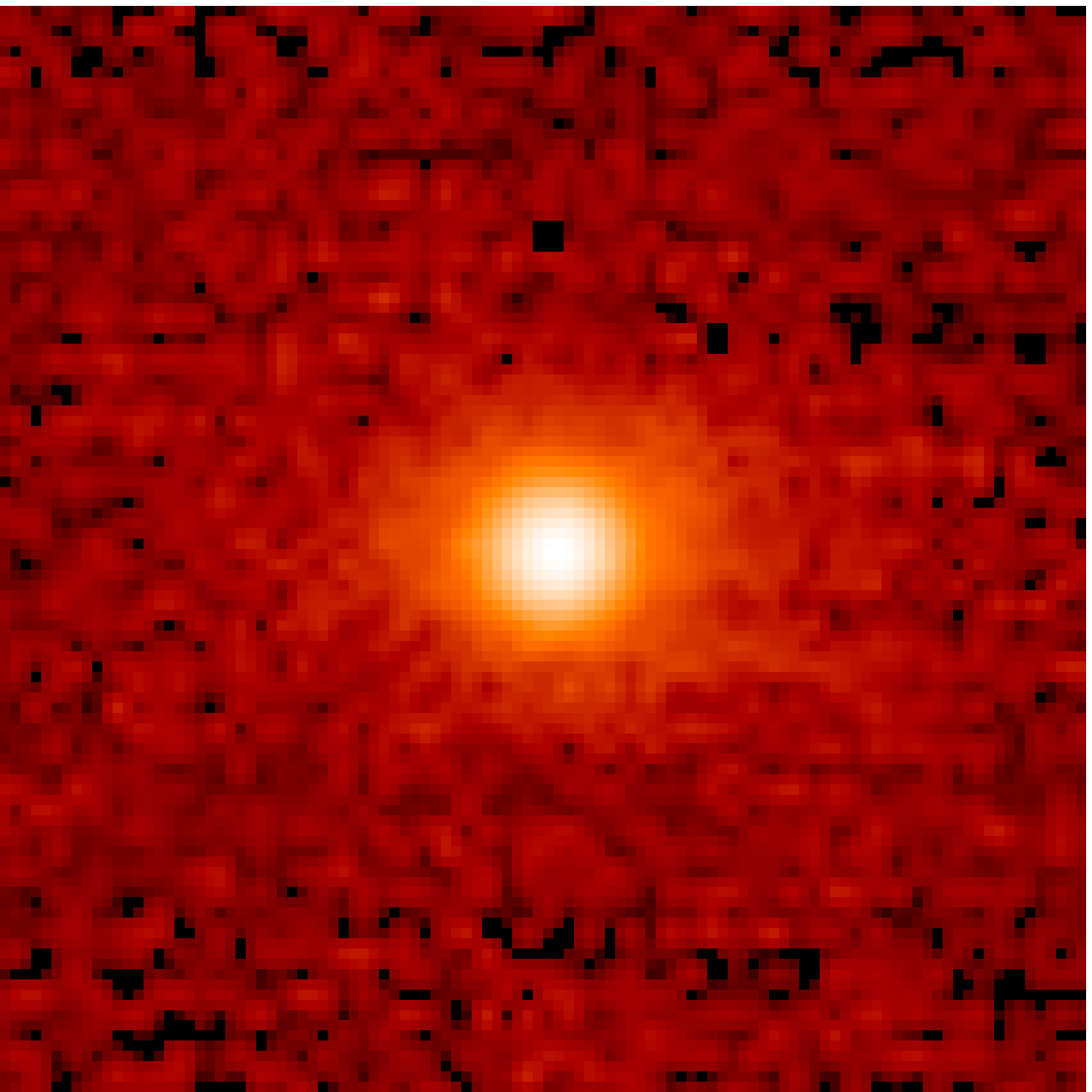}&
\includegraphics[width=2.5cm]{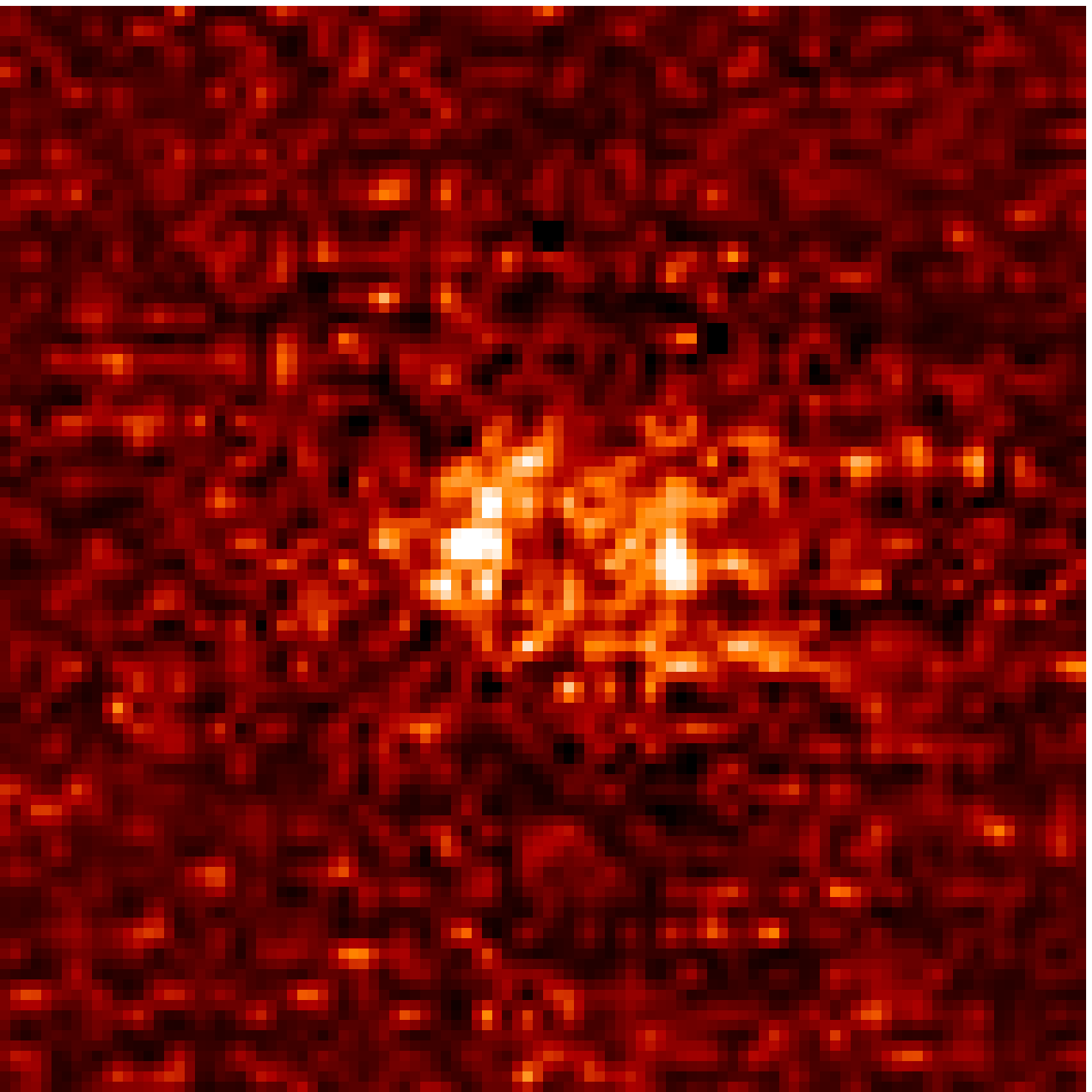}\\
\includegraphics[width=2.5cm]{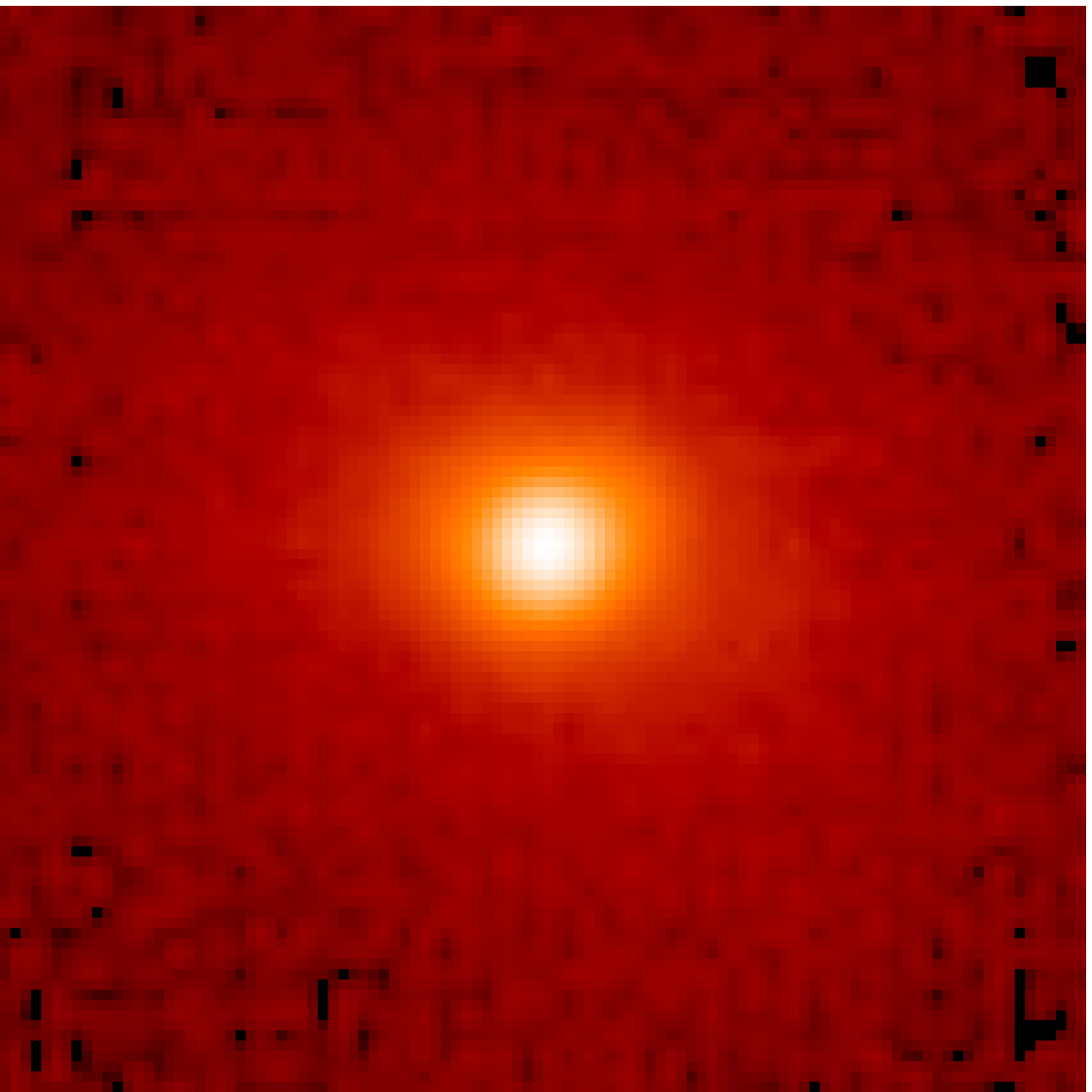}&
\includegraphics[width=2.5cm]{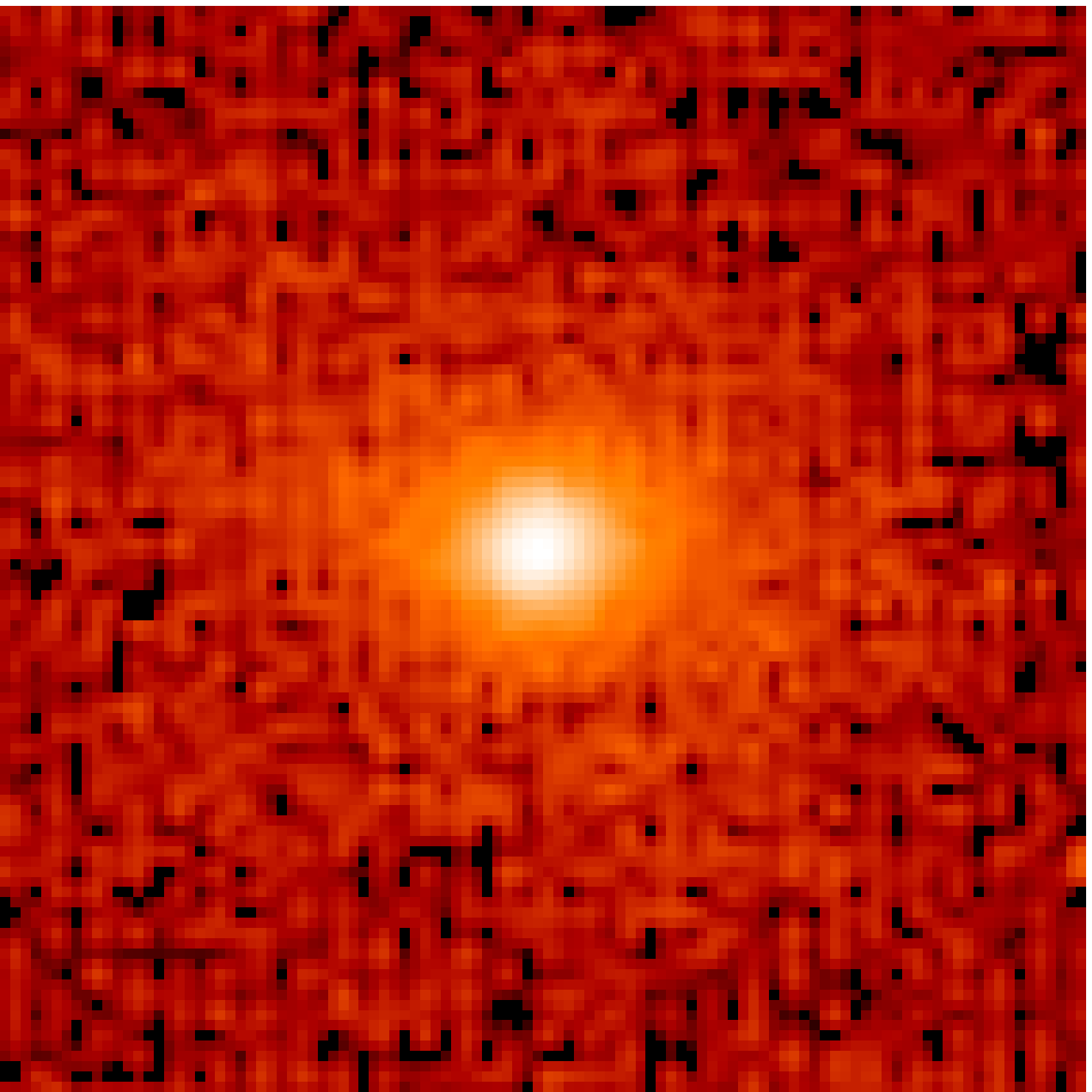}&
\includegraphics[width=2.5cm]{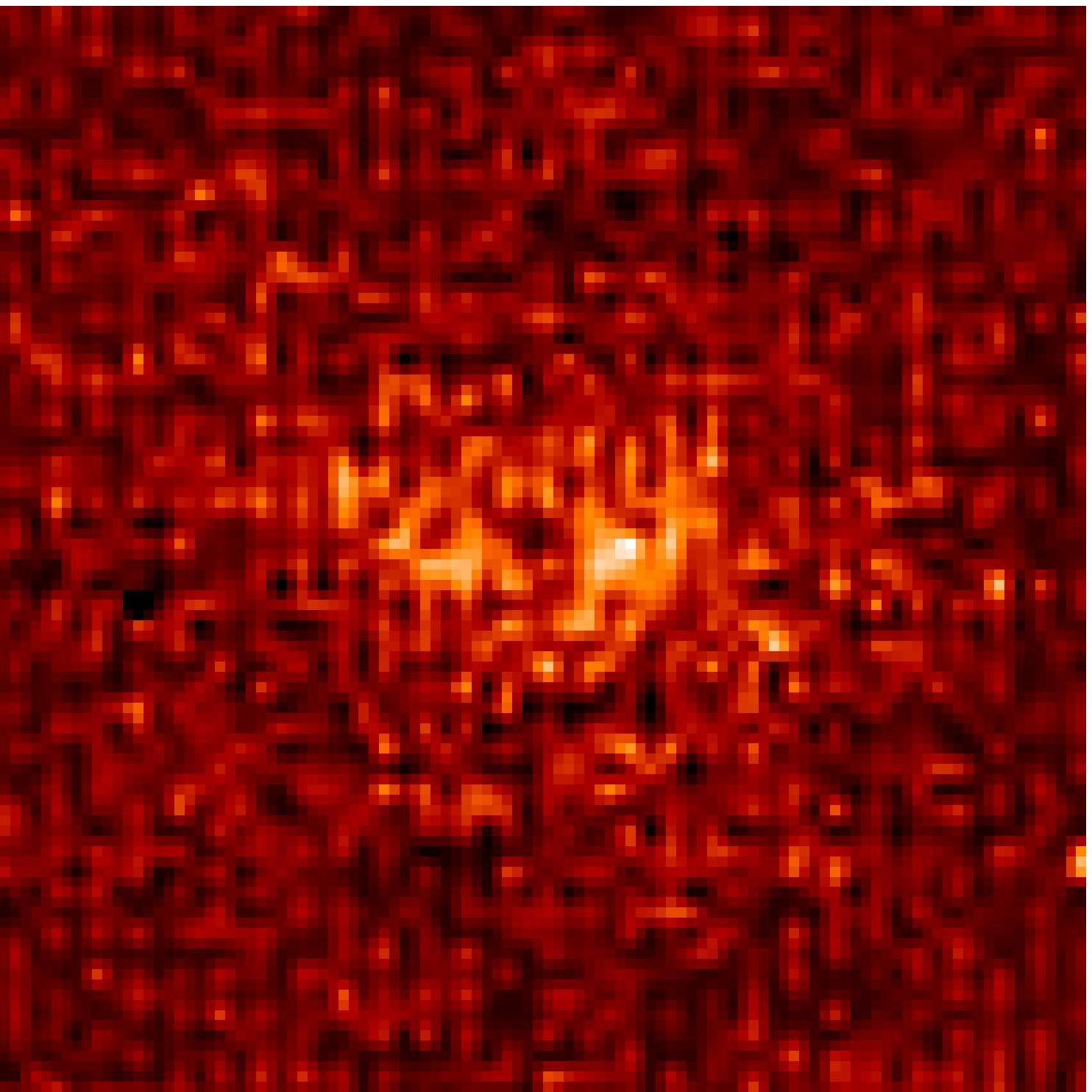}\\
\end{tabular}
\caption{Result of the PSF-subtraction of RW\,Aur\,B in Br$\gamma$ (top) and [FeII] (bottom). From left to right\,: primary and secondary (1/3 power stretch display)
and secondary minus primary (linear stretch, upper cut is 10-sigma of the residual noise). FOV is 1$\farcs$4$\times$1$\farcs$4.}
\label{fig:RW_Aur}
\end{figure}

{\bf RW Aur C}\,: The suspected companion of RW Aur B (0$\farcs$12 separation,  K-band flux ratio of 0.024), detected in NIR by Ghez et al. 
(\,\cite{Ghez_etal1993}), was not detected. Our 5-sigma detection limit in the PSF-subtracted frames at the given position are 
0.03, 0.02 and 0.12 in Br$\gamma$, H$_2$ and [FeII], respectively. This means that the companion would have been marginally detected 
in our data. As pointed out by Ghez et al. (\cite{Ghez_etal1997b}) and White \& Ghez (\cite{White_Ghez_2001}) who did not detect the close pair in their 
HST optical images either, several reasons can explain our non-detection. First, orbital motion may have decreased the separation below the 
resolution limit of NACO. The second explanation involves a drop of the flux of the companion since its detection in 1990 below our detection 
limit. Alternatively, the companion detected by Ghez et al. may in fact not be stellar but rather be the emission arising from shocked gas 
driven by a jet. RW Aur A is known to harbor an optical jet (Dougados et al.\,\cite{Dougados_etal2000}) but no Herbig-Haro object is known 
to originate from RW Aur B. 
Another possibility, favored by White \& Ghez (\cite{White_Ghez_2001}), is that this pair is spurious since the detection was obtained close to the detection limit 
and at relatively low SNR (Ghez et al.\,\cite{Ghez_etal1993}). The reason of this false detection may however be attributed to the presence of faint extended emission. 
In fact, the residuals of the PSF-subtraction clearly show some extended nebulosity around component B which could hardly be attributed to the difference of the 
AO-correction between the components of this 1$\farcs$45  pair (Fig.\,\ref{fig:RW_Aur}).

\noindent {\bf Sz 19 B}\,:  The faint companion of Sz 19 A (K-band flux ratio of 0.023, Ghez et al.\,\cite{Ghez_etal1997b}, Chelli et al.\,\cite{Chelli_etal1995}) 
is not detected at the expected separation and PA (4.9$\pm$0$\farcs$2, 202$\pm$3$^{\circ}$). Our 5-sigma detection limits at that position are 0.05, 0.04 and 0.04 
in Br$\gamma$, H$_2$ and [FeII], respectively. Instead, we detect a 3-sigma emission at $\sim$4$\farcs$6 separation, PA$\sim$202$^{\circ}$, which might be the 
companion. This is consistent with the values reported in RZ93 and the change of separation is within the uncertainties given by  Ghez et al. (\cite{Ghez_etal1997b}).

\noindent {\bf HK Tau B}\,:  This companion, which was resolved as an edge-on disk (Stapelfeldt et al.\,\cite{Stapelfeldt_etal1998}, Koresko \cite{Koresko_1998}), 
is only marginally detected in our data at the 3-5 sigma level. Only the northwest nebula is detected at separations and PAs, as determined by the component 
centroids, 2$\farcs$33, 171.6$^{\circ}$, 2$\farcs$29, 171.1$^{\circ}$ and 2$\farcs$31, 171.1$^{\circ}$ in Br$\gamma$, H$_2$ and [FeII], respectively. 
The position averaged over the three filters is 2$\farcs$31$\pm$0$\farcs$02 and PA=171.3$\pm$0.6$^{\circ}$. 
The southeast nebula is a factor $\sim$7 and $\sim$8 fainter in integrated flux than its northwest counterpart in H and K, 
respectively (Koresko \cite{Koresko_1998}), hence its non-detection.
The position of the northwest nebula reported here is likely to be very similar to the position of the centroid of both nebulae, given their flux ratio. 
This position is slightly smaller than that reported in the visible HST/WFPC2 imaging of Stapelfeldt et al. (\cite{Stapelfeldt_etal1998}), who measured 
2$\farcs$38, PA=172$^{\circ}$. Since the flux ratios between nebulae in optical are similar to those in the near-infrared, the centroid in the optical would be coincident 
with that in the near-infrared. Leinert et al. (\cite{Leinert_etal1993}) also measured a somewhat larger separation of 2$\farcs$4$\pm$0$\farcs$1 
and PA=175$\pm$2$^{\circ}$. In fact, we re-measured the position of the disk in the HST/WFPC2 images of Stapelfeldt et al. (\cite{Stapelfeldt_etal1998}) and find a 
better agreement with our derived NIR positions with separations 2$\farcs$32$\pm$0$\farcs$01 and PA=171.7$\pm$0.2$^{\circ}$. 
The derived uncertainties take into account the formal error between the two images as well as plate scale and detector orientation uncertainties.
We additionally confirmed our NACO separation and PA values with higher SNR NACO images in Ks which were available in the ESO 
archive and for which we derived a separation of 2$\farcs$318$\pm$0$\farcs$006 and PA=171.3$\pm$0.5$^{\circ}$. All these values are consistent with 
an unchanged position of HK Tau B with respect to the primary.
 
\noindent {\bf LkH$\alpha$ 263C}\,:  This edge-on disk (Chauvin et al. \cite{Chauvin_etal2002}, Jayawardhana et al. \cite{Jayawardhana_etal2002}) was undetected 
for sensitivity reasons. Our 5-sigma detection limits at the expected position are 0.06, 0.05 and 0.08 in Br$\gamma$, H$_2$ and [FeII], respectively. These are 
about an order of magnitude larger than the actual flux ratio of the AC pair, determined by Jayawardhana et al. to be 0.007 and 0.004 in H and K-band, respectively.

\noindent {\bf Sz15}\,:  Luhman (\cite{Luhman_2004}), who recently produced a census of the Cha I region using also a comprehensive list of cross 
identifications provided in Carpenter et al. (\cite{Carpenter_etal2002}), concluded from the lack of Li absorption of Sz\,15 (alias T19) and its low-extinction 
that it is actually a foreground star. Hartigan (1993) also noted the lack of H$\alpha$ which precludes its identification as a YSO. 

\noindent {\bf PH$\alpha$ 30 AB}\,: This system was indicated to be a background Be star based on an optical spectrum by Pettersson (\cite{Pettersson_1987}).

 \begin{figure}
\centering
\includegraphics[width=8.5cm]{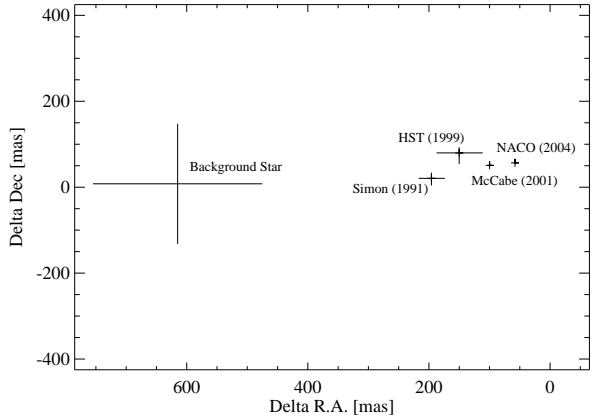}
\caption{Astrometric measurements of SR\,24\,C. Offset positions are shown with respect to SR\,24\,B. The expected position of a background star at the time 
of the NACO measurements (May 2004) is shown assuming it was located at the position measured by Simon et al. in August 1991.}
\label{fig:SR24N_orbit}
\end{figure}

\noindent {\bf SR\,24N\,: } By combining our new astrometric results with previous measurements we can study the relative motion of this close pair. 
We include the 1991 measurements of  Simon et al. (\cite{Simon_etal1995}) and measurements from 1999 unpublished archival HST/WFPC2 data 
in the FW606 and FW814 filters (HST proposal 7387, PI: K. Stapelfeldt). 
In addition, our measurements are confirmed by archival NACO data taken almost simultaneously to ours. The 1996 measurements of 
Costa et al. (\cite{Costa_etal2000}) are discarded from our analysis because of an obvious lack of accuracy. The locations of component C with 
respect to component B as a function of time are not consistent with a linear motion (Fig.\,\ref{fig:SR24N_orbit}). The NACO measurements unambiguously 
show that component C is in orbital motion around component B, rather than being an unrelated foreground/background object. Even if we don't consider the 1991 measurement, the resulting linear motion of velocity 19.0$\pm$6.5\,mas\,yr$^{-1}$ at position angle 255.9$^{\circ}$$\pm$13.1$^{\circ}$ is significantly different to the 
opposite of the proper motion of the system given by that of SR\,24S which is 33.0$\pm$22.7\,mas\,yr$^{-1}$ at position angle 271.7$^{\circ}$$\pm$39.3$^{\circ}$ 
(Ducourant et al.\,\cite{Ducourant_etal2005}). With the approach of the pericenter, it should be possible to derive a fairly accurate orbit in the next few years. 
Recently, McCabe et al. (\cite{McCabe_etal2006}) provide another measurement for an epoch intermediate between those of HST and NACO which fits 
quite nicely, although indicating that the marginally resolved HST measurement is of low quality.  

\noindent {\bf Sz\,41\,C}\,: Known as CHX 20a in the optical identification of Einstein observations (Feigelson \& Kriss\,\cite{Feigelson_Kriss_1989}), it was 
later identified by Walter (\cite{Walter_1992}) as a K0\,III star, with radial velocity and Li\,6707\,\AA~inconsistent with cloud membership.

\noindent {\bf VW\,Cha\,: } Like Brandeker et al. (\cite{Brandeker_etal2001}), we did not see the 2$\farcs$7 separated candidate infrared companion 
found by Ghez et al. (\cite{Ghez_etal1997b}). Our 5-sigma flux ratio detection limits at that position are $\sim$0.004 in all three narrow-band filters.

\begin{figure}
\centering
\includegraphics[width=9cm]{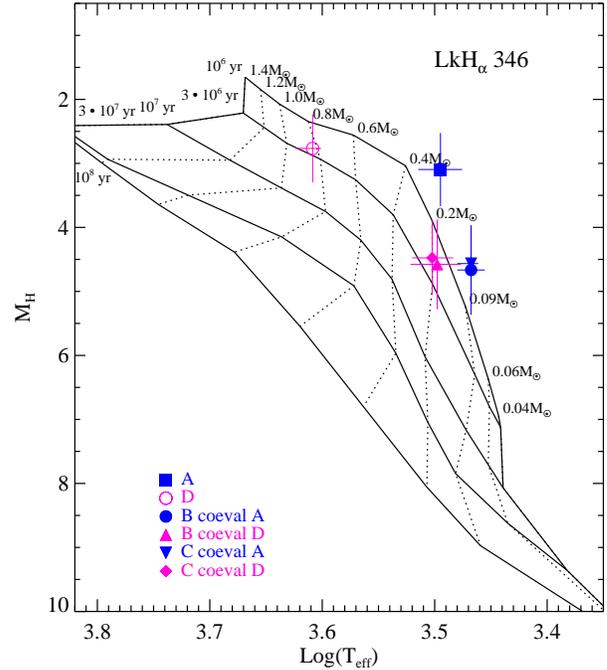}
\caption{Locus of the components of the LkH$\alpha$\,346 system in the (T$_{\mathrm{eff}}$, M$_{\mathrm{H}}$) plane together with the PMS evolutionary 
tracks and isochrones of Baraffe et al. (\cite{Baraffe_etal1998}). The components B and C are placed using their M$_{\mathrm{H}}$ and the assumption of 
coevality with either component A or D.}
\label{fig:HR_diag_LkHa346}
\end{figure}

\noindent {\bf LkH$\alpha$\,346\,: } We give here details of the derivation of estimated spectral types and masses for the two new candidate companions 
(B and C) using their H-mag and the assumption of coevality with the other components (A and D) of the system. For that purpose, we assume a distance of 
160\,pc and assign A$_V$=0.65 derived for LkH$\alpha$\,346\,SE (Cohen \& Kuhi \cite{Cohen_Kuhi_1979}, hereafter CK79) to components A, B, C and 
A$_V$=1.16 derived for LkH$\alpha$\,346\,NW  (CK79) to component D. 
The spectral type of component A is assumed to be M5, i.e. that of LkH$\alpha$\,346\,SE (CK79), while the spectral type of component D is assumed to 
be K7, as the one of LkH$\alpha$\,346\,NW (CK79). Converting to effective temperatures using the scale constructed by Sherry et al. (\cite{Sherry_etal2004}) 
gives T$_{\mathrm{eff}}$=3125\,K and 4060\,K for components A and D, respectively. 
The masses and ages of these two components are derived by comparison with the Baraffe et al. (\cite{Baraffe_etal1998}) stellar evolutionary models in the 
M$_{\mathrm{H}}$/T$_{\mathrm{eff}}$ plane (Fig.\,\ref{fig:HR_diag_LkHa346}), and are 0.28\,$^{+0.16}_{-0.12}$\,M$_{\odot}$ and $\la$\,1\,Myr for component A, 
0.8\,$^{+0.16}_{-0.04}$\,M$_{\odot}$ and 2\,$^{+5}_{-1}$\,Myr for component D. 
Because the component A falls above the youngest isochrone, we constructed an extrapolated isochrone going through that locus from which the mass was derived. 
If we assume that the new candidate companions are coeval with component A, to which they are closest, this leads to a mass of 
0.09\,$^{+0.03}_{-0.02}$\,M$_{\odot}$ and 0.09\,$^{+0.03}_{-0.01}$\,M$_{\odot}$ for B and C, respectively. The derived range of 
T$_{\mathrm{eff}}$ corresponds to a spectral type of M6-7 for both B and C. 
If we assume coevality with component D rather than with component A we obtain masses and spectral types of $\sim$\,0.2\,M$_{\odot}$ and $\sim$\,M4-6 for both 
components. 

Despite the large uncertainties involved in these mass and age estimates, a significant ($>$\,1-sigma) non-coevality between 
component A and D is indicated. Alternatively, the relative locus of component A with respect to component D could mean that the absolute magnitude is overestimated. 
This might be due in part to the H-band continuum excess which is not as negligable as the one in J-band (Greene \& Meyer \cite{Greene_Meyer_1995}).
However, estimating the luminosity of the original two components of the binary (SE \& NW, A and D component in our designation) 
via J magnitudes (or even visual magnitudes) and bolometric corrections leads to a similar age disparity. 
Using the luminosities derived by CK79 increases the age of the SE component ($\sim$\,2\,Myr) but gives an unrealistically large age for the 
NW component of $\sim$\,32\,Myr. On the other hand, only the formal errors are included, and it might well be that model uncertainties could be 
largely responsible for the discrepancy.
However, if one can consider the Baraffe et al. (\cite{Baraffe_etal1998}) set of PMS tracks accurate enough, an overestimate of the luminosity of 
component A would have the consequence for the candidate companions B and C to be located in the substellar regime. 
Another potential source of error is the spectral type of component A. It might well be that the close proximity of components B and C have 
biased the spectral classification of CK79 to a later-type. This could then reconcile the age estimate of component A with that of component D. 
Accurate spectral types for all components of this system are required to further investigate this point.
 
\noindent {\bf Hen\,3-600 (TWA\,3)\,: }  This binary exhibits a significant relative motion of $\sim$420\,mas at PA$\sim$143$^{\circ}$ since RZ93, which 
corresponds to $\sim$35.7\,mas\,yr$^{-1}$. This relative motion is not consistent with the proper-motion measured for the system of 137.8\,mas\,yr$^{-1}$ at 
PA=89.7$^{\circ}$ (Mamajek\,\cite{Mamajek_2005}) and must therefore originate from orbital motion. At the distance of the binary (50\,pc, Table\,\ref{Tab:sample}), 
the relative motion we measured corresponds to a linear relative motion of $\sim$8.5\,km\,s$^{-1}$, which is similar to a typical Keplerian velocity 
of 3.6\,km\,s$^{-1}$ for such separation ($\sim$70\,AU), considering a solar mass primary and a circular orbit in the plane of the sky. In addition, including the 
2000 measurement of Brandeker et al. (\cite{Brandeker_etal2003}, B03 hereafter) shows that there is a significant change in the PA of the relative motion from 
145$^{\circ}$ to 128$^{\circ}$, from RZ93 to B03 and B03 to our measurement, respectively. We therefore conclude that the pair is physical and exhibits 
orbital motion.


\section{Edge-on disks in multiple systems}
\label{sect:edge_on_disks}

Circumstellar disks in multiple T Tauri systems have only been imaged so far in the favorable detection case in which one disk is nearly seen edge-on. 
Three such systems exist\,: HK\,Tau\,B, LkH$\alpha$\,262/263\,C  and HV\,Tau\,C. The last two are in a system with more than 2 components 
(quadruple and triple, respectively). Disk inclinations are estimated to be $\sim$85$^{\circ}$, $\sim$87$^{\circ}$ and $\sim$84$^{\circ}$, while 
disk radius are $\sim$150\,AU, $\sim$150\,AU and $\sim$50\,AU, respectively. The following is an attempt to estimate lower and upper limits 
on the expected fraction of edge-on disks in systems and compare the latter to the observed fraction in our sample.

If we consider a random distribution of disk inclinations then the probability for a disk to be at inclination $\ga$\,85$^{\circ}$ is $\sim$9\%. 
Therefore, among all stars with a disk, 9\% should have their disk oriented edge-on, if disks in multiple systems are supposed to be independently oriented. 
In the case in which disks of a pair are {\it almost} coplanar, this fraction should be increased to roughly 9\,\% of the pairs, neglecting systems with exactly 
coplanar disks whose detection in the edge-on orientation suffers from strong selection effects. 
Indeed, pairs with both disks edge-on would be difficult to detect as these sources would be faint, but if the disks are slightly misaligned (a few degrees from the 
edge-on orientation is sufficient for the star to become visible) these pairs would appear to contain a faint companion (like e.g. HK\,Tau\,B) until high-angular resolution observations resolve its spatial extension.
After removal of non-cloud members and probable chance projections our survey counts 42 doubles, 7 triples and 7 quadruples, 
i.e. a total of 133 components and 77 pairs. Two of the known edge-on disks are included in our sample (HK\,Tau\,B and LkH$\alpha$\,262/263\,C), 
which leads to a fraction of components and pairs of 2/133=1.5$\pm$1.1\% and 2/77=2.6$\pm$1.8\%, respectively, i.e. more than 6- and 3-sigma lower 
than the above predicted fraction. 

However, the fraction of edge-on disks in our sample might be higher if we consider that\,: 
(1) they could not have been spatially resolved (this apply for the more distant objects and the smaller disk like the 50\,AU radius HV\,Tau\,C), 
(2) the sensitivity might not have been enough to detect the faintest one like LkH$\alpha$\,262/263\,C which was not detected here (while HK\,Tau\,C 
has been, Sect.\,\ref{sect:comments_ind_objects}), 
(3) the actual number of companions harboring a disk in pairs is a fraction of the total number of pairs, since T Tauri pairs are known to have a 
significant fraction of mixed T Tauri types, i.e. disk-less companions (see Sect.\,\ref{sect:disk_evolution}).   
On the other hand, although disks tend mostly to be coplanar with few exceptions as the latest polarimetric studies seems to indicate 
(Monin et al.\,\cite{Monin_etal_2006}), there are some indications for preferentially misaligned disks in high-order multiples 
(Monin et al.\,\cite{Monin_etal_2006_PPV}). This would have the effect to reduce the predicted upper-limit on the fraction of edge-on disks derived above 
assuming coplanar disks for all pairs, irrespective of their multiple system memberships ( the 9\,\% of the pairs). While the existence of a deficit of edge-on disk in 
our sample is difficult to establish firmly, the fact that 2/3 of the known edge-on disks are found in high-order multiple systems suggest that misaligned 
disks are more likely to occur in these systems compared to binaries because of internal dynamical evolution in unstable multiple systems (Monin et al.\,\cite{Monin_etal_2006_PPV}). 

It is also worth to remark that an infrared companion (IRC, e.g. Koresko et al.\,\cite{Koresko_etal1997}) candidate, namely FV\,Tau\,D, 
is included in our sample of high-order multiple systems. This is consistent with the frequency of $\sim$\,10\% with which this class of objects is 
thought to occur in the T Tauri multiple stars population. Together with V\,773\,Tau\,D, WL\,20\,S and T\,Tau\,Sa, there are an increasing number 
of IRCs\,\footnote{The total number of {\it bona fide} IRCs is only 9 so far (Duch\^{e}ne et al.\,\cite{Duchene_etal2003}).} that are members of high-order 
multiple systems, which may possibly hint at a relation between the IRC phenomenon and the degree of multiplicity. 
This might favor, at least for some of them, the scenario proposed some years ago by Reipurth (\cite{Reipurth_2000}).


\begin{acknowledgements}
We would like to thank the NACO service-mode observing team at Paranal Observatory. 
We are particularly grateful to R.D. Scholz for interesting discussions and helping with the translation 
from Russian of the Ambartsumian paper. The authors would also like to acknowledge A. Brandeker for 
fruitful discussions as well as T. Kouwenhoven and G. Duch\^{e}ne for useful comments on the manuscript. 
We also wish to thank an anonymous referee for useful comments. 
This research has made use of the SIMBAD database, operated at CDS, Strasbourg, France. 
This publication makes use of data products from the Two Micron All Sky Survey, which is a joint 
project of the University of Massachusetts and the Infrared Processing and Analysis Center/California 
Institute of Technology, funded by the National Aeronautics and Space Administration and the National 
Science Foundation.

\end{acknowledgements}



\begin{thebibliography}{}
\bibitem[2005]{Alencar_etal2005} Alencar, S.H.P., Basri, G., Hartmann, L. et al. 2005, A\&A, 440, 595.
\bibitem[1954]{Ambartsumian_1954} Ambartsumian, V.A. 1954, Soobscenija Bjurakanskoj Observatorii, 15, 3.
\bibitem[2005]{Andrews_Williams2005} Andrews, S.M. \& Williams, J.P. 2005, ApJ, 619, L175.
\bibitem[1983]{Appenzeller_1983}Appenzeller, I., Jankovics, I. and Krautter, J. 1983, Astr.Ap. Suppl. 53, 291.
\bibitem[1999]{Armitage_1999} Armitage, P.J., Clarke, C.J., Tout, C.A. 1999, MNRAS, 304, 425.
\bibitem[1998]{Baraffe_etal1998} Baraffe, I., Chabrier, G., Allard, F. et al. 1998, A\&A, 337, 403.
\bibitem[1997]{Bate_Bonnell1997} Bate, M.R. \& Bonnell, I.A. 1997, MNRAS, 285, 33.
\bibitem[2003]{Bate_etal2003} Bate, M.R., Bonnell, I.A., Bromm, V. 2003, MNRAS, 339, 577.
\bibitem[1973]{Batten_1973} Batten, A.H. 1973, Binary and Multiple Star Systems (Pergamon Press, Oxford).
\bibitem[1988]{Bessel_Brett1988} Bessel, M.S. \& Brett, J.M. 1988, PASP, 100, 1134.
\bibitem[2006]{Beust_Dutrey_2006} Beust, H. \& Dutrey, A. 2006, A\&A, 446, 137.
\bibitem[1995]{Bodenheimer_1995} Bodenheimer, P. 1995, ARA\&A, 33, 199.
\bibitem[1991]{Bonnell_Bastien1991} Bonnell, I. \& Bastien, P. 1991, ApJ, 374, 610.
\bibitem[1994]{Bonnell_1994} Bonnell, I.A. 1994, MNRAS, 269, 837.
\bibitem[1986]{Boss_1986} Boss, A.P. 1986, ApJS, 62, 519.
\bibitem[1992]{Bouvier_Appenzeller_1992} Bouvier, J. \& Appenzeller, I. 1992, A\&AS, 92, 481.
\bibitem[2001]{Brandeker_etal2001} Brandeker, A., Liseau, R., Artymowicz, P., Jayawardhana, R. 2001, ApJ, 561, L199.
\bibitem[2003]{Brandeker_etal2003} Brandeker, Jayawardhana, R., Najita, J. 2003, AJ, 126, 2009.
\bibitem[1996]{Brandner_etal1996} Brandner, W., Alcal\'a, J.M., Kunkel, M. et al. 1996, A\&A, 307, 121.
\bibitem[1997]{Brandner_Zinnecker_1997} Brandner, W. \& Zinnecker, H. 1997, A\&A, 321, 220.
\bibitem[2000]{Brandner_etal2000} Brandner, W., Zinnecker, H., Alcal\'a, J.M. et al. 2000, AJ, 120, 950.
\bibitem[2002]{Carpenter_etal2002} Carpenter, J.M., Hillenbrand, L.A., Skrutskie, M.F. et al. 2002, AJ, 124, 1001.
\bibitem[2002]{Chauvin_etal2002} Chauvin, G., M\'enard, F., Fusco, T. et al. 2002, A\&A, 394, 949.
\bibitem[2005]{Chauvin_etal2005} Chauvin, G., Lagrange, A.-M., Dumas, C. et al. 2005, A\&A, 438, L25.
\bibitem[1995]{Chelli_etal1995} Chelli, A., Cruz-Gonzalez, Reipurth, B. 1995, A\&AS, 114, 135. 
\bibitem[1979]{Cohen_Kuhi_1979} Cohen, M. \& Kuhi, L.V. 1979, Ap.J. Suppl. 41, 743.
\bibitem[2000]{Costa_etal2000} Costa, A., Jessop, N.E., Yun, J.L. et al. 2000, in Birth and Evolution of Binary Stars, 
							ed. B. Reipurth \& H. Zinnecker, 
							Poster Proceedings of IAU Symposium 200, 48.
\bibitem[2002]{Davis_etal2002} Davis, C.J., Whelan, E., Ray, T.P. et al. 2002, A\&A, 397, 693.
\bibitem[2003]{Delgado_Donate_etal2003} Delgado-Donate, E.J., Clarke, C.J., Bate, M.R. 2003, MNRAS, 342, 926.
\bibitem[2004]{Delgado_Donate_etal2004} Delgado-Donate, E.J., Clarke, C.J., Bate, M.R. et al. 2004, MNRAS, 351, 617.
\bibitem[2000]{Dougados_etal2000} Dougados, C., Cabrit, S., Lavalley, C., M\'enard, F. 2000, A\&A, 357, 61. 
\bibitem[1999]{Duchene_etal1999} Duch\^{e}ne, G., Monin, J.-L., Bouvier, J., M\'enard, F. 1999, A\&A, 351, 954. 
\bibitem[1999]{Duchene_1999} Duch\^{e}ne, G. 1999, A\&A, 341, 547. 
\bibitem[2002]{Duchene_etal2002} Duch\^{e}ne, G., Ghez, A.M., McCabe, C. et al. 2003, ApJ, 592, 288.
\bibitem[2003]{Duchene_etal2003} Duch\^{e}ne, G., Ghez, A., McCabe, C. 2003, in Open Issues in Local Star Formation, 
							ed. J. L\'epine and J. Gregorio-Hetem, Astrophysics and Space Science Library, Vol. 299, 
							Kluwer Academic Publishers, 223.
\bibitem[2006]{Duchene_etal_ppv} Duch\^{e}ne, G., Delgado-Donate, E., Haisch Jr., K.E. et al. 2006, PPV conference. 
\bibitem[2005]{Ducourant_etal2005} Ducourant, C., Teixeira, R., P\'eri\'e, J.P. et al. 2005, A\&A, 438, 769.
\bibitem[1994]{Dutrey_etal1994} Dutrey, A., Guilloteau, S., Simon, M. 1994, A\&A, 286, 149. 
\bibitem[1991]{Duquennoy_Mayor1991} Duquennoy, A. \& Mayor, M. 1991, A\&A, 248, 485, (DM91).
\bibitem[1987]{Edwards_1987} Edwards, S. et al. 1987, ApJ, 321, 473.
\bibitem[1993]{Edwards_etal1993} Edwards, S., Ray, T., Mundt, R. 1993, 
						        in Protostars and Planets III, ed. E. H. Levy \& J. I. Lunine (Tucson: Univ. Arizona Press), 567. 
\bibitem[1995]{Eggleton_Kiseleva_1995} Eggleton, P. \& Kiseleva, L. 1995, ApJ, 455, 640.
\bibitem[1989]{Feigelson_Kriss_1989} Feigelson, E.D. \& Kriss, G.A. 1989, ApJ, 338, 262.
\bibitem[1981]{Fekel_1981} Fekel, F.C. 1981, ApJ, 246, 879.
\bibitem[1999]{Gahm_etal1999} Gahm, G.F., Petrov, P.P., Duemmler, R. et al. 1999, A\&A, 352, L95.
\bibitem[1993]{Ghez_etal1993} Ghez, A. M., Neugebauer, G., Matthews, K. 1993, AJ, 106, 2005.
\bibitem[1997a]{Ghez_etal1997a} Ghez, A.M., McCarthy, D.W., Patience, J.L. et al. 1997, ApJ, 481, 378.
\bibitem[1997b]{Ghez_etal1997b} Ghez, A.M., White, R.J., Simon, M. 1997, ApJ, 490, 353.
\bibitem[2003]{Gomez_Mardones_2003} G\`{o}mez, M. \& Mardones, D. 2003, AJ, 125, 2134.
\bibitem[2004]{Goodwin_etal2004} Goodwin, S.P., Whitworth, A.P., Ward-Thompson, D. 2004, A\&A, 414, 633.
\bibitem[1998]{Gray_Corbally_1998} Gray, R.O., Corbally, C.J. 1998, AJ, 116, 2530.
\bibitem[1994]{Gredel_1994} Gredel, R. 1994, A\&A, 292, 580.
\bibitem[1995]{Greene_Meyer_1995} Greene, T.P. \& Meyer, M.R. 1995, ApJ, 450, 233.
\bibitem[1975]{Harrington_1975} Harrington, R.S. 1975, AJ, 80, 1081.
\bibitem[1994]{Hartigan_etal1994} Hartigan, P., Strom, K.M., Strom, S.E. 1994, ApJ, 427, 961.
\bibitem[2003]{Hartigan_Kenyon2003} Hartigan, P. \& Kenyon, S.J. 2003, ApJ, 583, 334.
\bibitem[2004]{Hartigan_etal2004} Hartigan, P., Edwards, S., Pierson, R. 2004, ApJ, 609, 261.
\bibitem[2000]{Hearty_etal2000} Hearty, T., Fern\'andez, M., Alcal\'a, J.M. et al. 2000, A\&A, 357, 68.
\bibitem[1988]{Herbig_Bell_1988} Herbig, G.H. \& Bell, K.R. 1988, Lick Obs. Bull., 1111, 1.
\bibitem[1975]{Herbst_1975} Herbst, W. 1975, AJ, 80, 683.
\bibitem[2004]{Herbst_etal2004} Herbst, W., Williams, E.C., Hawley, W.P. 2004, AJ, 127, 1594.
\bibitem[1989]{Heyer_Graham_1989} Heyer, M.H., Graham, J.A. 1989, PASP, 101, 816.
\bibitem[1994]{Hirth_etal1994} Hirth, G.A., Mundt, R., Solf, J. et al. 1994, ApJ, 427, L99.
\bibitem[1992]{Hughes_Hartigan_1992} Hughes, J. \& Hartigan, P. 1992, AJ, 104, 680.
\bibitem[1994]{Hughes_etal1994} Hughes, J., Hartigan, P., Krautter, J. et al. 1994, AJ, 108, 1071. 
\bibitem[2001]{Jayawardhana_etal2001} Jayawardhana, R., Wolk, S.J., Barrado y Navascu\'es, D. et al. 2001, ApJ, 550, L197.
\bibitem[2002]{Jayawardhana_etal2002} Jayawardhana, R., Luhman, K.L., D'Alessio, P. et al. 2002, ApJ, 571, L51.
\bibitem[1997]{Jensen_Mathieu_1997} Jensen, E.L.N. \& Mathieu, R.D. 1997, AJ, 114, 301.
\bibitem[1995]{Kenyon_Hartmann_1995} Kenyon, S.J. \& Hartmann, L. 1995, ApJS, 101, 117.
\bibitem[2005]{Kim_etal2005} Kim, J.S., Walter, F.M., Wolk, S.J. 2005, AJ 129, 1564. 
\bibitem[1998]{Koehler_Leinert1998} K\"{o}hler, R. \& Leinert, Ch. 1998, A\&A 331, 977. 
\bibitem[2000a]{Koehler_etal2000a} K\"{o}hler, R., Kasper, M., Herbst, T. 2000,  in Birth and Evolution of Binary Stars: 
							 Poster Proceedings of IAU Symp. 200, ed. B. Reipurth \& H. Zinnecker ( Potsdam: Astrophys. Inst.), 63. 
\bibitem[2000b]{Koehler_etal2000b} K\"{o}hler, R., Kunkel, M., Leinert, Ch. et al. 2000, A\&A 356, 541. 
\bibitem[2001]{Koehler_2001} K\"{o}hler, R. 2001, AJ 122, 3325. 
\bibitem[2006]{Koehler_etal2006} K\"{o}hler, R., Petr-Gotzens, M.G., McCaughrean, M.J. et al. 2006, A\&A in press, astro-ph/0607670. 
\bibitem[1997]{Koresko_etal1997} Koresko, C.D., Herbst, T.M., Leinert, Ch. 1997, ApJ, 480, 741.
\bibitem[1998]{Koresko_1998} Koresko, C.D. 1998, ApJ 507, L145.
\bibitem[2000]{Koresko_2000} Koresko, C.D. 2000, ApJ, 531, L147.
\bibitem[2002]{Koresko_2002} Koresko, C.D. 2002, AJ 124, 1082. 
\bibitem[1972]{Larson_1972} Larson, R.B. 1972, MNRAS, 156, 437.
\bibitem[2002]{Larson_2002} Larson, R.B. 2002, MNRAS, 332, L155.
\bibitem[1991]{Leinert_etal1991} Leinert, Ch., Haas, M., Mundt, R. et al. 1991, A\&A 250, 407.
\bibitem[1993]{Leinert_etal1993} Leinert, Ch., Zinnecker, H., Weitzel, N. et al. 1993, A\&A 278, 129.
\bibitem[1999]{Luhman_1999} Luhman, K.L. 1999, ApJ, 525, 466.
\bibitem[1999]{Luhman_Rieke_1999} Luhman, K.L. \& Rieke, G.H. 1999, ApJ, 525, 440.
\bibitem[2001]{Luhman_2001} Luhman, K.L. 2001, ApJ, 560, 287.
\bibitem[2004]{Luhman_2004} Luhman, K.L. 2004, ApJ, 602, 816.
\bibitem[2001]{Mardling_Aarseth_2001} Mardling, R.A. \& Aarseth, S.J. 2001, MNRAS, 321, 398.
\bibitem[1998]{Martin_1998} Mart\`{\i}n, E.L. 1998, AJ, 115, 351.
\bibitem[2003]{Masciadri_2003} Masciadri, E.,  Brandner, W., Bouy, H. et al. 2003, A\&A 411, 157.
\bibitem[1989]{Mathieu_etal1989} Mathieu, R.D., Walter, F.M., Myers, P.C. 1989, AJ, 98, 987.
\bibitem[1996]{Mathieu_etal1996} Mathieu, R.D., Mart\`{\i}n, E.L., Magazzu, A. 1996, BAAS, 188, 60.05.
\bibitem[2006]{McCabe_etal2006} McCabe, C., Ghez, A.M., Prato, L. et al. 2006, ApJ, 636, 932.
\bibitem[2004]{McGroarty_Ray_2004} McGroarty, F. \& Ray, T.P. 2004, A\&A, 420, 975.
\bibitem[2005]{Mamajek_2005} Mamajek, E.E. 2005, ApJ, 634, 1385.
\bibitem[2003]{Melo_2003} Melo, C.H.F. 2003, A\&A, 410, 269.
\bibitem[1993]{Meyer_etal1993} Meyer, M.R., Wilking, B.A., Zinnecker, H. 1993, AJ, 105, 619.
\bibitem[1997]{Meyer_etal1997} Meyer, M.R., Calvet, N., Hillenbrand, L.A. 1997, AJ, 114, 288.
\bibitem[1956]{Mestel_Spitzer1956} Mestel, L. \& Spitzer, L.Jr. 1956, MNRAS, 116, 503.
\bibitem[2000]{Monin_Bouvier_2000} Monin, J.-L. \& Bouvier, J. 2000, A\&A, 356, L75.
\bibitem[2006a]{Monin_etal_2006} Monin, J.-L., M\'enard, F., Peretto, N. 2006, A\&A, 446, 201.
\bibitem[2006b]{Monin_etal_2006_PPV} Monin, J.-L., Clarke, C.J., Prato, L. et al. 2006, PPV conference.  
\bibitem[1977]{Mouschovias_1977} Mouschovias, T. Ch. 1977, ApJ, 211, 147.
\bibitem[1998]{Mundt_Eisloeffel_1998} Mundt, R. \& Eisl\"{o}ffel 1998, AJ, 116, 860.
\bibitem[1924]{Opik_1924} \"{O}pik, E. 1924, Publ. Obs. Astron. Univ. Tartu, 25, 6.
\bibitem[2001]{Palla_Stahler2001} Palla, F. \& Stahler, S.W. 2001, ApJ, 553, 299.
\bibitem[2001]{Petrov_etal2001} Petrov, P.P., Gahm, G.F., Gameiro, J.F. et al. 2001, A\&A, 369, 993.
\bibitem[1987]{Pettersson_1987} Pettersson, B. 1987, A\&A, 171, 101.
\bibitem[2002]{Prato_etal2002} Prato, L., Simon, M., Mazeh, T et al. 2002, ApJ, 579, L99.
\bibitem[2003]{Prato_etal2003} Prato, L., Greene, T.P., Simon, M. 2003, ApJ, 584, 853.
\bibitem[1991]{Pringle_1991} Pringle, J.E. 1991, NATO-ASI Series C, Vol. 342, ed. CJ. Lada and N.D. Kylafis.
\bibitem[2005]{Pyo_etal2005} Pyo, T.-S., Hayashi, M., Naoto, K. et al. 2005, JKAS, 38, 249.
\bibitem[2005]{Ratzka_etal2005} Ratzka, T., K\"{o}hler, R., Leinert, Ch. 2005, A\&A 437, 611. 
\bibitem[1993]{Reipurth_Zinnecker1993} Reipurth, B. \& Zinnecker, H. 1993, A\&A, 278, 81, (RZ93).
\bibitem[1993]{Reipurth_Pettersson1993} Reipurth, B. \& Pettersson, B. 1993, A\&A, 267, 439.
\bibitem[2000]{Reipurth_2000} Reipurth, B. 2000, AJ, 120, 3177.
\bibitem[2002]{Reipurth_etal2002} Reipurth, B., Lindgren, H., Mayor, M. et al. 2002, AJ, 124, 2813.
\bibitem[2004]{Reipurth_Aspin_2004} Reipurth, B. \& Aspin, C. 2004, ApJ, 608, L65.
\bibitem[1985]{Rieke_Lebofsky_1985} Rieke, G.H. \& Lebofsky, M.J. 1985, ApJ, 288, 618.
\bibitem[2002]{Rousset_etal2002} Rousset, G., Lacombe, F., Puget, P. et al. 2002,
			in Adaptive Optical System Technologies II. ed. Wizinowich, P.L., Bonaccini, D., SPIE proc. Vol 4839, 140.
\bibitem[1980]{Rydgren_1980} Rydgren, A.E. 1980, Astr.J. 85, 444.
\bibitem[1987]{Sandell_etal_1987} Sandell, G., Reipurth, B., Gham, G. 1987, A\&A, 181, 283.
\bibitem[2004]{Sherry_etal2004} Sherry, W.H., Walter, F.M., Wolk, S.J. 2004, AJ, 128, 2316.
\bibitem[1987]{Shu_etal_1987} Shu, F.H., Adams, F.C. Lizano, S. 1987, ARA\&A, 25, 23.
\bibitem[1992]{Simon_etal1992} Simon, M., Chen, W.P., Howen, R.R. et al. 1992, AJ, 384, 212.
\bibitem[1995]{Simon_etal1995} Simon, M., Ghez, A., Leinert, Ch. et al. 1995, ApJ, 443, 625.
\bibitem[2004]{Simon_Prato_2004} Simon, M. \& Prato, L. 2004, ApJ, 613, 69.
\bibitem[1998]{Stapelfeldt_etal1998} Stapelfeldt, K.R., Krist, J.E., M\'enard, F. et al. 1998, ApJ, 502, L65.
\bibitem[2003]{Stapelfeld_etal2003} Stapelfeldt, K.R., M\''enard, F., Watson, A.M. et al. 2003, ApJ, 589, 410.
\bibitem[1998]{Sterzik_Durisen1998} Sterzik, M.F. \& Durisen, R.H. 1998, A\&A, 339, 95.
\bibitem[2002]{Sterzik_Tokovinin_2002} Sterzik, M.F. \& Tokovinin, A.A. 2002, 384, 1030.
\bibitem[2003]{Sterzik_Durisen2003} Sterzik, M.F. \& Durisen, R.H. 2003, A\&A, 400, 1031.
\bibitem[2003]{Sterzik_etal2003} Sterzik, M.F., Durisen, R.H., Zinnecker, H. 2003, A\&A, 411, 91.
\bibitem[2005]{Sterzik_etal2005} Sterzik, M.F., Melo, C.H.F., Tokovinin, A.A. et al. 2005, A\&A, 434, 671.
\bibitem[1987]{Stetson_1987} Stetson, P. B. 1987, PASP, 99, 191. 
\bibitem[2002]{Tohline_2002} Tohline, J.E. 2002, ARA\&A, 40, 349.
\bibitem[1997]{Tokovinin_1997} Tokovinin, A.A. 1997, A\&AS, 124, 75.
\bibitem[2001]{Tokovinin_2001} Tokovinin, A.A. 2001, in IAU Symp. 200, The Formation of Binary Stars, 
						ed. H. Zinnecker and R.D. Mathieu (San Francisco: ASP), 84.
\bibitem[2002]{Tokovinin_Smekhov_2002} Tokovinin, A.A. \& Smekhov, M.G. 2002, A\&A, 382, 118.
\bibitem[2006]{Tokovinin_etal2006} Tokovinin, A., Thomas, S., Sterzik, M. et al. 2006, A\&A, 450, 681.
\bibitem[2006]{Tokovinin_2006} Tokovinin, A. 2006, to appear in the proceedings of ESO Workshop "Multiple stars across the HR diagram", astro-ph/0601524.
\bibitem[2004]{Tokunaga_etal2004} Tokunaga, A.T, Reipurth, B., G\"{a}ssler, W. et al. 2004, AJ, 127, 444.
\bibitem[2003]{Torres_etal2003} Torres, G., Guenther, E.K., Marschall, L.A. et al. 2003, AJ, 125, 825.
\bibitem[1998]{Van_den_Ancker_etal1998} van den Ancker, M.E., de Winter, D., Tjin A Djie, H.R.E. 1998, A\&A, 330, 145.
\bibitem[2001]{VanDokkum_2001} Van Dokkum, P.G. 2001, PASP, 113, 1420. 
\bibitem[2003]{Viera_etal2003} Viera, S.L.A., Corradi, W.J.B., Alencar, S.H.P. 2003, AJ, 126, 297.
\bibitem[1992]{Walter_1992} Walter, F.M. 1992, AJ, 104, 758.
\bibitem[1999]{White_etal1999} White, R.J., Ghez, A.M., Reid, I.N., Schultz, G. 1999, ApJ, 520, 811.
\bibitem[2001]{White_Ghez_2001} White, R.J.  \& Ghez, A.M. 2001, ApJ, 556, 265.
\bibitem[2002]{White_etal2002} White, R.J., Hillenbrand, L., Metchev, S. et al. 2002, in BAAS, vol. 34, p. 1134. 
\bibitem[1974]{Whittet_1974} Whittet, D.C.B. 1974, MNRAS, 168, 371.
\bibitem[1997]{Whittet_etal1997} Whittet, D.C.B., Prusti, T., Franco, G.A.P., Gerakines, P.A. 1997, A\&A, 327, 1194.
\bibitem[1998]{Wichmann_etal1998} Wichmann, R., Bastian, U., Krautter, J. et al. 1998, MNRAS, 301, L39.
\bibitem[2001]{Woitas_etal2001} Woitas, J., Leinert, Ch., K\"{o}hler, R. 2001, A\&A, 376, 982.
\bibitem[1991]{Zinnecker_1991} Zinnecker, H. 1991, in IAU Symp. 147, Fragmentation of Molecular Clouds and Star Formation, 
							ed.  E. Falgarone, F. Boulanger, and G. Duvert, Kluwer Academic Publishers, 526.


\end{thebibliography}
\end{document}